\documentclass[final,3p,times,twocolumn,authoryear]{elsarticle}
\usepackage{graphicx}
\usepackage[T1]{fontenc}

\usepackage{hyperref}
\usepackage{ifthen}

\journal{New Astronomy}

\def\glc{{\sc Galacticus}}

\def\G{{\rm G}}
\def\clight{{\rm c}}
\def\d{{\rm d}}

\newcounter{AGNDone}
\def\AGN{\ifthenelse{\equal{\arabic{AGNDone}}{0}}{active galactic nuclei (AGN) \setcounter{AGNDone}{1}}{AGN}}

\newcounter{CDMDone}
\def\CDM{\ifthenelse{\equal{\arabic{CDMDone}}{0}}{cold dark matter (CDM) \setcounter{CDMDone}{1}}{CDM}}

\newcounter{CMBDone}
\def\CMB{\ifthenelse{\equal{\arabic{CMBDone}}{0}}{cosmic microwave background (CMB) \setcounter{CMBDone}{1}}{CMB}}

\newcounter{IMFDone}
\def\IMF{\ifthenelse{\equal{\arabic{IMFDone}}{0}}{initial mass function (IMF) \setcounter{IMFDone}{1}}{IMF}}

\newcounter{ISMDone}
\def\ISM{\ifthenelse{\equal{\arabic{ISMDone}}{0}}{interstellar medium (ISM) \setcounter{ISMDone}{1}}{ISM}}

\newcounter{IGMDone}
\def\IGM{\ifthenelse{\equal{\arabic{IGMDone}}{0}}{intergalactic medium (IGM) \setcounter{IGMDone}{1}}{IGM}}

\newcounter{ISCODone}
\def\ISCO{\ifthenelse{\equal{\arabic{ISCODone}}{0}}{innermost stable circular orbit (ISCO) \setcounter{ISCODone}{1}}{ISCO}}

\newcounter{NFWDone}
\def\NFW{\ifthenelse{\equal{\arabic{NFWDone}}{0}}{Navarro-Frenk-White (NFW) \setcounter{NFWDone}{1}}{NFW}}

\newcounter{SNeDone}
\def\SNe{\ifthenelse{\equal{\arabic{SNeDone}}{0}}{supernovae (SNe) \setcounter{SNeDone}{1}}{SNe}}

\newcounter{TdFDone}
\def\TdF{\ifthenelse{\equal{\arabic{TdFDone}}{0}}{\href{http://www.mso.anu.edu.au/2dFGRS/}{Two-degree Field Galaxy Redshift Survey} (2dFGRS) \setcounter{TdFDone}{1}}{2dFGRS}}

\newcounter{TMASSDone}
\def\TMASS{\ifthenelse{\equal{\arabic{TMASSDone}}{0}}{\href{http://www.ipac.caltech.edu/2mass/}{Two-Micron All Sky Survey} (2MASS) \setcounter{TMASSDone}{1}}{2MASS}}

\newcounter{SDSSDone}
\def\SDSS{\ifthenelse{\equal{\arabic{SDSSDone}}{0}}{\href{http://www.sdss.org/}{Sloan Digital Sky Survey} (SDSS) \setcounter{SDSSDone}{1}}{SDSS}}

\begin{document}

\begin{frontmatter}



\title{{\sc Galacticus}: A Semi-Analytic Model of Galaxy Formation}


\author{Andrew J. Benson}

\address{California Institute of Technology, MC350-17, 1200 E. California Blvd., Pasadena, CA 91125, U.S.A.}
\ead{abenson@caltech.edu}

\begin{abstract}
We describe a new, free and open source semi-analytic model of galaxy formation, \protect\glc. The \protect\glc\ model was designed to be highly modular to facilitate expansion and the exploration of alternative descriptions of key physical ingredients. We detail the \protect\glc\ engine for evolving galaxies through a merging hierarchy of dark matter halos and give details of the specific implementations of physics currently available in \protect\glc. Finally, we show results from an example model that is in reasonably good agreement with several observational datasets. We use this model to explore numerical convergence and to demonstrate the types of information which can be extracted from \protect\glc.
\end{abstract}


\end{frontmatter}

\section{Introduction}

The physics of galaxy formation is rich and complex, and has provided a challenge for theorists to understand and explain ever since it became clear that the Universe is filled with galaxies \citep{shapley__1921}. Various theoretical tools have been brought to bear on this problem, ranging from direct analytic reasoning to large scale numerical N-body simulations. An intermediate and highly successful approach is that known as ``semi-analytic'' galaxy formation modeling. In this approach the numerous complex non-linear physics involved are solved using a combination of analytic approximations and empirical calibrations from more detailed, numerical solutions. Historically, such models were first contemplated by \cite{white_core_1978} and have since been developed further, notably by \cite{white_galaxy_1991}, \cite{kauffmann_formation_1993}, \cite{cole_hierarchical_2000}, \cite{hatton_galics-_2003}, \cite{monaco_morgana_2007}, \cite{somerville_semi-analytic_2008} among others. Models of this type aim to begin with the initial state of the Universe (specified shortly after the Big Bang) and apply physical principles to determine the properties of galaxies in the Universe at later times, including the present day. Typical properties computed include the mass of stars and gas in each galaxy, broad structural properties (e.g. radii, rotation speeds, geometrical shape etc.), dark matter and black hole contents, and observable quantities such as luminosities, chemical composition etc. As with N-body/hydrodynamical simulations, the degree of approximation varies considerably with the complexity of the physics being treated, ranging from precision-calibrated estimates of dark matter merger rates to empirically motivated scaling functions with large parameter uncertainty (e.g. in the case of star formation and feedback).

The primary advantage of the semi-analytic approach is that it is computationally inexpensive compared to N-body/hydrodynamical simulations. This facilitates the construction of samples of galaxies orders of magnitude larger than possible with N-body techniques and for the rapid exploration of parameter space \citep{henriques_monte_2009,benson_galaxy_2010-1,bower_parameter_2010} and model space (i.e. adding in new physics and assessing the effects). The primary disadvantage is that they involve a greater degree of approximation. The extent to which this actually matters has not yet been well assessed. Comparison studies of semi-analytic vs. N-body/hydro calculations have shown overall quite good agreement (at least on mass scales well above the resolution limit of the simulation) but have been limited to either simplified physics (e.g. hydrodynamics and cooling only; \citealt{benson_comparison_2001,yoshida_gas_2002,helly_comparison_2003,lu_algorithms_2010}) or to simulations of individual galaxies \citep{stringer_analytic_2010}.

In this work, we describe a new, free and open source semi-analytic model, \glc. The \glc\ model solves the physics describing how galaxies evolve in a merging hierarchy of dark matter halos in a cold dark matter universe. \glc\ has much in common with other semi-analytic models, such as the range of physical processes included and the type of quantities that it can predict, but has some key distinguishing features. In designing \glc\ our main goal was to make the code flexible, modular and easily extensible. Much greater priority was placed on making the code easy to use and to modify than on making it fast. We believe that a modular and extensible nature is crucial as galaxy formation is an evolving science. In particular, key design features are:
\begin{description}
 \item [Extensible methods for all functions:] Essentially all functions within \glc\ are designed to be extensible, meaning that anyone can write their own version and easily insert it into \glc. For example, suppose an improved functional form for the \CDM\ halo mass function is derived. A user of \glc\ can simply write a short module conforming to a specified template that computes this mass function and which includes a short directive in the code which explains to \glc's build system how to incorporate this module. A recompile of the code will then incorporate the new function. The user is absolved from having to understand the details of the inner workings of the code, instead simply being required to conform to a standard interface.

 \item [Extensible components for tree nodes:] The basic structure in \glc\ is a merger tree, which consists of a set of linked tree nodes which have various properties. \glc\ works by evolving the nodes forward in time subject to a collection of differential equations and other rules. Each node can contain an arbitrary number of \emph{components}. For example, a component may be a dark matter halo, a galactic disk, a black hole etc. Each component may have an arbitrary number of \emph{properties} (some of which may be evolving, others of which can be fixed). \glc\ makes it easy to add additional components. For example, suppose that a user wanted to add a ``stellar halo'' component (consisting of stars stripped from satellite galaxies). To do this, they would write a module which specifies the following for this component:
 \begin{itemize}
  \item Number of properties;
  \item Interfaces to set and get property values and rates of change;
  \item ``Pipes'' which allow for flows of mass/energy/etc. from one component to another (and which facilitate interaction between components without the need for knowledge of the specific implementation of any component);
  \item Functions describing the differential equations which govern the evolution of the properties;
  \item Functions describing how the component responds to various events (e.g. the node becoming a satellite, a galaxy-galaxy merger, etc.);
  \item Auxiliary routines for handling outputs etc.
 \end{itemize}
 Short directives embedded in this module explain to the \glc\ build system how to incorporate the new component. A recompile will then build the new component into \glc. Typically, a new component can be created quickly by copying an existing one and modifying it as necessary. Furthermore, multiple implementations of a component are allowed. For example, \glc\ contains a component which is a Hernquist spheroid. One could add a de Vaucouler's spheroid component and an input parameter then allows one to simply select which implementation will be used in a given run.

 \item [Centralized ODE solver:] \glc\ evolves nodes in merger trees by calling an ODE solver which integrates forward in time to solve for the evolution of the properties of each component in a node. This means that it is not necessary to provide explicit solutions for ODEs (in many cases such solutions are not available anyway) and time-stepping is automatically handled to achieve a specified level of precision. The ODE solver allows for the evolution to be interrupted. A component may trigger an interrupt at any time and may do so for a number of reasons. A typical use is to actually create a component within a given node---for example when gas first begins to cool and inflow in a node a galactic disk component may be created. Other uses include interrupting evolution when a merging event occurs.
\end{description}

To summarize, the \glc\ code is therefore highly modular. Every part of it consists of a simple and well-defined interface into which an alternative implementation of a calculation can easily be added. Similarly, the physical description of galaxies is extremely flexible. Each galaxy has a set of components which can be created/destroyed as needed, each of which has a set of properties. These components are also modular, and any such module can be trivially replaced by another that performs the calculations differently if required. The actual formation and evolution of galaxies is treated by simply defining a set of differential equations for each galaxy. Then, these are all fed in to the ODE solver which evolves them to a specified accuracy.

The goal of this article is to describe the technical and physical implementation of \glc. Detailed examination of its scientific predictions and their consequences will be deferred to a future work. The remainder of this article is arranged as follows. In \S\ref{sec:Technical} we describe the technical implementation of \glc, while in \S\ref{sec:ImplementationComponents} we describe the current specific implementation of galactic components, and in \S\ref{sec:ImplementationPhysics} the current specific implementation of numerous physical processes and properties. In \S\ref{sec:Example} we show the results of an example calculation. Finally, in \S\ref{sec:Summary} we summarize the information presented in this article. The \glc\ model is available for download from \href{http://sites.google.com/site/galacticusmodel/}{{\tt http://sites.google.com/site/galacticusmodel}}. A full manual documenting both the physics and technical implementation of \glc\ can also be found on the same website.

\section{Technical Implementation}\label{sec:Technical}

In the following section we describe the technical and practical aspects of \glc. The \glc\ code is free and open source and was designed to depend only on free and open source compilers and libraries to maximize its portability. In addition to the \href{http://gcc.gnu.org/}{GNU compilers}, \glc\ requires the \href{http://www.gnu.org/software/gsl/}{GNU Scientific Library} (GSL) and the \href{http://www.lrz-muenchen.de/services/software/mathematik/gsl/fortran/}{FGSL} wrapper for GSL, the \href{http://uszla.me.uk/space/software/FoX/}{FoX} \href{http://www.w3.org/XML/}{XML} parser and the \href{http://hdf4.org/HDF5/}{HDF5} library. Additionally, the analysis codes provided with 
\glc\ make extensive use of \href{http://www.perl.org/}{Perl} and \href{http://pdl.perl.org/}{PDL}, although \glc\ output can be just as easily analyzed with other tools.

\subsection{Running}

The behavior of \glc\ is controlled by a set of parameters which are given in a file specified as a command line argument. (\glc\ will in fact provide sensible default values for all parameters if no parameter file is specified, or if some parameters are missing from the file). The parameter file is an XML file, which allows it to be generated easily using a variety of pre-existing XML tools and libraries. Scripts are provided with \glc\ to generate parameter files and run grids of models which span a range of parameter values.

When run, \glc\ proceeds by performing a set of predefined tasks in order until all tasks are done. As with all other aspects of \glc, these tasks are modular and extensible, allowing new tasks to be added in as desired. Typically, however, the tasks will consist of some initialization, followed by creation and evolution of one or more merger trees.

\subsection{Output}

\glc\ stores its output in an HDF5 file which allows this output to be viewed and manipulated using a variety of tools in addition to the standard HDF5 C API including:
\begin{description}
 \item[\href{http://www.hdfgroup.org/hdf-java-html/hdfview/}{{\sc HDFView}}] A graphical viewer for exploring the contents of HDF5 files;
 \item[\href{http://www.hdfgroup.org/products/hdf5_tools/index.html\#h5dist}{HDF5 Command Line Tools}] A set of tools which can be used to extract data from HDF5 files (\href{http://www.hdfgroup.org/HDF5/doc/RM/Tools.html#Tools-Dump}{{\tt h5dump}} and \href{http://www.hdfgroup.org/HDF5/doc/RM/Tools.html#Tools-Ls}{{\tt h5ls}} are particularly useful);
 \item[\href{http://www.hdfgroup.org/HDF5/doc/RM/RM_H5Front.html\#F90andCPPlus}{C++ and Fortran 90 APIs}] Allow access to and manipulation of data in HDF5 files from within C++ and Fortran90 codes;
 \item[\href{http://code.google.com/p/h5py/}{{\sc h5py}}] A Python interface to HDF5 files;
 \item[\href{http://search.cpan.org/~cerney/PDL-IO-HDF5-0.1/HDF5/Dataset.pm}{PDL::IO::HDF5}] A Perl interface to HDF5 files.
\end{description}

\subsubsection{Output Datasets}

In addition to the properties of galaxies and their associated dark matter halos in all merger trees, each \glc\ output file stores a full record of all parameter values (either input or default) that were used for the particular run. These can easily be extracted to an XML file suitable for re-input into \glc.
The output also contains a record of the \glc\ version used for this model, storing the major and minor version numbers, and the revision number along with the time at which the model was run.

Optionally, \glc\ will compute and store volume averaged properties of the entire galaxy population at a set of snapshot times. Currently, the properties stored are star formation rate density, stellar mass density, \ISM\ gas density and the density in resolved dark matter halos, all as a function of cosmic time.

Galaxy data can be output at one or more snapshot times, specified as input parameters to \glc. At each output time, each merger tree is stored separately within the HDF5 file to facilitate easy access. Each such merger tree group contains all data on a single merger tree, and consists of a collection of datasets each of which lists a property of all nodes in the tree which exist at the output time. Additionally, a weight (in Mpc$^{-3}$) which should be assigned to this tree (and all nodes in it) to create a volume-averaged sample is stored.

Optionally, a mass accretion history (i.e. mass as a function of time) for the main branch\footnote{``Main branch'' is defined by starting from the root node of a tree and repeatedly stepping back to the most massive progenitor of the branch. This does not necessarily select the most massive progenitor at a given time.} in each merger tree can be output, which lists the mass of the main branch as a function of time.

Finally, \glc\ can output the full structure of merger trees prior to any evolution. This is useful to permit the same trees to be used in other codes, or to examine how galaxy properties depend on merging history for example. If such output is requested the mass of each node in the tree is recorded, along with the cosmic time at which it exists, indices describing relationships between parent, child and sibling nodes and, if desired, quantities such as the virial radius, virial velocity and dark matter scale radius.

Additionally, \glc\ allows for arbitrary additional outputs to be easily implemented.

\subsubsection{Post-processing of Galaxy Properties}

\glc\ is provided with a Perl module that allows for easy extraction of datasets from a \glc\ output file together with a straightforward way to implement derived properties (i.e. properties computed by post-processing the output data). Implementations for simple dust-extinction calculations are provided which utilize this framework, together with modules that convert output luminosities into absolute magnitudes in AB or Vega systems. Any such derived properties can be stored back to the \glc\ output file.

\subsection{Node Evolution Engine}

\glc's main task is to evolve galaxies (and their associated dark matter halos) through a complex merging hierarchy. It begins with a pre-constructed dark matter halo merger tree as depicted in the ``Stage 1'' panel of Fig.~\ref{fig:Engine}. In this figure, colored circles represent nodes in the tree (larger circles implying greater mass total mass) and the $y$-axis represents time increasing upwards (such that $t_1>t_2>t_3$ etc.). Solid lines connect main progenitor nodes (typically the most massive) to their parents, while dotted lines connect other progenitor nodes (ones that will become subhalos) to their parents. Each node in the tree is given a unique ID number. \glc\ begins with the root node (number 1) and checks to see if it can be evolved forward in time. Since it has some progenitors, it cannot. Therefore, \glc\ steps to the primary progenitor (indicated by the dashed red arrow) and applies the same condition. This eventually leads it to node 4 which has no progenitor and so can be evolved forward in time (indicated by the dotted green arrow).

Once node 4 reaches the time at which its parent, node 3, is defined, node 4 has effectively become node 3 and is promoted, replacing node 3. This is a \emph{node promotion event} in \glc\ parlance. Since it still has no progenitors, \glc\ will continue to evolve node 4 until it reaches node 2 and is promoted to replace it, as indicated in ``Stage 2'' in Fig.~\ref{fig:Engine}. Since node 4 now has progenitors it cannot be evolved and \glc\ steps back to node 6, which it then evolves forward in time, until it is promoted to replace node 5. Following this, node 7 is evolved until it reaches node 6. Since node 7 was not the primary progenitor of node 6, it will become a subhalo within node 6 (subhalos are indicated by dot-dashed circles in Fig.~\ref{fig:Engine}). This is termed a \emph{node merger event} in \glc\ parlance, and is shown in ``Stage 3'' of Fig.~\ref{fig:Engine}. Nodes 6 and 7 are then evolved in lockstep until they reach node 4. \glc\ is able to handle nested substructure hierarchies, but the current implementation forces there to be a single level of subhalos. Therefore, both nodes 6 and 7 now become subhalos within node 4. 

Nodes 4, 6 and 7 now evolve in lockstep until ``Stage 4'' of Fig.~\ref{fig:Engine} is reached. Here, subhalo node 6 has reached the center of its host node 4 (due to the action of dynamical friction) and so any galaxies residing in nodes 4 and 6 will merge. This is a \emph{satellite merger event} in \glc\ parlance. After this, nodes 4 and 7 evolve in lockstep until node 4 is promoted to replace node 1 as shown in ``Stage 5'' of Fig.~\ref{fig:Engine}.

Finally, \glc\ descends the remaining branch of the merger tree and evolves it forward in time until ``Stage 6'' of Fig.~\ref{fig:Engine} is reached. Since all nodes are now at the final time, $t_1$, \glc\ stops evolving them and outputs any properties.

\begin{figure*}
 \begin{center}
 \begin{tabular}{cc}
 \includegraphics[width=80mm]{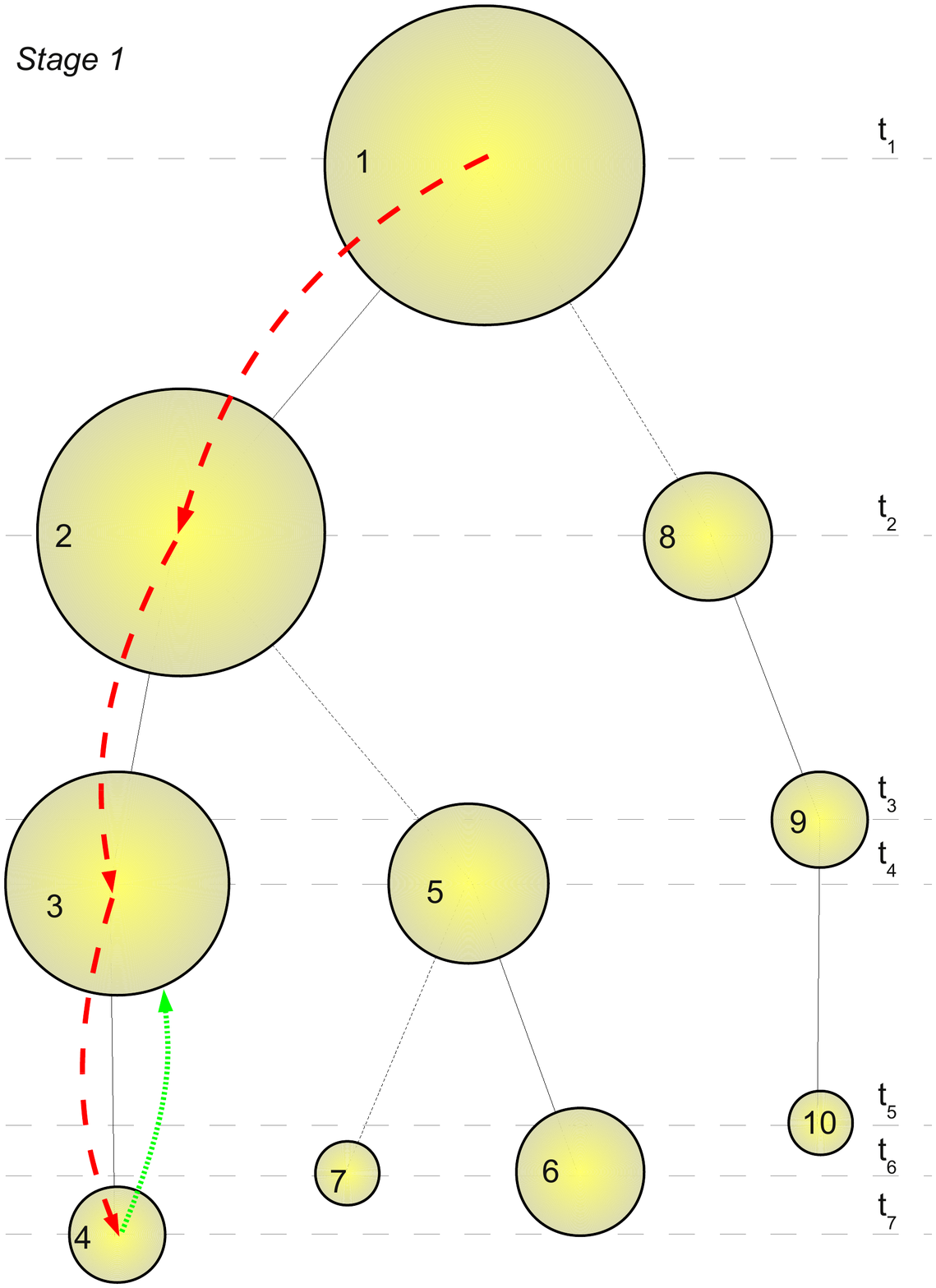} &
 \includegraphics[width=80mm]{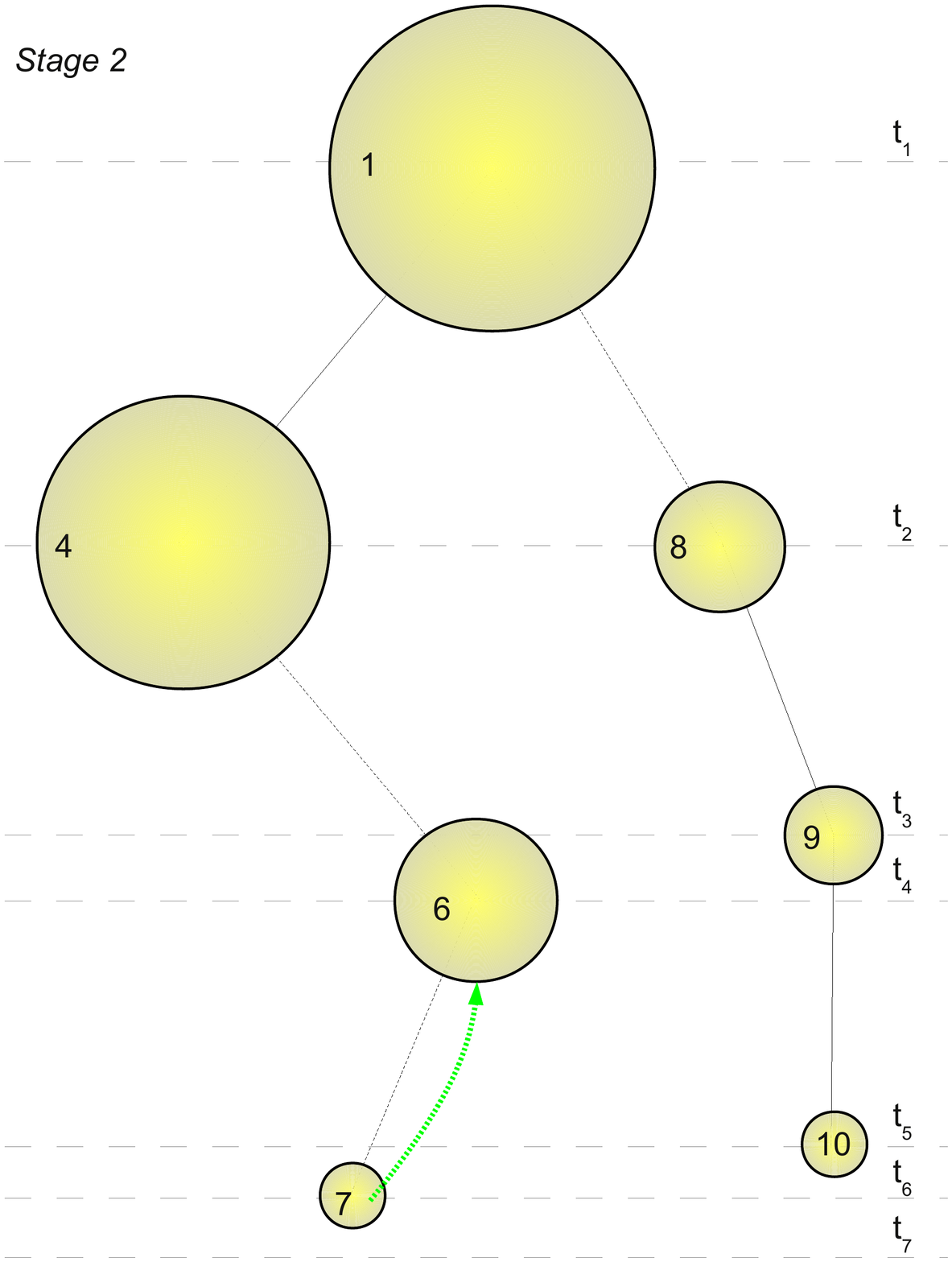} \\
 \includegraphics[width=80mm]{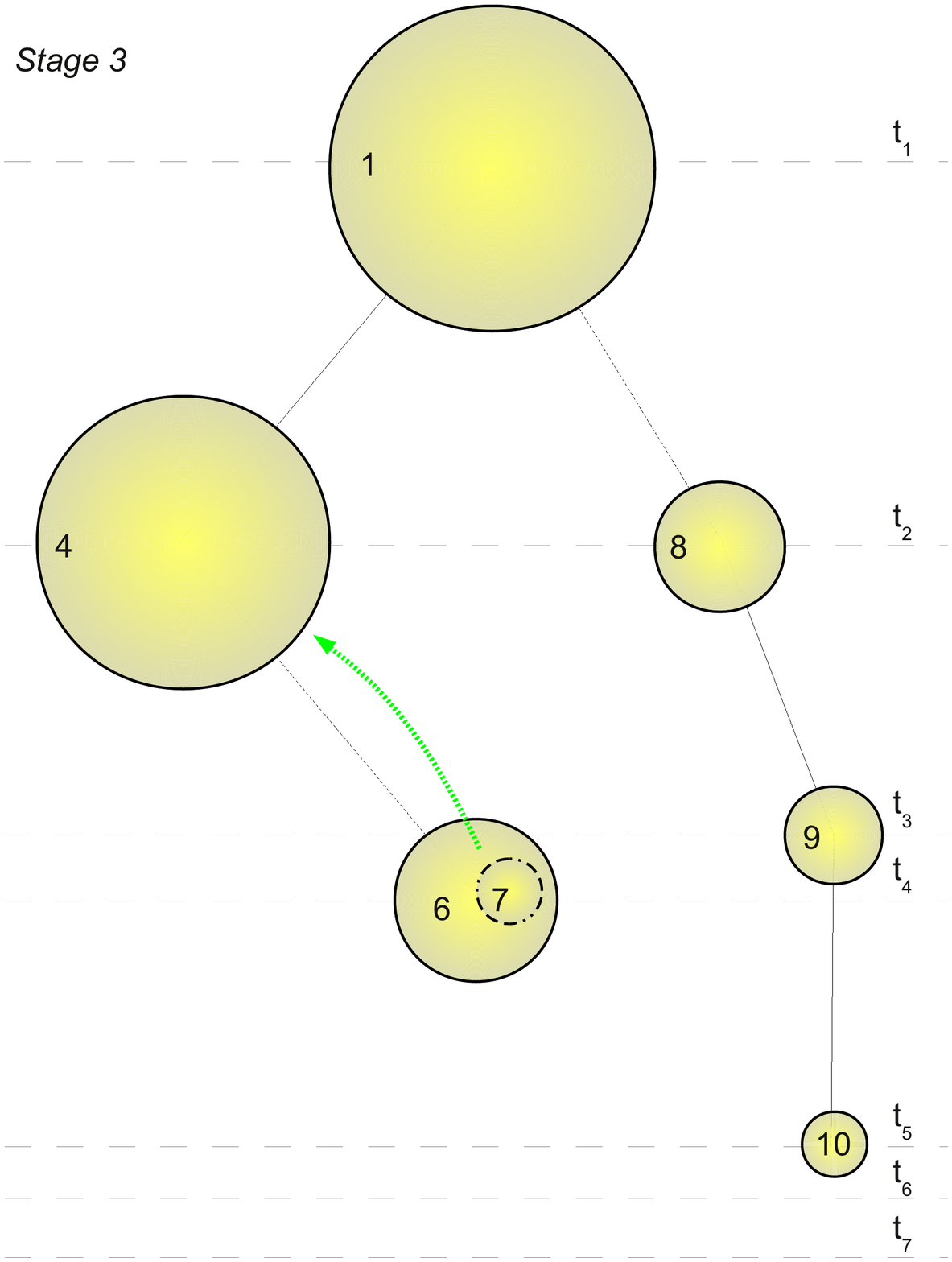} &
 \includegraphics[width=80mm]{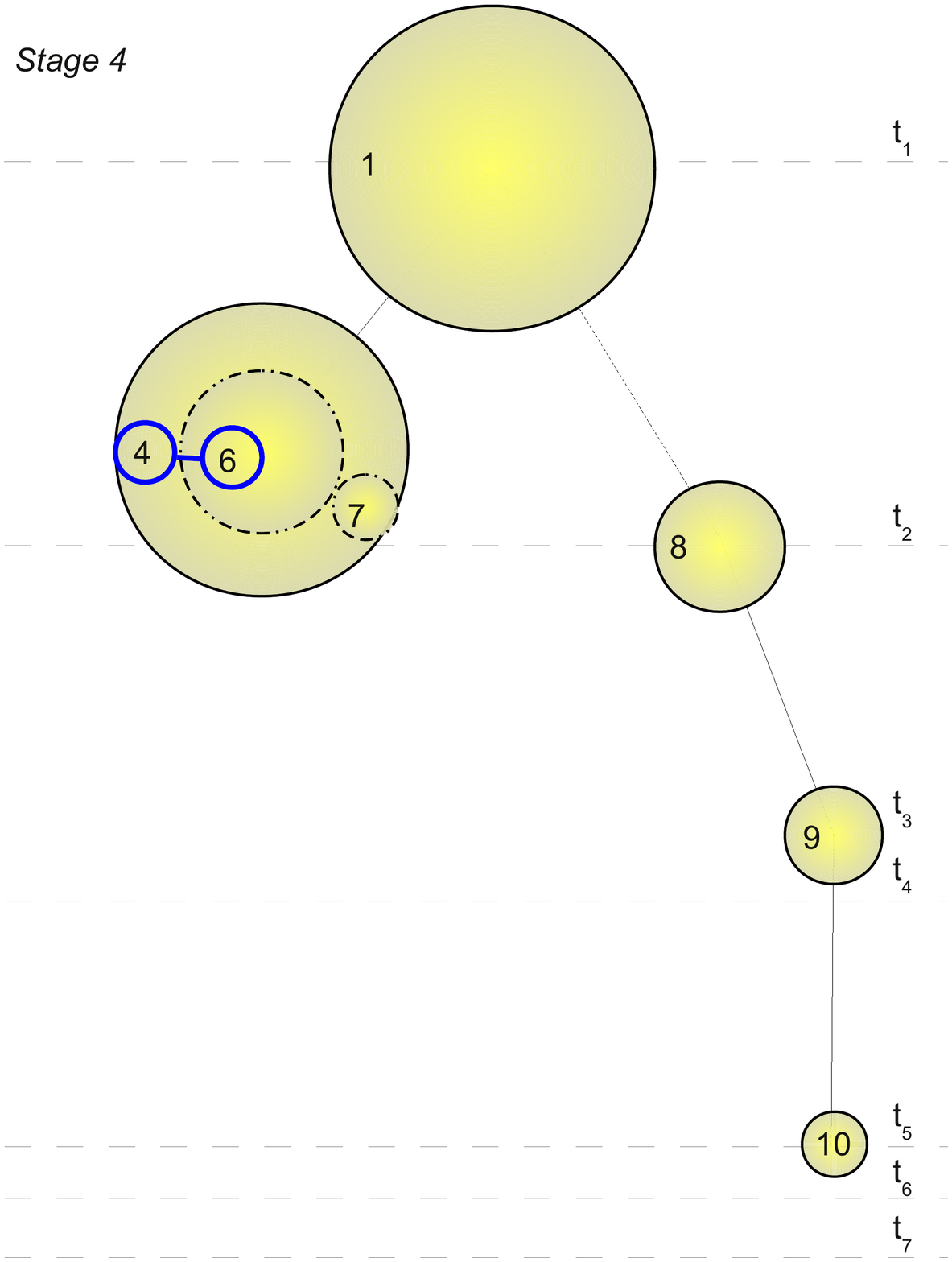}
 \end{tabular}
 \end{center}
 \caption{\emph{Description on following page.}}
 \label{fig:Engine}
\end{figure*}

\begin{figure*}
 \begin{center}
 \begin{tabular}{cc}
 \includegraphics[width=80mm]{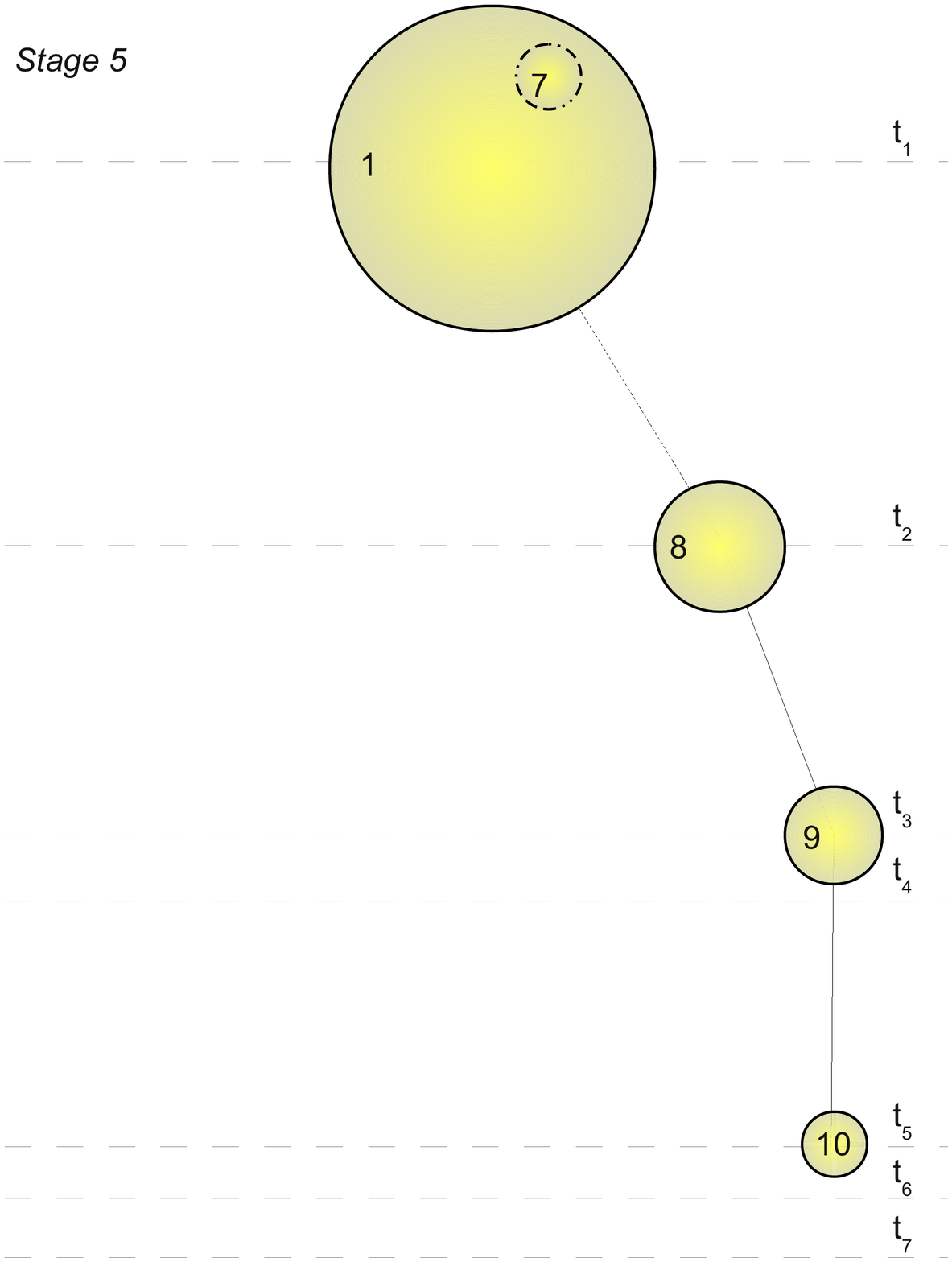} &
 \includegraphics[width=80mm]{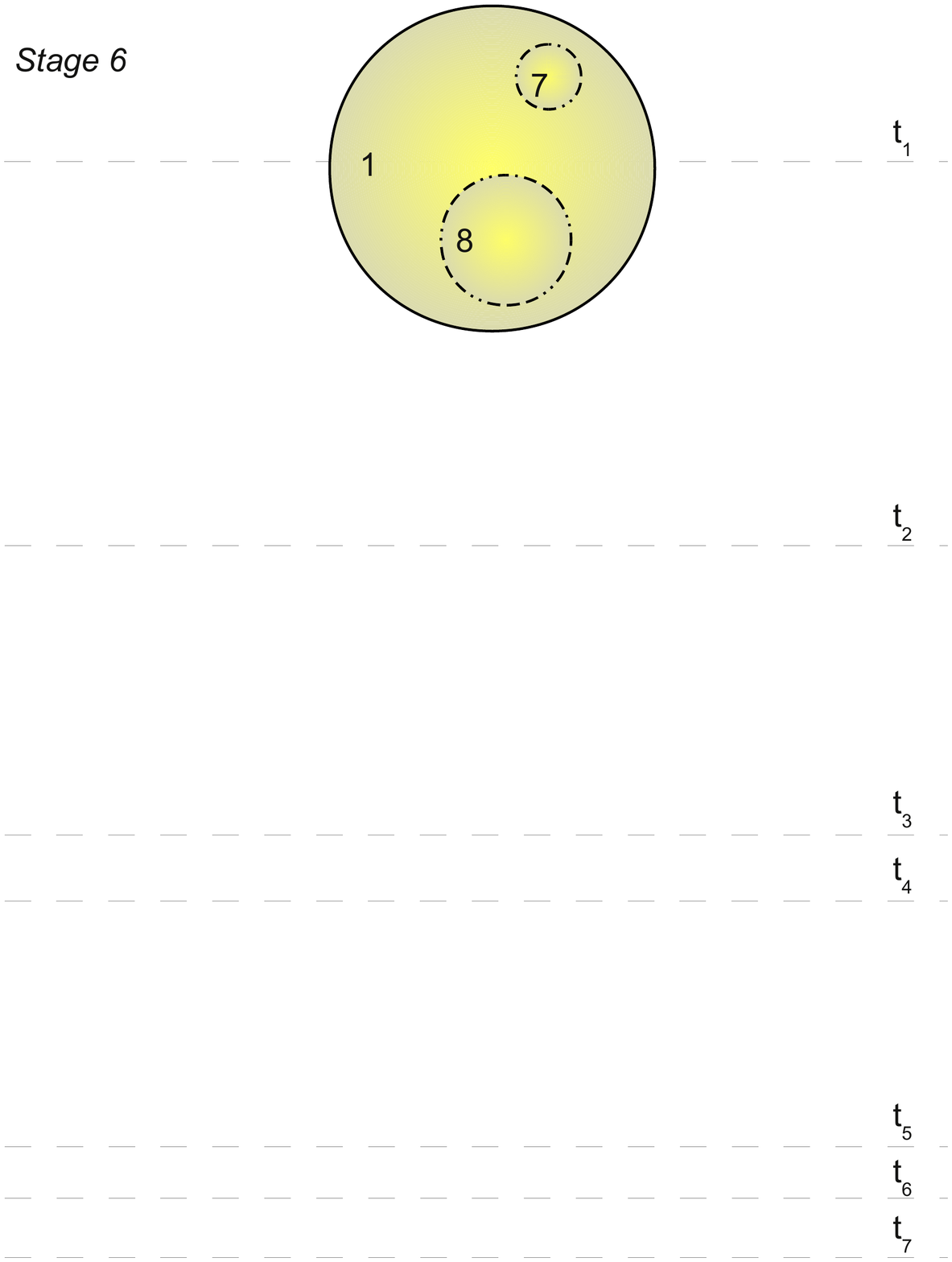}
 \end{tabular}
 \end{center}
 \addtocounter{figure}{-1}
 \caption{\emph{(cont.)} A schematic representation of the evolution of a merger tree within \protect\glc. Colored circles represent nodes in the merger tree (with larger circles indicating greater mass). Solid circles are isolated halos, while dashed circles represent subhalos inside a larger halo. Time progresses from bottom to top (i.e. $t_1>t_2>t_3$\ldots). Red dashed arrows indicate \protect\glc\ stepping backward through the merger tree to find a halo which can be evolved. Green dotted arrows indicate a halo being evolved forward in time. Finally, linked blue circles indicate a merger between a subhalo and its host halo and any galaxies that they contain. The evolution of the merger tree is depicted at six key stages, each of which is discussed in the text.}
\end{figure*}

It should be noted that, under this algorithm, at any given point in the calculation different branches of the tree may have been evolved to different times (as in ``Stage 5'' of Fig.~\ref{fig:Engine} where the left-branch has been evolved to $t_1$, but the right-branch has yet to be evolved at all). This is acceptable providing that galaxies in different branches do not interact. If such interactions were included it would be necessary to impose a synchronicity condition on all branches so that no one branch could evolve significantly past any other. By design, \glc\ makes it easy to add in additional time-stepping criteria such as would be needed to implement such an algorithm.

The evolution of nodes forward in time is carried out by \glc's ODE solver, which uses GSL's embedded Runge-Kutta-Fehlberg (4, 5) method ({\tt gsl\_odeiv\_step\_rkf45}). However, any number of events are allowed to trigger an ``interrupt'' which causes ODE solving to cease at a specified time. When an interrupt is generated, a function is specified which will be called to handle the interrupt. The function can manipulate the node being evolved as necessary before passing it back to the ODE solver. Node promotion, node merging and satellite merging events all trigger interrupts. Interrupts are also used, for example, to create/destroy components as required.

\section{Implemented Components}\label{sec:ImplementationComponents}

In this section we describe the currently implemented node components available within \glc. We emphasize that a key feature of \glc\ is the ability to easily add new components, or replace the current components with alternative implementations. In several cases we include ``null'' implementations which effectively switch off a given component. Throughout, input parameters are identified using the form {\tt [inputParameter]}. For each component we describe the following:
\begin{description}
 \item[Properties] Any variables (e.g. mass, angular momentum, etc.) that describe the component;
 \item[Initialization] How the component properties are initialized prior to any evolution of the merger tree;
 \item[Differential Evolution] The equations governing the smooth evolution of the component;
 \item[Event Evolution] How the component changes in response to one of the following events:
 \begin{description}
  \item[Node mergers] Triggered whenever one dark matter halo becomes a subhalo within a large halo;
  \item[Satellite merging] Triggered whenever a subhalo merges with its host halo (or potentially another subhalo);
  \item[Node promotion] Triggered when the main progenitor of a given node in the merger tree has evolved to the time at which its parent node is defined, and so must be promoted to become that node.
 \end{description}
\end{description}

\subsection{(Supermassive) Black Hole}

\subsubsection{``Null'' Implementation}

The null black hole implementation defines the same properties as all other black hole implementations, but implements dummy functions for all black hole properties. It can be used to effectively switch off black holes. Of course, this is not safe if any of the other active components expect to get or set black hole properties (or if they rely on a sensible implementation of black hole evolution).

\subsubsection{``Standard'' Implementation}\label{sec:BlackHoleStandard}

\emph{Properties:} The standard black hole implementation defines the following properties:
\begin{itemize}
 \item The mass of the black hole: $M_\bullet$;
 \item The spin of the black hole, $j_\bullet$.
\end{itemize}

\emph{Initialization:} Black holes are not initialized, they are created (with a seed mass given by {\tt [blackHoleSeedMass]} and zero spin) as needed.

\emph{Differential Evolution:} The mass and spin evolve as:
\begin{eqnarray}
\dot{M}_\bullet &=& (1-\epsilon_{\rm radiation}) \dot{M}_0 \\
\dot{\jmath}_\bullet &=& \dot{\jmath}(M_\bullet,j_\bullet,\dot{M}_0),
\end{eqnarray}
where $\dot{M}_0$ is the rest mass accretion rate, $\epsilon_{\rm radiation}$ is the radiative efficiency of the accretion flow feeding the black hole and $\dot{\jmath}(M_\bullet,j_\bullet,\dot{M}_0)$ is the spin-up function of that accretion flow (see \S\ref{sec:CircumnuclearDisks}). The rest mass accretion rate is computed assuming Bondi-Hoyle-Lyttleton accretion (see, for example, \citealt{edgar_review_2004}) from the spheroid gas reservoir (with an assumed temperature of {\tt [bondiHoyleAccretionTemperatureSpheroid]}) enhanced by a factor of {\tt [bondiHoyleAccretionEnhancementSpheroid]} and from the host halo (with whatever temperature the hot halo temperature profile specifies; see \S\ref{sec:HotHaloTemperature}) enhanced by a factor of {\tt [bondiHoyleAccretionEnhancementHotHalo]}. The rest mass accretion rate is removed (as a mass sink; see \S\ref{sec:SpheroidHernquist}) from the spheroid component. The black hole is assumed to cause feedback in two ways:
\begin{description}
 \item [Radio-mode] Any jet power from the black hole-accretion disk system (see \S\ref{sec:CircumnuclearDisks}) is included in the hot halo heating rate providing that the halo is in the slow cooling regime (i.e. if the cooling radius is smaller than the virial radius; see, for example, \citealt{benson_cold_2010});
 \item [Quasar-mode] A mechanical wind luminosity of \citep{ostriker_momentum_2010}
\begin{equation}
 L_{\rm wind} = \epsilon_{\bullet, wind} \dot{M}_0 \clight^2,
\end{equation}
where $\epsilon_{\bullet, wind}=${\tt [blackHoleWindEfficiency]} is the black hole wind efficiency, is added to the gas component of the spheroid (which, presumably, will respond with an outflow for example) if and only if the wind pressure (at the spheroid characteristic radius) is less than the typical thermal pressure in the spheroid gas \citep{ciotti_feedbackcentral_2009}, i.e.
\begin{eqnarray}
 P_{\rm wind} &<& P_{\rm ISM} \nonumber \\
 \frac{1}{2}\rho_{\rm wind} V_{\rm wind}^2 &<& {3 {\rm k_B} T_{\rm ISM} \langle \rho_{\rm ISM}\rangle \over 2 m_{\rm H}}.
\end{eqnarray}
Since $\Omega r^2 \rho_{\rm wind} V_{\rm wind}^3 = L_{\rm wind}$ where $\Omega$ is the solid angle of the wind flow, this can be rearranged to give $\langle\rho_{\rm ISM}\rangle > \rho_{\rm wind, critical}$ where
\begin{equation}
\rho_{\rm wind,critical} = {2 m_{\rm H} L_{\rm wind} \over 3 \Omega r^2 V_{\rm wind} {\rm k_B} T_{\rm ISM}}.
\end{equation}
This critical wind density is computed at the characteristic radius of the spheroid, $r_{\rm spheroid}$, assuming $V_{\rm wind}=10^4$km/s, $T_{\rm ISM}=10^4$K and $\Omega=\pi$, and the \ISM\ density is approximated by
\begin{equation}
 \langle\rho_{\rm ISM}\rangle = {3 M_{\rm gas, spheroid} \over 4 \pi} r_{\rm spheroid}^3.
\end{equation}
For numerical ease, the fraction, $f_{\rm wind}$, of the wind luminosity added to the spheroid is adjusted smoothly through the $\rho_{\rm ISM}\approx\rho_{\rm wind,critical}$ region according to
\begin{equation}
 f_{\rm wind} = \left\{ \begin{array}{ll} 0 & \hbox{ if } x < 0, \\ 3x^2-2x^3 & \hbox{ if } 0 \le x \le 1, \\ 1 & \hbox{ if } x > 1, \end{array} \right.
\end{equation}
where $x=\rho_{\rm ISM}/\rho_{\rm wind,critical}-1/2$.
\end{description}

\emph{Event Evolution $\rightarrow$ Node mergers:} None.

\emph{Event Evolution $\rightarrow$ Satellite merging:} The black holes in the two merging galaxies are instantaneously merged. Properties are computed using the selected black hole binary merger method (see \S\ref{sec:BlackHoleBinaryMergers}).

\emph{Event Evolution $\rightarrow$ Node promotion:} None.

\subsection{Hot Halo}

\subsubsection{``Null'' Implementation}

The null hot halo implements dummy functions for all hot halo properties. It can be used to effectively switch off hot halos. Of course, this is not safe if any of the other active components expect to get or set hot halo properties (or if they rely on a sensible implementation of hot halo evolution).

\subsubsection{``Standard'' Implementation}\label{sec:HotHaloStandard}

\emph{Properties:} The standard hot halo implementation defines the following properties:
\begin{itemize}
 \item The mass of gas which failed to accrete in to the hot halo\footnote{By ``failed to accrete'' we mean any mass which would accrete into the halo in the absence of baryonic physics, such as pressure and heating. This provides a way to accrete this mass later if, for example, the dark matter halo potential well deepens and becomes more effective at accreting mass.}: $M_{\rm failed}$;
 \item The mass of gas in the hot halo: $M_{\rm hot}$;
 \item The angular momentum of the gas in the hot halo, $J_{\rm hot}$;
 \item The mass(es) of heavy elements in gas in the hot halo, $M_{Z, {\rm hot}}$;
 \item The mass of gas from outflows in the hot halo: $M_{\rm outflowed}$;
 \item The angular momentum of the outflowed gas in the hot halo, $J_{\rm outflowed}$;
 \item The mass(es) of heavy elements in outflowed gas, $M_{Z, {\rm outflowed}}$,
\end{itemize}
and the following pipes:
\begin{description}
 \item [Energy Input] Energy sent through this pipe is added to the hot halo and used to offset the cooling rate (see below).
 \item [Cooling Gas] The net cooling rate of gas mass (and metal content and angular momentum) is sent through this pipe. Any component may claim this pipe and connect to it, allowing it to receive the cooling gas. For example, in the current implementation, the exponential disk component (see \S\ref{sec:DiskExponential}) claims and connects to this pipe.
 \item [Gas Outflow] Galactic components that wish to expel gas due to an outflow can send that mass (plus metals and angular momentum) through this pipe, where it will be received into the hot halo component. 
 \item [Gas Mass Sink] Removes gas (and proportionate amounts of angular momentum and elements) from the hot gas halo.
\end{description}

\emph{Initialization:} At initialization, any nodes with no progenitors are assigned a hot halo mass, and failed accreted mass as dictated by the baryonic accretion method (see \S\ref{sec:AccretionBaryonic}) and angular momentum based on the accreted mass and the halo spin parameter.

\emph{Differential Evolution:} In the standard hot halo implementation the hot gas mass and heavy element mass(es) evolves as:
\begin{eqnarray}
 \dot{M}_{\rm failed} &=& \dot{M}_{\rm failed~accretion} \\
 \dot{M}_{\rm hot} &=& \dot{M}_{\rm accretion} - \dot{M}_{\rm cooling} \nonumber \\
 & & + \dot{M}_{\rm outflow,return}, \\
 \dot{M}_{Z, {\rm hot}} &=& - \dot{M}_{\rm cooling} {M_{Z, {\rm hot}}\over M_{\rm hot}} \nonumber \\
 & &  + \dot{M}_{Z, {\rm outflow,return}}, \\
\end{eqnarray}
where $\dot{M}_{\rm accretion}$ is the rate of growth of the hot component due to accretion from the \IGM, $\dot{M}_{\rm failed~accretion}$ is the rate of failed accretion from the \IGM\ (these may include a component due to transfer of mass from the failed to hot reservoirs) and $\dot{M}_{\rm cooling}$ is the rate of mass loss from the hot halo due to cooling (see \S\ref{sec:CoolingRate}) minus any heating rate defined as
\begin{equation}
 \dot{M}_{\rm heating} = \dot{E}{\rm input} / V_{\rm virial}^2,
\end{equation}
where $\dot{E}$ is the rate at which energy is being sent through the ``energy input'' pipe and $V_{\rm virial}$ is the virial velocity of the halo.\footnote{The net cooling rate is never allowed to drop below zero. If the mass heating rate exceeds the mass cooling rate then the excess energy is not used.} The angular momentum of the hot gas evolves as:
\begin{eqnarray}
 \dot{J}_{\rm hot} &=& \dot{M}_{\rm accretion} {\dot{J}_{\rm node} \over \dot{M}_{\rm node}} - \dot{M}_{\rm cooling} r_{\rm cool} V_{\rm rotate} \nonumber \\
  & & + \dot{J}_{\rm outflow,return},
\end{eqnarray}
where $\dot{M}_{\rm node}$ and $\dot{J}_{\rm node}$ are defined in \S\ref{sec:ComponentBasicProperties}, $r_{\rm cool}$ is the cooling radius (see \S\ref{sec:CoolingRadius}) and $V_{\rm rotate}$ is the effective rotation speed of the halo for its angular momentum (see \S\ref{sec:DarkMatterDensityProfile}). For the outflowed components:
\begin{eqnarray}
 \dot{M}_{\rm outflowed} &=& - \dot{M}_{\rm outflow,return} + \dot{M}_{\rm outflows}, \\
 \dot{M}_{Z, {\rm outflowed}} &=& - \dot{M}_{Z, {\rm outflow,return}} + \dot{M}_{Z, {\rm outflows}}, \\
\end{eqnarray}
and:
\begin{equation}
 \dot{J}_{\rm outflowed} = - \dot{J}_{\rm outflow,return} + \dot{J}_{\rm outflows}.
\end{equation}
In the above
\begin{equation}
 \dot{M}|\dot{M}_Z|\dot{J}_{\rm outflow,return} = \alpha_{\rm outflow~return~rate} {M|M_Z|J_{\rm outflowed}\over \tau_{\rm dynamical, halo}},
\end{equation}
where $\alpha_{\rm outflow~return~rate}=${\tt [hotHaloOutflowReturnRate]} is an input parameter controlling the rate at which gas flows from the outflowed to hot reservoirs, and $\dot{M}|\dot{M}_Z|\dot{J}_{\rm outflows}$ are the net rates of outflow from any components in the node.

\emph{Event Evolution $\rightarrow$ Node mergers:} If the {\tt [starveSatellites]} parameter is true, then any hot halo properties of the minor node are added to those of the major node and the hot halo component is removed from the minor node. Additionally in this case, any material outflowed from the the satellite galaxy to its hot halo is transferred to the hot halo of the host dark matter halo after each timestep.

\emph{Event Evolution $\rightarrow$ Satellite merging:} If the {\tt [starveSatellites]} parameter is false, then any hot halo properties of the satellite node are added to those of the host node and the hot halo component is removed from the satellite node.

\emph{Event Evolution $\rightarrow$ Node promotion:} Any hot halo properties of the parent node are added to those of the node prior to promotion.

\subsection{Galactic Disk}

\subsubsection{``Null'' Implementation}

The null disk implementation implements dummy functions for all disk properties. It can be used to effectively switch off disks. Of course, this is not safe if any of the other active components expect to get or set disk properties (or if they rely on a sensible implementation of disk evolution).

\subsubsection{``Exponential'' Implementation}\label{sec:DiskExponential}

This implementation assumes a disk with an exponential surface density profile in which stars trace gas.

\emph{Properties:} The exponential galactic disk implementation defines the following properties:
\begin{itemize}
 \item The mass of gas in the disk: $M_{\rm disk, gas}$;
 \item The masses of elements in the gaseous disk: $M_{Z, {\rm disk, gas}}$;
 \item The mass of stars in the disk: $M_{\rm disk, stars}$;
 \item The masses of elements in the stellar disk: $M_{Z, {\rm disk, stars}}$;
 \item The luminosities (in multiple bands) of the stellar disk: $L_{\rm disk, stars}$;
 \item The angular momentum of the disk, $J_{\rm disk}$;
 \item The radial scale length of the disk, $R_{\rm disk}$;
 \item The circular velocity of the disk at $R_{\rm disk}$, $V_{\rm disk}$.
\end{itemize}

\emph{Initialization:} No initialization is performed---disks are created as needed.

\emph{Differential Evolution:} In the exponential galactic disk implementation the gas mass evolves as:
\begin{eqnarray}
 \dot{M}_{\rm disk, gas} &=& \dot{M}_{\rm cooling} - \dot{M}_{\rm outflow, disk} - \dot{M}_{\rm stars, disk} \nonumber \\ 
 & & - M_{\rm disk, gas}/\tau_{\rm bar},
\end{eqnarray}
where the rate of change of stellar mass is
\begin{equation}
 \dot{M}_{\rm disk, stars} = \Psi - \dot{R} - M_{\rm disk, stars}/\tau_{\rm bar},
\end{equation}
with
\begin{equation}
 \Psi = {M_{\rm disk, gas} \over \tau_{\rm disk, star~formation}},
\end{equation}
where $\tau_{\rm disk, star~formation}$ is the star formation timescale (see \S\ref{sec:StarFormationTimescales}), $\dot{R}$ is the rate of mass recycling from stars and $\tau_{\rm bar}$ is a bar instability timescale (see \S\ref{sec:DiskStability}). The mass removed from the disk by the bar instability mechanism is added to the active spheroid component. Element abundances (including total metals) evolve according to:
\begin{eqnarray}
  \dot{M}_{Z, {\rm disk, gas}} &=& \dot{M}_{Z, {\rm cooling}} - \dot{M}_{Z, {\rm outflow, disk}} \nonumber \\
 & & - \dot{M}_{Z, {\rm stars, disk}} + \dot{y},
\end{eqnarray}
and
\begin{equation}
 \dot{M}_{Z, {\rm stars, disk}} = \Psi {M_{Z, {\rm disk, gas}} \over M_{\rm disk, gas}} - \dot{R}_Z
\end{equation}
where $\dot{y}$ is the rate of element yield from stars and $\dot{R}_Z$ is the rate of element recycling. Recycling rates and yields are discussed in \S\ref{sec:StellarIMF} and \S\ref{sec:StellarPopulationProperties}. The angular momentum evolves as:
\begin{eqnarray}
 \dot{J}_{\rm disk} &=& \dot{J}_{\rm cooling} \nonumber \\ 
 & & - \left[ \dot{M}_{\rm outflow, disk} + {M_{\rm disk, gas}  + M_{\rm disk, stars} \over \tau_{\rm bar}}\right] \nonumber \\
 & & \times {J_{\rm disk} \over M_{\rm disk, gas} + M_{\rm disk, stars}}.
\end{eqnarray}
The outflow rate, $\dot{M}_{\rm outflow, disk}$, is computed for the current star formation rate and gas properties by the stellar properties subsystem (see \S\ref{sec:StellarPopulationProperties}).

\emph{Event Evolution $\rightarrow$ Node mergers:} None.

\emph{Event Evolution $\rightarrow$ Satellite merging:} Disks may be destroyed (or, potentially, created or otherwise modified) as the result of a satellite merging event, as dictated by the selected merger remnant mass movement method (see \S\ref{sec:MergingMassMove}).

\emph{Event Evolution $\rightarrow$ Node promotion:} None.

\subsection{Galactic Spheroid}

\subsubsection{``Null'' Implementation}

The null spheroid implements dummy functions for all spheroid properties. It can be used to effectively switch off spheroids. Of course, this is not safe if any of the other active components expect to get or set spheroid properties (or if they rely on a sensible implementation of spheroid evolution).

\subsubsection{``Hernquist'' Implementation}\label{sec:SpheroidHernquist}

This implementation assumes a Hernquist profile \citep{hernquist_analytical_1990} for the spheroidal component of a galaxy in which stars trace gas.

\emph{Properties:} The Hernquist galactic spheroid implementation defines the following properties:
\begin{itemize}
 \item The mass of gas in the spheroid: $M_{\rm spheroid, gas}$;
 \item The masses of elements in the gaseous spheroid: $M_{Z, {\rm spheroid, gas}}$;
 \item The mass of stars in the spheroid: $M_{\rm spheroid, stars}$;
 \item The masses of elements in the stellar spheroid: $M_{Z, {\rm spheroid, stars}}$;
 \item The luminosities (in multiple bands) of the stellar spheroid: $L_{\rm spheroid, stars}$;
 \item The pseudo-angular momentum\footnote{Effectively the angular momentum that the spheroid would have, were it rotationally supported rather than pressure supported.} of the spheroid, $J_{\rm spheroid}$;
 \item The radial scale length of the spheroid, $r_{\rm spheroid}$;
 \item The circular velocity of the spheroid at $r_{\rm spheroid}$, $V_{\rm spheroid}$.
\end{itemize}
and the following pipes:
\begin{description}
 \item [Energy Input] Energy sent through this pipe is added to the gas of the spheroid and will result in an outflow (see below).
 \item [Gas Mass Sink] Removes gas (and proportionate amounts of angular momentum and elements) from the spheroid gas.
\end{description}

\emph{Initialization:} No initialization is performed---spheroids are created as needed.

\emph{Differential Evolution:} In the Hernquist galactic spheroid implementation the gas mass evolves as\footnote{There may be an additional contribution to the mass and angular momentum rates of change in the spheroid due to material transferred from the disk component via the bar instability mechanism (see \S\protect\ref{sec:DiskExponential}). This is not included here as it is not intrinsic to this specific spheroid implementation---it is handled explicitly by the disk component and so applies equally to any spheroid component implementation.}:
\begin{equation}
 \dot{M}_{\rm spheroid, gas} = - \dot{M}_{\rm outflow, spheroid} - \dot{M}_{\rm stars, spheroid},
\end{equation}
where the rate of change of stellar mass is
\begin{equation}
 \dot{M}_{\rm stars, spheroid} = \Psi - \dot{R}
\end{equation}
with
\begin{equation}
 \Psi = {M_{\rm spheroid, gas} \over \tau_{\rm spheroid, star~formation}}
\end{equation}
with $\tau_{\rm spheroid, star~formation}$ being the star formation timescale (see \S\ref{sec:StarFormationTimescales}) and $\dot{R}$ is the rate of mass recycling from stars.
Element abundances (including total metals) evolve according to:
\begin{equation}
  \dot{M}_{Z, {\rm spheroid, gas}} = - \dot{M}_{Z, {\rm outflow, spheroid}} - \dot{M}_{Z, {\rm stars, spheroid}} + \dot{y},
\end{equation}
and
\begin{equation}
 \dot{M}_{Z, {\rm stars, spheroid}} = \Psi {M_{Z, {\rm spheroid, gas}} \over M_{\rm spheroid, gas}} - \dot{R}_Z
\end{equation}
where $\dot{y}$ is the rate of element yield from stars and $\dot{R}_Z$ is the rate of element recycling. Recycling rates and yields are discussed in \S\ref{sec:StellarIMF} and \S\ref{sec:StellarPopulationProperties}.  The angular momentum evolves as:
\begin{equation}
 \dot{J}_{\rm spheroid} = \dot{M}_{\rm outflow, spheroid} {J_{\rm spheroid} \over M_{\rm spheroid, gas} + M_{\rm spheroid, stars}}.
\end{equation}
The outflow rate, $\dot{M}_{\rm outflow, disk}$, is computed for the current star formation rate and gas properties by the stellar properties subsystem (see \S\ref{sec:StellarPopulationProperties}), with an additional contribution given by
\begin{equation}
 \dot{M}_{\rm outflow, spheroid} = \beta_{\rm spheroid, energy} {\dot{E}_{\rm gas, spheroid} \over V_{\rm spheroid}^2}
\end{equation}
where $\beta_{\rm spheroid, energy}$ is set by the {\tt [spheroidEnergeticOutflowMassRate]} input parameter, and $\dot{E}_{\rm gas,spheroid}$ is any input energy sent through the ``Energy Input'' pipe.

\emph{Event Evolution $\rightarrow$ Node mergers:} None.

\emph{Event Evolution $\rightarrow$ Satellite merging:} Spheroids may be created as the result of a satellite merging event, as dictated by the selected merger remnant mass movement method (see \S\ref{sec:MergingMassMove}).

\emph{Event Evolution $\rightarrow$ Node promotion:} None.

\subsection{Basic Properties}\label{sec:ComponentBasicProperties}

Basic properties are the total mass of a node and the cosmic time at which it currently exists.

\subsubsection{``Simple'' Implementation}

\emph{Properties:} The simple basic properties implementation defines the following properties:
\begin{itemize}
 \item The total mass of the node: $M_{\rm node}$;
 \item The time at which the node is defined: $t_{\rm node}$;
 \item The time at which the node was last an isolated halo (i.e. not a subhalo).
\end{itemize}

\emph{Initialization:} All basic properties are required to be initialized by the merger tree construction routine (see \S\ref{sec:MergerTreeConstruct}).

\emph{Differential Evolution:} Properties are evolved according to:
\begin{eqnarray}
 \dot{M}_{\rm node} &=& \left\{\begin{array}{ll}{M_{\rm node, parent} - M_{\rm node} \over t_{\rm node, parent} - t_{\rm node}} & \hbox{if primary progenitor} \\ 0 & \hbox{otherwise}, \end{array} \right. \\
 \dot{t}_{\rm node} &=& 1,
\end{eqnarray}
where the ``parent'' subscript indicates a property of the parent node in the merger tree.

\emph{Event Evolution $\rightarrow$ Node mergers:} None.

\emph{Event Evolution $\rightarrow$ Satellite merging:} None.

\emph{Event Evolution $\rightarrow$ Node promotion:} $M_{\rm node}$ is updated to the node mass of the parent prior to promotion.

\subsection{Satellite Node Orbit}

This component tracks the orbital properties of subhalos.

\subsubsection{``Simple'' Implementation}

\emph{Properties:} The simple satellite orbit implementation defines the following properties:
\begin{itemize}
 \item The time until the satellite will merge with its host: $t_{\rm satellite, merge}$.
\end{itemize}

\emph{Initialization:} None.

\emph{Differential Evolution:} Properties are evolved according to:
\begin{equation}
 \dot{t}_{\rm satellite, merge} = -1.
\end{equation}

\emph{Event Evolution $\rightarrow$ Node mergers:} The component is created and the time to merging is assigned a value (see \S\ref{sec:MergingTimescales}).

\emph{Event Evolution $\rightarrow$ Satellite merging:} None.

\emph{Event Evolution $\rightarrow$ Node promotion:} Not applicable (component only exists for satellite nodes).

\subsection{Dark Matter Halo Spin}

This component stores and tracks the spin parameters of dark matter halos.

\subsubsection{``Null'' Implementation}

The null spin component implements dummy functions for all spin properties. It can be used to effectively switch off spins. Of course, this is not safe if any of the other active components expect to get or set spin properties (or if they rely on a sensible implementation of spin evolution).

\subsubsection{``Random'' Implementation}\label{sec:SpinsRandom}

\emph{Properties:} The random dark matter halo spin implementation defines the following properties:
\begin{itemize}
 \item The spin parameter of the halo: $\lambda$.
\end{itemize}

\emph{Initialization:} The spin parameter of each node, if not already assigned, is selected at random from a distribution of spin parameters. This value is assigned to the earliest progenitor of the halo traced along its primary branch. The value is then propagated forward along the primary branch until the node mass exceeds that of the node for which the spin was selected by a factor of {\tt [randomSpinResetMassFactor]}, at which point a new spin is selected at random, and the process repeated until the end of the branch is reached, in a manner similar to the algorithm used by \cite{cole_hierarchical_2000}. 

\emph{Differential Evolution:} The spin parameter does not evolve.

\emph{Event Evolution $\rightarrow$ Node mergers:} None

\emph{Event Evolution $\rightarrow$ Satellite merging:} None.

\emph{Event Evolution $\rightarrow$ Node promotion:} The spin is updated to equal that of the parent node. (The two will differ only if this is a case where the new halo node was sufficiently more massive than the node for which a spin was last selected that a new spin value was chosen.)

\subsection{Dark Matter Profile}

This component stores dynamic properties associated with dark matter halo density profiles.

\subsubsection{``Null'' Implementation}

The null profile implements dummy functions for all profile properties. It can be used to effectively switch off profiles. Of course, this is not safe if any of the other active components expect to get or set profile properties (or if they rely on a sensible implementation of profile evolution).

\subsubsection{``Scale'' Implementation}\label{sec:ComponentDarkMatterProfileScale}

\emph{Properties:} The scale dark matter profile implementation defines the following properties:
\begin{itemize}
 \item The scale length of the density profile.
\end{itemize}

\emph{Initialization:} The scale length of each node, if not already assigned, is assigned using the concentration parameter function (see \S\ref{sec:DmConcentration}), but is not allowed to drop below {\tt [darkMatterProfileMinimumConcentration]}, such that the scale length is equal to the virial radius divided by that concentration. The value is propagated in both directions along the primary child branch from the node.

\emph{Differential Evolution:} The scale radius does not evolve.

\emph{Event Evolution $\rightarrow$ Node mergers:} None.

\emph{Event Evolution $\rightarrow$ Satellite merging:} None.

\emph{Event Evolution $\rightarrow$ Node promotion:} None.

\section{Specific Physical Implementation}\label{sec:ImplementationPhysics}

In this section we describe specific implementations of the numerous physical processes that are currently available in \glc. In each subsection we briefly describe what physical process/property is being discussed and then describe the specific implementations available within \glc.

\subsection{Accretion of Gas into Halos}\label{sec:AccretionBaryonic}\index{accretion!baryonic}

The accretion rate of gas from the \IGM\ into a dark matter halo is expected to depend on (at least) the rate at which that halo's mass is growing, the depth of its potential well and the thermodynamical properties of the accreting gas. \glc\ implements the following calculations of gas accretion from the \IGM.

\emph{Simple Method:} Currently the only option, this method sets the accretion rate of baryons into a halo to be:
\begin{equation}
 \dot{M}_{\rm accretion} = \left\{ \begin{array}{ll} {\Omega_{\rm b} \over \Omega_0} \dot{M}_{\rm halo} & \hbox{ if } V_{\rm virial} > V_{\rm reionization} \\ & \,\,\,\hbox{ or } z > z_{\rm reionization} \\ 0 & \hbox{ otherwise,}\end{array} \right.
\end{equation}
where $\dot{M}_{\rm halo}$ is the total rate of growth of the node mass, $z_{\rm reionization}=${\tt [reionizationSuppressionRedshift]} is the redshift at which the Universe is reionized and $V_{\rm reionization}=${\tt [reionizationSuppressionVelocity]} is the virial velocity below which accretion is suppressed after reionization. Setting $V_{\rm reionization}$ to zero will effectively switch off the effects of reionization on the accretion of baryons. This algorithm attempts to offer a simple prescription for the effects of reionization and has been explored by multiple authors (e.g. \citealt{benson_effects_2002}). In particular, \cite{font_modelingmilky_2010} show that it produces results in good agreement with more elaborate treatments of reionization. For halos below the accretion threshold, any accretion rate that would have otherwise occurred is instead placed into the ``failed'' accretion rate. For halos which can accrete, and which have some mass in their ``failed'' reservoir, that mass will be added to the regular accretion rate at a rate equal to the mass of the ``failed'' reservoir times the specific growth rate of the halo.

\subsection{Background Cosmology}\label{sec:Cosmology}\index{cosmology}

The background cosmology describes the evolution of an isotropic, homogeneous Universe within which galaxy formation calculations are carried out. For the purposes of \glc, the background cosmology is used to relate expansion factor/redshift to cosmic time and to compute the density of various components (e.g. dark matter, dark energy, etc.) at different epochs.

\emph{Matter + Lambda:} In this implementation, cosmological relations are computed assuming a universe that contains only collisionless matter and a cosmological constant.

\subsection{Circumnuclear Accretion Disks}\label{sec:CircumnuclearDisks}\index{accretion disks}\index{accretion!disk}

Circumnuclear accretion disks surrounding supermassive black holes at the centers of galaxies influence the evolution of both the black hole (via accretion rates of mass and angular momentum and possibly by extracting rotational energy from the black hole) and the surrounding galaxy if they lead to energetic outflows (e.g. jets) from the nuclear region. Current implementations of accretion disks are as follows and are selected via {\tt [accretionDisksMethod]}.

\emph{Shakura-Sunyaev Geometrically Thin, Radiatively Efficient Disks:} This implementation assumes that accretion disks are always described by a radiatively efficient, geometrically thin accretion disk as described by \cite{shakura_black_1973}. The radiative efficiency of the flow is computed assuming that material falls into the black hole without further energy loss from the \ISCO, while the spin-up rate of the black hole is computed assuming that the material enters the black hole with the specific angular momentum of the \ISCO\ (i.e. there are no torques on the material once it begins to fall in from the \ISCO; \citealt{bardeen_kerr_1970}). For these thin disks, jet power is computed, using the expressions from \citeauthor{meier_association_2001}~(\citeyear{meier_association_2001}; his equations 4 and 5).

\emph{Advection Dominated, Geometrically Thick, Radiatively Inefficient Flows (ADAFs):} This implementation assumes that accretion is via an advection dominated accretion flow \citep{narayan_advection-dominated_1994} which is radiatively inefficient and geometrically thick. The radiative efficiency of the flow, which will be zero for a pure ADAF, can be set via the input parameter {\tt [adafRadiativeEfficiency]}. The spin up rate of the black hole and the jet power produced as material accretes into the black hole are computed using the method of \cite{benson_maximum_2009}. The energy of the accreted material can be set equal to the energy at infinity (as expected for a pure ADAF) or the energy at the \ISCO\ by use of the {\tt [adafEnergyOption]} parameter. The ADAF structure is controlled by the adiabatic index, $\gamma$, and viscosity parameter, $\alpha$, which are specified via the {\tt [adafAdiabaticIndex]} and {\tt [adafViscosityOption]} input parameters respectively. {\tt [adafViscosityOption]} may be set to ``{\tt fit}'', in which case the fitting function for $\alpha$ as a function of black hole spin from \cite{benson_maximum_2009} will be used.

\emph{``Switched'' Disks:} This method allows for accretion disks to switch between radiatively efficient (Shakura-Sunyaev) and inefficient (ADAF) modes. Which mode is used is determine by the accretion rate onto the disk:
\begin{itemize}
 \item Radiatively efficient accretion if $\dot{M}/\dot{M}_{\rm Eddington}>${\tt [accretionRateThinDiskMinimum]} and \newline $\dot{M}/\dot{M}_{\rm Eddington}<${\tt [accretionRateThinDiskMaximum]};
 \item Radiatively inefficient accretion otherwise.
\end{itemize}

\subsection{Cold Dark Matter Structure Formation}\index{structure formation}\index{cold dark matter}

A variety of functions are used to describe structure formation in cold dark matter dominated universes. These are described below.

\subsubsection{Primordial Power Spectrum}\label{sec:PrimordialPowerSpectrum}\index{power spectrum!primordial}

The functional form of the primordial dark matter power spectrum. The power spectrum is computed from the specified primordial power spectrum and the transfer function (see \S\ref{sec:TransferFunction}) and normalized to a value of $\sigma_8$ specified by {\tt [sigma\_8]}.

\emph{(Running) Power Law Spectrum:} This method implements a primordial power spectrum of the form:
\begin{equation}
 P(k) \propto k^{n_{\rm eff}(k)},
\end{equation}
where
\begin{equation}
 n_{\rm eff}(k) = n_{\rm s} + {\d n \over \d \ln k} {k \over k_{\rm ref}},
\end{equation}
where $n_{\rm s}=${\tt [powerSpectrumIndex]} is the power spectrum index at wavenumber $k_{\rm ref}=${\tt [powerSpectrumReferenceWavenumber]} and $\d n / \d \ln k=${\tt [powerSpectrumRunning]} describes the running of this index with wavenumber.

\subsubsection{Transfer Function}\label{sec:TransferFunction}\index{transfer function}

The functional form of the cold dark matter transfer function is selected by {\tt [transferFunctionMethod]}. The power spectrum is computed from the specified transfer function and the primordial power spectrum (see \S\ref{sec:PrimordialPowerSpectrum}) and normalized to a value of $\sigma_8$ specified by {\tt [sigma\_8]}.

\emph{BBKS:} This method uses the fitting function of \cite{bardeen_statistics_1986} to compute the \CDM\ transfer function.

\emph{Eisenstein \& Hu:} This method uses the fitting function of \cite{eisenstein_power_1999} to compute the \CDM\ transfer function. It requires that the effective number of neutrino species be specified via the {\tt [effectiveNumberNeutrinos]} parameter and summed mass of all neutrino species (in eV) be specified via the {\tt [summedNeutrinoMasses]} parameter.

\begin{figure}
 \begin{center}
 \includegraphics[width=80mm]{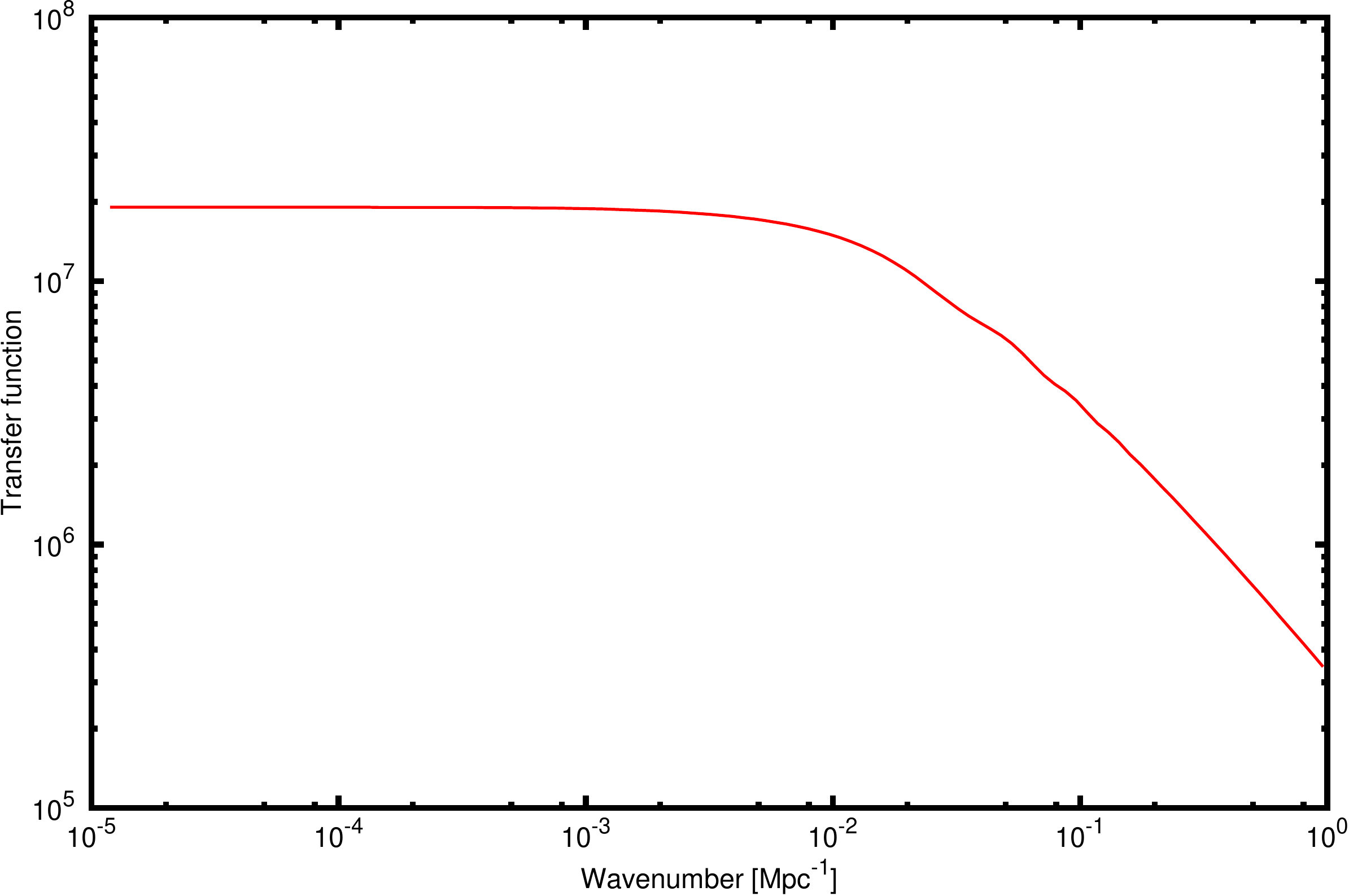}
 \end{center}
 \caption{The transfer function, $T(k)$, computed automatically by \protect\glc\ by downloading, compiling and running {\sc CMBFast}. The transfer function is computed for a cosmological model with $\Omega_{\rm M}=0.25$, $\Omega_\Lambda=0.75$, $\Omega_{\rm b}=0.045$ and $H_0=75$~km/s/Mpc.}
 \label{fig:transferFunctionCMBFast}
\end{figure}

\emph{{\sc CMBFast}:} This method uses the {\sc CMBFast} code to compute the \CDM\ transfer function. It requires that the mass fraction of helium in the early Universe be specified via the {\tt [Y\_He]} parameter. {\sc CMBFast} will be downloaded and run if the transfer function needs to be computed. The transfer function will then be stored in a file for future reference. An example of a transfer function computed in this way is shown in Fig.~\ref{fig:transferFunctionCMBFast}.

\emph{File:} This method reads a tabulated transfer function from an XML file, interpolating between tabulated points.

\subsubsection{Linear Growth Function}\index{linear growth}

The function describing the amplitude of linear perturbations.

\emph{Simple:} This method calculates the growth of linear perturbations using standard perturbation theory in a Universe consisting of collisionless matter and a cosmological constant.

\subsubsection{Critical Overdensity}\index{density!critical}

The method used to compute the critical linear overdensity at which overdense regions virialize.

\emph{Spherical Collapse (Matter + Cosmological Constant):} This method calculates critical overdensity using a spherical top-hat collapse model assuming a Universe which contains collisionless matter and a cosmological constant (see, for example, \citealt{percival_cosmological_2005}).

\subsubsection{Virial Density Contrast}\label{sec:VirialDensityConstrast}\index{density!virial}

The method used to compute the mean density contrast of virialized dark matter halos is specified via {\tt [virialDensityContrastMethod]}.

\emph{Bryan \& Norman Fitting Function:} This method calculates virial density contrast using the fitting functions given by \cite{bryan_statistical_1998}. As such, it is valid only for $\Omega_\Lambda=0$ or $\Omega_M+\Omega_\Lambda=1$ cosmologies and will abort on other cosmologies.

\emph{Spherical Collapse (Matter + Cosmological Constant):} This method calculates virial density contrast using a spherical top-hat collapse model assuming a Universe which contains collisionless matter and a cosmological constant (see, for example, \citealt{percival_cosmological_2005}).

\subsubsection{Halo Mass Function}\label{sec:HaloMassFunction}\index{halo mass function}\index{dark matter halos!mass function}

The dark matter halo mass function (i.e. the number of halos per unit volume per unit mass interval). Which mass function to use is specified by {\tt [haloMassFunctionMethod]}.

\emph{Press-Schechter:} This method uses the functional form proposed by \cite{press_formation_1974} to compute the halo mass function.

\emph{Sheth-Tormen:} This method uses the functional form proposed by \cite{sheth_ellipsoidal_2001} to compute the halo mass function.

\emph{Tinker:} This method uses the functional form proposed by \cite{tinker_large_2010} to compute the halo mass function. The mass function is computed at the appropriate virial overdensity (see \S\ref{sec:VirialDensityConstrast}).

\subsection{Cooling of Gas Inside Halos}\index{cooling}

The cooling of gas within dark matter halos is controlled by a number of different algorithms which will be described below.

\subsubsection{Cooling Function}\label{sec:CoolingFunction}\index{cooling function}\index{cooling!cooling function}

The cooling function of gas, $\Lambda(\rho,T,{\bf Z})$, where $\rho$ is gas density, $T$ is temperature and ${\bf Z}$ is a vector of elemental abundances, is selected by {\tt [coolingFunctionMethod]}. Multiple such methods may be specified and are cumulative (i.e. the net cooling function is the sum over all specified cooling functions).

\begin{figure}
 \begin{center}
 \includegraphics[width=80mm]{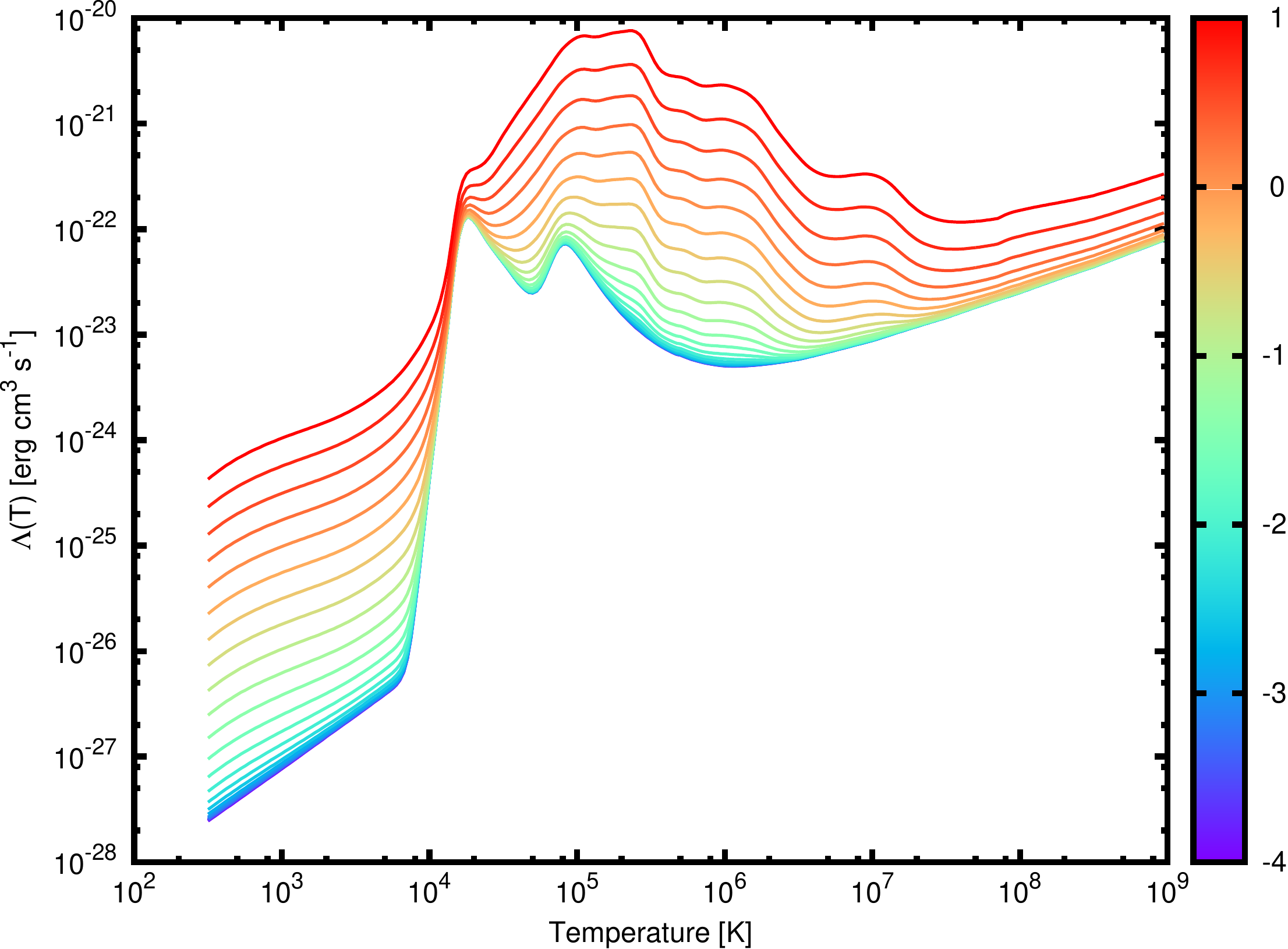}
 \end{center}
 \caption{The cooling function for atomic gas in collisional ionization equilibrium computed using {\sc Cloudy} 08.00 (automatically downloaded, compiled and run by \protect\glc) as a function of gas temperature. The color scale indicates the gas metallicity. Solar abundance ratios are assumed, except for zero metallicity calculations which use {\sc Cloudy}'s ``primordial'' metallicity.}
 \label{fig:atomicCIECloudyCoolingFunction}
\end{figure}

\emph{Atomic Collisional Ionization Equilibrium Using {\sc Cloudy}:} This method computes the cooling function using the {\sc Cloudy} code and under the assumption of collisional ionization equilibrium with no molecular contribution. Abundances are Solar, except for zero metallicity calculations which use {\sc Cloudy}'s ``primordial'' metallicity. The helium abundance for non-zero metallicity is scaled linearly with metallicity between primordial and Solar values. The {\sc Cloudy} code will be downloaded and run to compute the cooling function as needed, which will then be stored for future use. As this process is slow, a precomputed table is provided with \glc. If metallicities outside the range tabulated in this file are required it will be regenerated with an appropriate range. Fig.~\ref{fig:atomicCIECloudyCoolingFunction} shows an example of a cooling function computed automatically by \glc\ using {\sc Cloudy},

\emph{Collisional Ionization Equilibrium From File:} This method the cooling function is read from a file. The cooling function is assumed to be computed under conditions of collisional ionization equilibrium and therefore to scale as $\rho^2$.

\emph{CMB Compton Cooling:} This method computes the cooling function due to Compton scattering off of \CMB\ photons:
\begin{equation}
\Lambda = {4 \sigma_{\rm T} {\rm a} {\rm k}_{\rm B} n_{\rm e } \over m_{\rm e} \clight} T_{\rm CMB}^4 \left( T - T_{\rm CMB} \right),
\end{equation}
where $\sigma_{\rm T}$ is the Thompson cross-section, $a$ is the radiation constant, ${\rm k}_{\rm B}$ is Boltzmann's constant, $n_{\rm e}$ is the number density of electrons, $m_{\rm e}$ is the electron mass, $\clight$ is the speed of light, $T_{\rm CMB}$ is the \CMB\ temperature at the current cosmic epoch and $T$ is the temperature of the gas. The electron density is computed from the selected ionization state method (see \S\ref{sec:IonizationState}).

\subsubsection{Cooling Rate}\label{sec:CoolingRate}\index{cooling!rate}

The algorithm used to compute the rate at which gas drops out of the hot halo due to cooling.

\emph{White \& Frenk:} This method computes the cooling rate using the expression given by \cite{white_galaxy_1991}, namely
\begin{equation}
\dot{M}_{\rm cool} = 4 \pi r_{\rm cool}^2 \rho(r_{\rm cool}) \dot{r}_{\rm cool},
\end{equation}
where $r_{\rm cool}$ is the cooling radius (see \S\ref{sec:CoolingRadius}) in the hot halo and $\rho(r)$ is the density profile of the hot halo (see \S\ref{sec:HotHaloDensity}).

\subsubsection{Cooling Radius}\label{sec:CoolingRadius}\index{cooling radius}\index{cooling!radius}

The algorithm used to compute the cooling radius in hot halo gas.

\emph{Simple:} This method computes the cooling radius by seeking the radius at which the time available for cooling (see \S\ref{sec:CoolingTimeAvailable}) equals the cooling time (see \S\ref{sec:CoolingTime}). The growth rate is determined consistently based on the slope of the density profile, the density dependence of the cooling function and the rate at which the time available for cooling is increasing. This method assumes that the cooling time is a monotonic function of radius.

\subsubsection{Cooling Time}\label{sec:CoolingTime}\index{cooling time}\index{cooling!time}

The algorithm used to compute the time taken for gas to cool (i.e. the cooling time).

\emph{Simple:} This method assumes that the cooling time is simply
\begin{equation}
 t_{\rm cool} = {N \over 2} {{\rm k}_{\rm B} T n_{\rm tot} \over \Lambda},
\end{equation}
where $N=${\tt [coolingTimeSimpleDegreesOfFreedom]} is the number of degrees of freedom in the cooling gas which has temperature $T$ and total particle number density $n_{\rm tot}$ and $\Lambda$ is the cooling function (see \S\ref{sec:CoolingFunction}).

\subsubsection{Time Available for Cooling}\label{sec:CoolingTimeAvailable}\index{cooling!time available}

The method used to determine the time available for cooling (i.e. the time for which gas in a halo has been able to cool).

\emph{White \& Frenk:} This method assumes that the time available for cooling is equal to
\begin{equation}
 t_{\rm available} = \exp\left[ f \ln t_{\rm Universe} + (1-f)\ln t_{\rm dynamical} \right],
\end{equation}
where $f=${\tt [coolingTimeAvailableAgeFactor]} is an interpolating factor, $t_{\rm Universe}$ is the age of the Universe and $t_{\rm dynamical}$ is the dynamical time in the halo. The original \cite{white_galaxy_1991} algorithm corresponds to $f=1$.

\subsection{Dark Matter Halos}

Several algorithms are used to implement dark matter halos. These are described below.

\subsubsection{Density Profile}\label{sec:DarkMatterDensityProfile}

The method used to compute density profiles of dark matter halos is selected via the {\tt [darkMatterProfileMethod]} parameter.

\emph{Isothermal:} Under this method the density profile is given by:
\begin{equation}
 \rho_{\rm node}(r) \propto r^{-2},
\end{equation}
normalized such that the total mass of the node is enclosed with the virial radius.

\emph{NFW:} Under this method the \NFW\ density profile \citep{navarro_universal_1997} is used
\begin{equation}
  \rho_{\rm node}(r) \propto \left({r\over r_{\rm s}}\right)^{-1} \left[1 + {r\over r_{\rm s}} \right]^{-2},
\end{equation}
normalized such that the total mass of the node is enclosed with the virial radius and with scale length $r_{\rm s} = r_{\rm virial}/c$ where $c$ is the halo concentration (see \S\ref{sec:DmConcentration}).

\subsubsection{Dark Matter Density Profile Concentration}\label{sec:DmConcentration}

The method used to compute the concentration of dark matter halos.

\emph{Gao2008:}  Under this method the concentration is computed using a fitting function from \cite{gao_redshift_2008}:
\begin{equation}
\log_{10} c = A \log_{10} M_{\rm halo} + B.
\end{equation}
The parameters are a function of expansion factor, $a$. We use the following fits to the \cite{gao_redshift_2008} results:
\begin{eqnarray}
A &=& -0.14 \exp\left[-\left({\log_{10}a+0.05\over0.35}\right)^2\right], \\
B &=&  2.646 \exp\left[-\left({\log_{10}a\over0.50}\right)^2\right].
\end{eqnarray}

\subsubsection{Spin Parameter Distribution}\label{sec:SpinParameterDistribution}

The method used to compute the distribution of dark matter halo spin parameters is selected via the {\tt [haloSpinDistributionMethod]} parameter.

\emph{Lognormal:} Under this method the spin is drawn from a lognormal distribution.

\emph{Bett2007:} Under this method the spin is drawn from the distribution found by \cite{bett_spin_2007}. The $\lambda_0$ and $\alpha$ parameter of Bett et al.'s distribution are set by the {\tt [spinDistributionBett2007Lambda0]} and {\tt [spinDistributionBett2007Alpha]} input parameters.

\subsection{Disk Stability/Bar Formation}\label{sec:DiskStability}\index{disks!stability}\index{bar instability}

The method uses to compute the bar instability timescale for galactic disks is selected by {\tt [barInstabilityMethod]}.

\emph{Efstathiou, Lake \& Negroponte:} This method uses the stability criterion of \cite{efstathiou_stability_1982} to estimate when disks are unstable to bar formation:
\begin{equation}
 \epsilon \left( \equiv {V_{\rm peak} \over \sqrt{\G M_{\rm disk}/r_{\rm disk}}} \right) < \epsilon_{\rm c},
\end{equation}
for stability, where $V_{\rm peak}$ is the peak velocity in the rotation curve (computed here assuming an isolated exponential disk), $M_{\rm disk}$ is the mass of the disk and $r_{\rm disk}$ is its scale length (assuming an exponential disk). The value of $\epsilon_{\rm c}$ is linearly interpolated in the disk gas fraction between values for purely stellar and gaseous disks as specified by {\tt [stabilityThresholdStellar]} and {\tt [stabilityThresholdGaseous]} respectively. For disks which are judged to be unstable, the timescale for bar formation is taken to be
\begin{equation}
 t_{\rm bar} = t_{\rm disk} {\epsilon_{\rm c} - \epsilon_{\rm iso} \over \epsilon_{\rm c} - \epsilon},
\end{equation}
where $\epsilon_{\rm iso}$ is the value of $\epsilon$ for an isolated disk and $t_{\rm disk}$ is the disk dynamical time, defined as $r/V$ (where $V$ is the circular velocity) at one scale length. This form gives an infinite timescale at the stability threshold, reducing to a dynamical time for highly unstable disks.

\subsection{Galactic Structure}\label{sec:GalacticStructure}\index{galactic structure}

The algorithm to be used when solving for galactic structure (specifically, finding radii of galactic components) is specified via {\tt [galacticStructureRadiusSolverMethod]}.

\emph{Simple:} This method determines the sizes of galactic components by assuming that their self-gravity is negligible (i.e. that the gravitational potential well is dominated by dark matter) and that, therefore, baryons do not modify the dark matter density profile. The radius of a given component is then found by solving
\begin{equation}
 j = \sqrt{\G M_{\rm DM}(r) r},
\end{equation}
where $j$ is the specific angular momentum of the component (at whatever point in the profile is to be solved for), $r$ is radius and $M(r)$ is the mass of dark matter within radius $r$.

\emph{Adiabatic:} This method takes into account the baryonic self-gravity of all galactic components when solving for structure and additionally accounts for back reaction of the baryons on the dark matter density profile using the adiabatic contraction algorithm of \cite{gnedin_response_2004}. The parameters $A$ and $\omega$ of that model are specified via input parameters {\tt [adiabaticContractionGnedinA]} and {\tt [adiabaticContractionGnedinOmega]} respectively. Solution proceeds via an iterative procedure to find equilibrium radii for all galactic components in a consistently contracted halo. The method used follows that described by \cite{benson_galaxy_2010-1}.

\subsection{Galaxy Merging}\index{galaxy!merging}\index{merging!galaxy}

The process of merging two galaxies currently involves two algorithms: one which decides how the merger causes mass components from both galaxies to move and one which determines the size of the remnant galaxy.

\subsubsection{Mass Movements}\label{sec:MergingMassMove}

The movement of mass elements in the merging galaxies. In the following, $M_1$ and $M_2$ are the baryonic masses of the satellite and central galaxies respectively that are about to merge.

\emph{Simple:} This method implements mass movements according to:
\begin{itemize}
 \item If $M_1 > f_{\rm major} M_2$ then all mass from both satellite and central galaxies moves to the spheroid component of the central galaxy;
 \item Otherwise: Gas from the satellite moves to the component of the central specified by the {\tt [minorMergerGasMovesTo]} parameter (either ``{\tt disk}'' or ``{\tt spheroid}''), stars from the satellite move to the spheroid of the central and mass in the central does not move.
\end{itemize}
Here, $f_{\rm major}=${\tt [majorMergerMassRatio]} is the mass ratio above which a merger is considered to be ``major''.

\subsubsection{Remnant Sizes}\label{sec:MergingSizes}\index{merging!remnant size}

The method used to calculate the sizes of merger remnants.

\emph{Null:} This is a null method which does nothing at all. It is useful, for example, when running \glc\ to study dark matter only (i.e. when no galaxy properties are computed).

\emph{Cole et al. (2000):} This method uses the algorithm of \cite{cole_hierarchical_2000} to compute merger remnant spheroid sizes. Specifically
\begin{equation}
\frac{(M_1+M_2)^2}{ r_{\rm new}} =
\frac{M_1^2}{r_1} + \frac{M_2^2}{r_2} + \frac{ f_{\rm orbit}}{c}
\frac{M_1 M_2}{r_1+r_2},
\end{equation}
where $M_1$ and $M_2$ are the baryonic masses of the merging galaxies and $r_1$
and $r_2$ are their half mass radii, $r_{\rm new}$ is the half mass radius of the spheroidal component of the remnant galaxy and $c$ is a constant which depends on the distribution of the mass. For a Hernquist spheroid $c=0.40$ can be found by numerical integration while for a exponential disk $c=0.49$. For simplicity a value of $c=0.5$ is adopted for all components. The parameter $f_{\rm orbit}=${\tt [mergerRemnantSizeOrbitalEnergy]} depends on the orbital parameters of the galaxy pair. For example, a value of $f_{\rm orbit} = 1$ corresponds to point mass galaxies in circular orbits about their center of mass. 

\subsection{Hot Halo Density Profile}\label{sec:HotHaloDensity}\index{hot halo!density profile}\index{density profile!hot halo}

The hot halo density profile is used, for example, in calculations of cooling rates.

\emph{Cored Isothermal:} This method adopts a spherically symmetric cored-isothermal density profile for the hot halo. Specifically,
\begin{equation}
 \rho_{\rm hot~halo}(r) \propto \left[ r^2 + r_{\rm core}^2 \right]^{-1},
\end{equation}
where the core radius, $r_{\rm core}$, is set to be a fixed fraction of the virial radius, that fraction being given by the input parameter {\tt [isothermalCoreRadiusOverVirialRadius]}. The profile is normalized such that the current mass in the hot gas profile is contained within the virial radius.

\subsection{Hot Halo Temperature Profile}\label{sec:HotHaloTemperature}\index{hot halo!temperature profile}\index{temperature profile!hot halo}

The hot halo temperature profile is used, for example, in calculations of cooling rates.

\emph{Virial Temperature:} This method assumes an isothermal halo with a temperature equal to the virial temperature of the halo.

\subsection{Initial Mass Functions}\label{sec:StellarIMF}\index{initial mass function}

The stellar \IMF{}subsystem within \glc\ supports multiple IMFs and extensible algorithms to select which \IMF\ to use based on the physical conditions of star formation.

\subsubsection{Initial Mass Function Selection}\label{sec:ImfSelect}\index{initial mass function!selection}

The method to use for selecting which \IMF\ to use.

\emph{Fixed:} This method uses a fixed \IMF\ irrespective of physical conditions. The \IMF\ to use is specified by the {\tt [imfSelectionFixed]} parameter (e.g. setting this parameter to {\tt Salpeter} selects the Salpeter \IMF).

\subsubsection{Initial Mass Functions}\label{sec:physicsIMF}\index{initial mass function}

A variety of different \IMF s are available. Each \IMF\ supplies a recycled fraction and metal yield for use in the instantaneous recycling approximation. These can be set via the parameters {\tt imf\{imfName\}RecycledInstantaneous} and {\tt imf\{imfName\}YieldInstantaneous} where {\tt \{imfName\}} is the name of the \IMF.

\emph{Chabrier:} The {\tt Chabrier} \IMF\ is defined by \citep{chabrier_galactic_2001}:
\begin{equation}
 \phi(M) \propto \left\{ \begin{array}{ll}
 M^{-1} {\rm e}^{-[\left.\log_{10}({M\over M_{\rm c}})\right/\sigma_{\rm c}]^2/2} & \hbox{ for } 0.1M_\odot \\ & \hbox{  }< M < 1M_\odot \\
 M^{-2.3} & \hbox{ for } 1M_\odot \\ & \hbox{  }< M < 125M_\odot \\
 0 & \hbox {otherwise,} \end{array} \right.
\end{equation}
where $\sigma_{\rm c}=0.69$ and $M_{\rm c}=0.08M_\odot$.

\emph{Kennicutt:} The {\tt Kennicutt} \IMF\ is defined by \citep{kennicutt_rate_1983}:
\begin{equation}
 \phi(M) \propto \left\{ \begin{array}{ll}
 M^{-1.25} & \hbox{ for } 0.10M_\odot < M < 1.00M_\odot \\
 M^{-2.00} & \hbox{ for } 1.00M_\odot < M < 2.00M_\odot \\
 M^{-2.30} & \hbox{ for } 2.00M_\odot < M < 125M_\odot \\
 0 & \hbox {otherwise.} \end{array} \right.
\end{equation}

\emph{Kroupa:} The {\tt Kroupa} \IMF\ is defined by \citep{kroupa_variation_2001}:
\begin{equation}
 \phi(M) \propto \left\{ \begin{array}{ll}
 M^{-0.3} & \hbox{ for } 0.01M_\odot < M < 0.08M_\odot \\ 
 M^{-1.8} & \hbox{ for } 0.08M_\odot < M < 0.5M_\odot \\ 
 M^{-2.7} & \hbox{ for } 0.5M_\odot < M < 1M_\odot \\ 
 M^{-2.3} & \hbox{ for } 1M_\odot < M < 125M_\odot \\ 
0 & \hbox {otherwise.} \end{array} \right.
\end{equation}

\emph{Miller-Scalo:} The {\tt Miller-Scalo} \IMF\ is defined by \citep{miller_initial_1979}:
\begin{equation}
 \phi(M) \propto \left\{ \begin{array}{ll}
 M^{-1.25} & \hbox{ for } 0.10M_\odot < M < 1.00M_\odot \\
 M^{-2.00} & \hbox{ for } 1.00M_\odot < M < 2.00M_\odot \\
 M^{-2.30} & \hbox{ for } 2.00M_\odot < M < 10.0M_\odot \\
 M^{-3.30} & \hbox{ for } 10.0M_\odot < M < 125M_\odot \\
 0 & \hbox {otherwise.} \end{array} \right.
\end{equation}

\emph{Salpeter:} The {\tt Salpeter} \IMF\ is defined by \citep{salpeter_luminosity_1955}:
\begin{equation}
 \phi(M) \propto \left\{ \begin{array}{ll} M^{-2.35} & \hbox{ for } 0.1M_\odot < M < 125M_\odot \\ 0 & \hbox {otherwise.} \end{array} \right.
\end{equation}

\emph{Scalo:} The {\tt Scalo} \IMF\ is defined by \citep{scalo_stellar_1986}:
\begin{equation}
 \phi(M) \propto \left\{ \begin{array}{ll}
 M^{+1.60} & \hbox{ for } 0.10M_\odot < M < 0.18M_\odot \\
 M^{-1.01} & \hbox{ for } 0.18M_\odot < M < 0.42M_\odot \\
 M^{-2.75} & \hbox{ for } 0.42M_\odot < M < 0.62M_\odot \\
 M^{-2.08} & \hbox{ for } 0.62M_\odot < M < 1.18M_\odot \\
 M^{-3.50} & \hbox{ for } 1.18M_\odot < M < 3.50M_\odot \\
 M^{-2.63} & \hbox{ for } 3.50M_\odot < M < 125M_\odot \\
 0 & \hbox {otherwise.} \end{array} \right.
\end{equation}

\subsection{Ionization State}\label{sec:IonizationState}\index{ionization state}

The ionization state of gas, including, for example, the electron density, as a function of the gas density, composition and temperature.

\emph{Atomic Collisional Ionization Equilibrium Using {\sc Cloudy}:} This method computes the ionization state using the {\sc Cloudy} code and under the assumption of collisional ionization equilibrium with no molecular contribution. Abundances are Solar, except for zero metallicity calculations which use {\sc Cloudy}'s ``primordial'' metallicity. The helium abundance for non-zero metallicity is scaled linearly with metallicity between primordial and Solar values. The {\sc Cloudy} code will be downloaded and run to compute the cooling function as needed, which will then be stored for future use. As this process is slow, a precomputed table is provided with \glc. If metallicities outside the range tabulated in this file are required it will be regenerated with an appropriate range.

\emph{Collisional Ionization Equilibrium From File:} In this method the ionization state is read from a file. The ionization state is assumed to be computed under conditions of collisional ionization equilibrium and therefore densities of all species scale as the total density, $\rho$.

\subsection{Merger Tree Construction}\label{sec:MergerTreeConstruct}\index{merger trees}

Merger trees are ``constructed\footnote{By ``construct'' we mean any process of creating a representation of a merger tree within \protect\glc.}'' using a method specified via {\tt [mergerTreeConstructMethod]}.

\emph{Read From File:} This method reads merger tree structures from an HDF5 file.

\emph{Build:} This method first creates a distribution of tree root halo masses at a specified final redshift and then builds a merger tree using the selected build algorithm (see \S\ref{sec:MergerTreeBuild}). The root halo masses are selected to lie within a user specified range, with a specified average number of trees per decade of halo mass. The distribution of halo masses is such that the mass of the $i^{\rm th}$ halo is
\begin{eqnarray}
 M_{\rm halo,i} &=& \exp\left[ \ln(M_{\rm halo,min}) \right. \nonumber \\
 & & \left. + \ln\left({M_{\rm halo,max}/M_{\rm halo,min}}\right) x_i^{1+\alpha} \right].
\end{eqnarray}
Here, $x_i$ is a number between 0 and 1 and $\alpha$ is an input parameter that controls the relative number of low and high mass tree produced. The distribution of $x$ is controlled by an input parameter with options:
\begin{description}
 \item [{\tt uniform}] $x$ is distributed uniformly between 0 and 1;
 \item [{\tt quasi}] $x$ is distributed using a quasi-random sequence.
\end{description}

\subsection{Merger Tree Branching}\label{sec:MergerTreeBranching}\index{merger trees!branching}

The method to be used for computing branching probabilities in merger trees when trees are constructed using Monte Carlo techniques.

\begin{figure*}
 \begin{center}
 \includegraphics[height=160mm,angle=90]{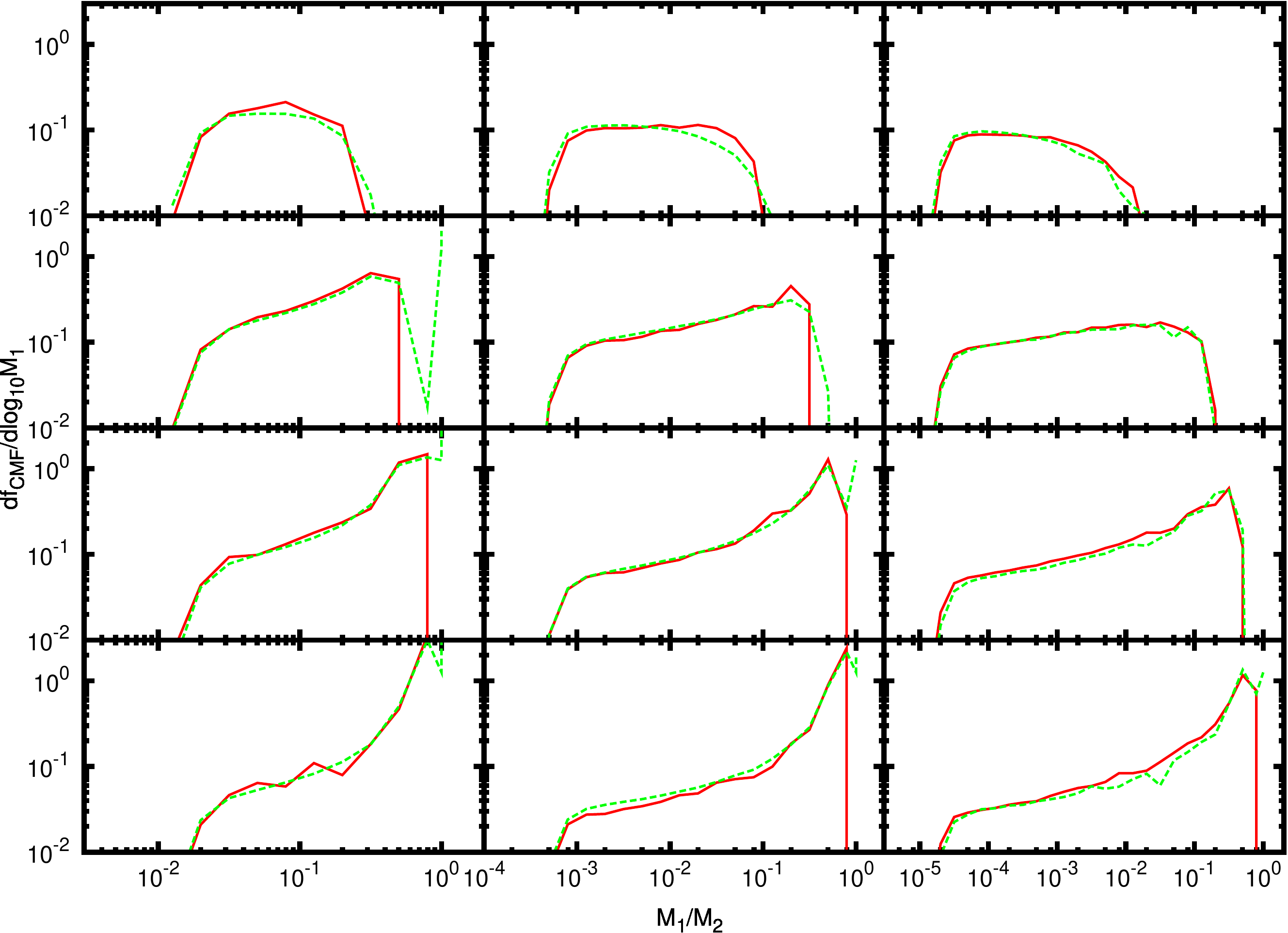}
 \end{center}
 \caption{Progenitor mass functions at redshifts $z=0.5$, 1, 2 and 4 (bottom to top) for halos of mass $10^{12\pm0.151}$, $10^{13.5\pm0.151}$ and $10^{15\pm0.151}h^{-1}M_\odot$ (left to right) at $z=0$ are shown. Here, $M_1$ is the mass of the progenitor halo and $M_2$ the mass of the $z=0$ halo. Green lines were measured from the Millennium Simulation \protect\citep{springel_simulations_2005} by \protect\cite{parkinson_generating_2008}, while red lines are computed using \glc's merger tree building routines (with the \cite{parkinson_generating_2008} branching algorithm and the \cite{cole_hierarchical_2000} tree building algorithm).}
 \label{fig:PCH_Progenitor_MFs}
\end{figure*}

\emph{Modified Press-Scheduler:} This method uses the algorithm of \cite{parkinson_generating_2008} to compute branching ratios. The parameters $G_0$, $\gamma_1$ and $\gamma_2$ of their algorithm are specified by the input parameters {\tt [modifiedPressSchechterG0]}, {\tt [modifiedPressSchechterGamma1]} and {\tt [modifiedPressSchechterGamma2]} respectively. Additionally, the parameter {\tt [modifiedPressSchechterFirstOrderAccuracy]} limits the step in the critical linear theory overdensity for collapse, $\delta_{\rm crit}$, so that it never exceeds {\tt [modifiedPressSchechterFirstOrderAccuracy]} times $\sqrt{2[\sigma^2(M_2/2)-\sigma^2(M_2)]}$, where $M_2$ is the mass of the halo being considered for branching and $\sigma(M)$ is the \CDM\ mass variance computed by filtering the power spectrum using top-hat spheres. This ensures that the first order expansion of the merging rate that is assumed is accurate. Progenitor mass functions computed by this branching algorithm when used in the \cite{cole_hierarchical_2000} merger tree building algorithm are shown in Fig.~\ref{fig:PCH_Progenitor_MFs} where they are compared with equivalent mass functions measured from the Millennium Simulation \citep{springel_simulations_2005}. Clearly there is excellent agreement with the N-body results.

\subsection{Merger Tree Building}\label{sec:MergerTreeBuild}\index{merger trees!building}

The method to be used for building merger trees.

\emph{Cole et al. (2000) Algorithm:} This method uses the algorithm described by \cite{cole_hierarchical_2000}, with a branching probability method selected via the {\tt treeBranchingMethod} parameter. This action of this algorithm is controlled by the following parameters:
\begin{description}
 \item [{\tt [mergerTreeBuildCole2000MergeProbability]}] The maximum probability for a binary merger allowed in a single timestep. This allows the probability to be kept small, such the the probability for multiple mergers within a single timestep is small.
 \item [{\tt [mergerTreeBuildCole2000AccretionLimit]}] The maximum fractional change in mass due to sub-resolution accretion allowed in any given timestep when building the tree.
 \item [{\tt [mergerTreeBuildCole2000MassResolution]}] The minimum halo mass that the algorithm will follow. Mass accretion below this scale is treated as smooth accretion and branches are truncated once they fall below this mass.
\end{description}

\emph{Smooth Accretion:} This method builds a branchless merger tree with a smooth accretion history using the fitting formula of \cite{wechsler_concentrations_2002}. The tree is defined by a final mass at a specified redshift and is continued back in time by decreasing the halo mass by a specified factor at each new node until a specified mass resolution is reached. The fitting formula of \cite{wechsler_concentrations_2002} has one free parameter, the formation redshift. The formation redshift can either be computed automatically using the method of \cite{bullock_profiles_2001}, or specified by an input parameter.

\subsection{Star Formation Timescales}\label{sec:StarFormationTimescales}\index{star formation!timescale}

The methods for computing star formation timescales in disks and spheroids.

\emph{Power Law:} This method (as used by \cite{cole_hierarchical_2000} for example) computes the star formation timescale to be:
\begin{equation}
 \tau_\star = \epsilon_\star^{-1} \tau_{\rm dynamical} \left( {V \over 200\hbox{km/s}} \right)^{\alpha_\star},
\end{equation}
where $\epsilon_\star=${\tt [starFormation\{Component\}Efficiency]} and $\alpha_\star=${\tt [starFormation\{Component\}VelocityExponent]} are input parameters (with {\tt \{Component\}} being a placeholder for {\tt Disk} or {\tt Spheroid}), $\tau_{\rm dynamical}\equiv r/V$ is the dynamical timescale of the component and $r$ and $V$ are the characteristic radius and velocity respectively of the component. The timescale is not allowed to fall below a minimum value specified by {\tt [starFormation\{Component\}MinimumTimescale]}.

\subsection{Stellar Population Properties}\label{sec:StellarPopulationProperties}\index{stellar populations}

The algorithm for determining stellar population properties---essentially the rates of change of stellar and gas mass and abundances given a star formation rate and fuel abundances (and perhaps a historical record of star formation in the component)--is specified by {\tt [stellarPopulationPropertiesMethod]}.

\emph{Instantaneous:} This method uses the instantaneous recycling approximation. Specifically, given a star formation rate $\phi$, this method assumes a rate of increase of stellar mass of $\dot{M}_\star=(1-R)\phi$, and a corresponding rate of decrease in fuel mass. The rate of change of the metal content of stars follows from the fuel metallicity, while that of the fuel changes according to
\begin{equation}
 \dot{M}_{fuel,Z} = - (1-R) Z_{\rm fuel} \phi + y \phi.
\end{equation}
In the above $R$ is the instantaneous recycled fraction and $y$ is the yield, both of which are supplied by the \IMF\ subsystem (see \S\ref{sec:StellarIMF}). The rate of energy input from the stellar population is computed assuming that the canonical amount of energy from a single stellar population (as defined by the {\tt feedbackEnergyInputAtInfinityCanonical}) is input instantaneously.

\emph{Non-instantaneous:} This method assumes fully non-instantaneous recycling and metal enrichment. Recycling and metal production rates from simple stellar populations are computed, for any given \IMF, from stellar evolution models. The rates of change are then:
\begin{eqnarray}
 \dot{M}_\star &=& \phi - \int_0^t \phi(t^\prime) \dot{R}(t-t^\prime;Z_{\rm fuel}[t^\prime]) \d t^\prime, \\
 \dot{M}_{\rm fuel} &=& -\phi + \int_0^t \phi(t^\prime) \dot{R}(t-t^\prime;Z_{\rm fuel}[t]) \d t^\prime, \\
 \dot{M}_{\star,Z} &=& Z_{\rm fuel} \phi \nonumber \\ 
 & & - \int_0^t \phi(t^\prime) Z_{\rm fuel}(t^\prime) \nonumber \\
 & & \times \dot{R}(t-t^\prime;Z_{\rm fuel}[t^\prime]) \d t^\prime, \\
 \dot{M}_{{\rm fuel},Z} &=& -Z_{\rm fuel} \phi \nonumber \\
 & & + \int_0^t  \phi(t^\prime) \{ Z_{\rm fuel}(t^\prime) \dot{R}(t-t^\prime;Z_{\rm fuel}[t^\prime]) \nonumber \\
 & & + \dot{p}(t-t^\prime;Z_{\rm fuel}[t^\prime]) \} \d t^\prime,
\end{eqnarray}
where $\dot{R}(t;Z)$ and $\dot{p}(t;Z)$ are the recycling and metal yield rates respectively from a stellar population of age $t$ and metallicity $Z$. The energy input rate is computed self-consistently from the star formation history.

\subsection{Stellar Population Spectra}

Stellar population spectra are used to construct integrated spectra of galaxies.

\emph{Conroy, White \& Gunn:} This method uses v2.1 of the \href{http://www.cfa.harvard.edu/~cconroy/FSPS.html}{{\tt FSPS}} code of \cite{conroy_propagation_2009} to compute stellar spectra. If necessary, the {\tt FSPS} code will be downloaded, patched\footnote{Patching is not to fix any bugs in FSPS, but merely to insert code for reading a tabulated \protect\IMF\ output by \protect\glc.} and compiled and run to generate spectra. These tabulations are then stored to file for later retrieval.

\emph{File:} This method reads stellar population spectra from an HDF5 file.

\subsection{Stellar Population Spectra Post-processing}

Stellar population spectra are post-processed (to handle, for example, absorption by the \IGM) by any number of the following algorithms.

\emph{Meiksin (2006) IGM Attenuation:} This method post-processes spectra through absorption by the \IGM\ using the results of \cite{meiksin_colour_2006}.

\emph{Madau (1995) IGM Attenuation:} This method post-processes spectra through absorption by the \IGM\ using the results of \cite{madau_radiative_1995}.

\emph{Null Method:} This method performs no post-processing.

\subsection{Stellar Astrophysics}

Various properties related to stellar astrophysics are required by \glc. The following documents their implementation.

\subsubsection{Basics}

This subset of properties include recycled mass, metal yield and lifetime.

\emph{File:} This method reads properties of individual stars of different initial mass and metallicity from an XML file and interpolates in them. The stars can be irregularly spaced in the plane of initial mass and metallicity.

\subsubsection{Stellar Winds}

Energy input to the \ISM\ from stellar winds is used in calculations of feedback efficiency.

\emph{Leitherer et al. (1992):} This method uses the fitting formulae of \cite{leitherer_deposition_1992} to compute stellar wind energy input from the luminosity and effective temperature of a star (see \S\ref{sec:StellarTracks}).

\subsubsection{Stellar Tracks}\label{sec:StellarTracks}

The method used to compute stellar evolutionary tracks.

\emph{File:} This method luminosities and effective temperatures of stars are computed from a tabulated set of stellar tracks. \glc\ is supplied with a suitable file constructed from the tracks of \cite{bertelli_scaled_2008} and \cite{bertelli_scaled_2009}.

\subsubsection{Supernovae Type Ia}\index{supernovae!Type Ia}

Properties of Type Ia supernovae, including the cumulative number occurring and metal yield.

\emph{Nagashima et al. (2005) Prescription:} This method uses the prescriptions from \cite{nagashima_metal_2005} to compute the numbers and yields of Type Ia supernovae.

\subsubsection{Population III Supernovae}\index{supernovae!Population III}\index{Population III!supernovae}

Properties of Population III specific supernovae, in particular the energy released.

\emph{Heger \& Woosley (2002):} This method computes the energies of pair instability supernovae from the results of \cite{heger_nucleosynthetic_2002}.

\subsubsection{Stellar Feedback}

Aspects of stellar feedback, such as the energy input to the surrounding \ISM.

\emph{Standard:} This method assumes that the cumulative energy input from a stellar population is equal to the total number of (Type II and Type Ia) supernovae multiplied by a specified energy per \SNe\ plus any Population III-specific supernovae energy plus the integrated energy input from stellar winds. The number of Type II \SNe\ is computed automatically from the \IMF\ and a specified minimum mass required for a  Type II supernova.

\subsection{Substructure and Merging}\index{merging!substructure}\index{substructure}

Substructures and merging of nodes/substructures is controlled by several algorithms which are described below:

\subsubsection{Merging Timescales}\label{sec:MergingTimescales}\index{merging!dynamical friction}

The method used to compute merging timescales of substructures is specified via the {\tt [satelliteMergingMethod]} parameter.

\emph{Dynamical Friction: Lacey \& Cole:} This method computes merging timescales using the dynamical friction calculation of \cite{lacey_merger_1993}. Timescales are multiplied by the value of the {\tt [mergingTimescaleMultiplier]} input parameter.

\emph{Dynamical Friction: Jiang (2008):} This method computes merging timescales using the dynamical friction calibration of \cite{jiang_fitting_2008}.

\emph{Dynamical Friction: Boylan-Kolchin (2008):} This method computes merging timescales using the dynamical friction calibration of \cite{boylan-kolchin_dynamical_2008}.

\subsubsection{Virial Orbits}\label{sec:VirialOrbits}

The algorithm to be used to determine orbital parameters of substructures when they first enter the virial radius of their host.

\emph{Benson (2005):} This method selects orbital parameters randomly from the distribution given by \cite{benson_orbital_2005}.

\subsubsection{Node Merging}

The algorithm to be used to process nodes when they become substructures.

\emph{Single Level Hierarchy:} This method maintains a single level hierarchy of substructure, i.e. it tracks only substructures, not sub-substructures or deeper levels. When a node first becomes a satellite it is appended to the list of satellites associated with its host halo. If the node contains its own satellites they will be detached from the node and appended to the list of satellites of the new host (and assigned new merging times; see \S\ref{sec:MergingTimescales}).

\subsection{Supernovae Feedback Models}\label{sec:SNeFeedback}\index{supernovae!feedback}\index{feedback}

The supernovae feedback driven outflow rate for disks and spheroids.

\emph{Power Law:} This method assumes an outflow rate of:
\begin{equation}
 \dot{M}_{\rm outflow} = \left({V_{\rm outflow} \over V}\right)^{\alpha_{\rm outflow}} {\dot{E} \over E_{\rm canonical}},
\end{equation}
where $V_{\rm outflow}=${\tt [\{component\}OutflowVelocity]} and $\alpha_{\rm outflow}=${\tt [\{component\}OutflowExponent]} are input parameters, $V$ is the characteristic velocity of the component, $\dot{E}$ is the rate of energy input from stellar populations and $E_{\rm canonical}$ is the total energy input by a canonical stellar population normalized to $1 M_\odot$ after infinite time.

\subsection{Supermassive Black Holes Binary Mergers}\label{sec:BlackHoleBinaryMergers}\index{supermassive black holes!mergers}

The method to be used for computing the effects of binary mergers of supermassive black holes.

\emph{Rezzolla et al. (2008):} This method uses the fitting function of \cite{rezzolla_final_2008} to compute the spin of the black hole resulting from a binary merger. The mass of the resulting black hole is assumed to equal the sum of the mass of the initial black holes (i.e. there is negligible energy loss through gravitational waves).

\section{Example Calculations}\label{sec:Example}

In this section we show results from an example \glc\ model. The parameters of this model are listed in Table~\ref{tb:ExampleParams}. It should be noted that the values of many of these parameters were chosen on the basis of previous experience with similar models---we have not performed an extensive search of the available parameter space, either manually or in an automated manner. As such, we do not claim that this example model represents the best match to observational data that can be obtained with the standard implementation of \glc---indeed it is certainly not so, although it does achieve reasonably good matches to many $z\approx 0$ observational datasets. Instead, this example model merely serves to illustrate the type of predictions which can be extracted from \glc. A detailed search of the available parameter space will be presented in a future paper.

\begin{table*}
\caption{Values of parameters used in the example model. Parameters selecting between different implementations of physical processes or components are only listed where more than one non-null implementation currently exists within \protect\glc.}
\label{tb:ExampleParams}
\begin{center}
\begin{tabular}{lll}
\hline \\
Parameter & Value & Reference \\
\hline \\
{\tt [H\_0]} & 70.2~km/s & \S\ref{sec:Cosmology}; \citep{komatsu_seven-year_2010} \\
{\tt [Omega\_0]} & 0.2725 & \S\ref{sec:Cosmology}; \citep{komatsu_seven-year_2010} \\
{\tt [Omega\_DE]} & 0.7275 & \S\ref{sec:Cosmology}; \citep{komatsu_seven-year_2010} \\
{\tt [Omega\_b]} & 0.0455 & \S\ref{sec:Cosmology}; \citep{komatsu_seven-year_2010} \\
{\tt [T\_CMB]} & 2.72548~K & \S\ref{sec:Cosmology}; \citep{komatsu_seven-year_2010} \\
{\tt [accretionDisksMethod]} & ADAF & \S\ref{sec:CircumnuclearDisks} \\
{\tt [adafAdiabaticIndex]} & 1.444 & \S\ref{sec:CircumnuclearDisks} \\
{\tt [adafEnergyOption]} & pure ADAF & \S\ref{sec:CircumnuclearDisks} \\
{\tt [adafRadiativeEfficiency]} & 0.01 & \S\ref{sec:CircumnuclearDisks} \\
{\tt [adafViscosityOption]} & fit & \S\ref{sec:CircumnuclearDisks} \\
{\tt [adiabaticContractionGnedinA]} & 0.8 & \S\ref{sec:GalacticStructure} \\
{\tt [adiabaticContractionGnedinOmega]} & 0.77 &\S\ref{sec:GalacticStructure} \\
{\tt [barInstabilityMethod]} & ELN & \S\ref{sec:DiskStability} \\
{\tt [blackHoleSeedMass]} & 100 & \S\ref{sec:BlackHoleStandard} \\
{\tt [blackHoleWindEfficiency]} & 0.001 & \S\ref{sec:BlackHoleStandard} \\
{\tt [bondiHoyleAccretionEnhancementHotHalo]} & 1 & \S\ref{sec:BlackHoleStandard} \\
{\tt [bondiHoyleAccretionEnhancementSpheroid]} & 1 & \S\ref{sec:BlackHoleStandard} \\
{\tt [bondiHoyleAccretionTemperatureSpheroid]} & 100 & \S\ref{sec:BlackHoleStandard} \\
{\tt [coolingFunctionMethod]} & atomic CIE Cloudy & \S\ref{sec:CoolingFunction} \\
{\tt [coolingTimeAvailableAgeFactor]} & 0 & \S\ref{sec:CoolingTimeAvailable} \\
{\tt [coolingTimeSimpleDegreesOfFreedom]} & 3 & \S\ref{sec:CoolingTime} \\
{\tt [darkMatterProfileMethod]} & NFW & \S\ref{sec:DarkMatterDensityProfile} \\
{\tt [darkMatterProfileMinimumConcentration]} & 4 & \S\ref{sec:ComponentDarkMatterProfileScale} \\
{\tt [diskOutflowExponent]} & 2 & \S\ref{sec:SNeFeedback} \\
{\tt [diskOutflowVelocity]} & 200~km/s & \S\ref{sec:SNeFeedback} \\
{\tt [effectiveNumberNeutrinos]} & 4.34 & \S\ref{sec:TransferFunction} \\
{\tt [galacticStructureRadiusSolverMethod]} & adiabatic & \S\ref{sec:GalacticStructure} \\
{\tt [haloMassFunctionMethod]} & Tinker2008 & \S\ref{sec:HaloMassFunction} \\
{\tt [haloSpinDistributionMethod]} & Bett2007 & \S\ref{sec:SpinParameterDistribution} \\
{\tt [hotHaloOutflowReturnRate]} & 1.26 & \S\ref{sec:HotHaloStandard} \\
{\tt [imfSalpeterRecycledInstantaneous]} & 0.39 & \S\ref{sec:physicsIMF} \\
{\tt [imfSalpeterYieldInstantaneous]} & 0.02 & \S\ref{sec:physicsIMF} \\
{\tt [imfSelectionFixed]} & Salpeter & \S\ref{sec:ImfSelect} \\
{\tt [isothermalCoreRadiusOverVirialRadius]} & 0.1 & \S\ref{sec:HotHaloDensity} \\
\hline
\end{tabular}
\end{center}
\end{table*}

\begin{table*}
\addtocounter{table}{-1}
\caption{\emph{(cont.)} Values of parameters used in the example model. Parameters selecting between different implementations of physical processes or components are only listed where more than one non-null implementation currently exists within \protect\glc.}
\begin{center}
\begin{tabular}{lll}
\hline \\
Parameter & Value & Reference \\
\hline \\
{\tt [majorMergerMassRatio]} & 0.1 & \S\ref{sec:MergingMassMove} \\
{\tt [mergerRemnantSizeOrbitalEnergy]} & 1 & \S\ref{sec:MergingSizes} \\
{\tt [mergerTreeBuildCole2000AccretionLimit]} & 0.1 & \S\ref{sec:MergerTreeBuild} \\
{\tt [mergerTreeBuildCole2000MassResolution]} & $5\times10^9M_\odot$ & \S\ref{sec:MergerTreeBuild} \\
{\tt [mergerTreeBuildCole2000MergeProbability]} & 0.1 & \S\ref{sec:MergerTreeBuild} \\
{\tt [mergerTreeConstructMethod]} & build & \S\ref{sec:MergerTreeConstruct} \\
{\tt [minorMergerGasMovesTo]} & spheroid & \S\ref{sec:MergingMassMove} \\
{\tt [modifiedPressSchechterFirstOrderAccuracy]} & 0.1 & \S\ref{sec:MergerTreeBranching} \\
{\tt [modifiedPressSchechterG0]} & 0.57 & \S\ref{sec:MergerTreeBranching} \\
{\tt [modifiedPressSchechterGamma1]} & 0.38 & \S\ref{sec:MergerTreeBranching} \\
{\tt [modifiedPressSchechterGamma2]} & -0.01 & \S\ref{sec:MergerTreeBranching} \\
{\tt [powerSpectrumIndex]} & 0.961 & \S\ref{sec:PrimordialPowerSpectrum}; \citep{komatsu_seven-year_2010} \\
{\tt [powerSpectrumReferenceWavenumber]} & 1~Mpc$^{-1}$ & \S\ref{sec:PrimordialPowerSpectrum}; \citep{komatsu_seven-year_2010} \\
{\tt [powerSpectrumRunning]} & 0 & \S\ref{sec:PrimordialPowerSpectrum}; \citep{komatsu_seven-year_2010} \\
{\tt [randomSpinResetMassFactor]} & 2 & \S\ref{sec:SpinsRandom} \\
{\tt [reionizationSuppressionRedshift]} & 9 & \S\ref{sec:AccretionBaryonic} \\
{\tt [reionizationSuppressionVelocity]} & 30~km/s & \S\ref{sec:AccretionBaryonic} \\
{\tt [satelliteMergingMethod]} & Jiang2008 & \S\ref{sec:MergingTimescales} \\
{\tt [sigma\_8]} & 0.807 & \S\ref{sec:PrimordialPowerSpectrum} \& \S\ref{sec:TransferFunction} \\
{\tt [spheroidEnergeticOutflowMassRate]} & 1 & \S\ref{sec:SpheroidHernquist} \\
{\tt [spheroidOutflowExponent]} & 2 & \S\ref{sec:SNeFeedback} \\
{\tt [spheroidOutflowVelocity]} & 50~km/s & \S\ref{sec:SNeFeedback} \\
{\tt [spinDistributionBett2007Alpha]} & 2.509 & \S\ref{sec:SpinParameterDistribution} \\
{\tt [spinDistributionBett2007Lambda0]} & 0.04326 & \S\ref{sec:SpinParameterDistribution} \\
{\tt [stabilityThresholdGaseous]} & 0.9 & \S\ref{sec:DiskStability} \\
{\tt [stabilityThresholdStellar]} & 1.1 & \S\ref{sec:DiskStability} \\
{\tt [starFormationDiskEfficiency]} & 0.01 & \S\ref{sec:StarFormationTimescales} \\
{\tt [starFormationDiskMinimumTimescale]} & 0.001~Gyr & \S\ref{sec:StarFormationTimescales} \\
{\tt [starFormationDiskVelocityExponent]} & -1.5 & \S\ref{sec:StarFormationTimescales} \\
{\tt [starFormationSpheroidEfficiency]} & 0.1 & \S\ref{sec:StarFormationTimescales} \\
{\tt [starFormationSpheroidMinimumTimescale]} & 0.001~Gyr & \S\ref{sec:StarFormationTimescales} \\
{\tt [starveSatellites]} & true & \S\ref{sec:HotHaloStandard} \\
{\tt [stellarPopulationPropertiesMethod]} & instantaneous & \S\ref{sec:StellarPopulationProperties} \\
{\tt [summedNeutrinoMasses]} & 0 & \S\ref{sec:TransferFunction} \\
{\tt [transferFunctionMethod]} & Eisenstein + Hu & \S\ref{sec:TransferFunction} \\
{\tt [virialDensityContrastMethod]} & spherical top hat & \S\ref{sec:VirialDensityConstrast} \\
\hline
\end{tabular}
\end{center}
\end{table*}

\subsection{Numerical Results}

The numerical solution of ODEs in \glc\ is controlled by various parameters which affect the accuracy of solution. In this subsection we explore the numerical convergence of \glc\ with respect to these results. Additionally, we examine the time taken to evolve merger trees in the example model.

\subsubsection{Convergence}

\begin{figure}
 \begin{center}
 \includegraphics[width=80mm]{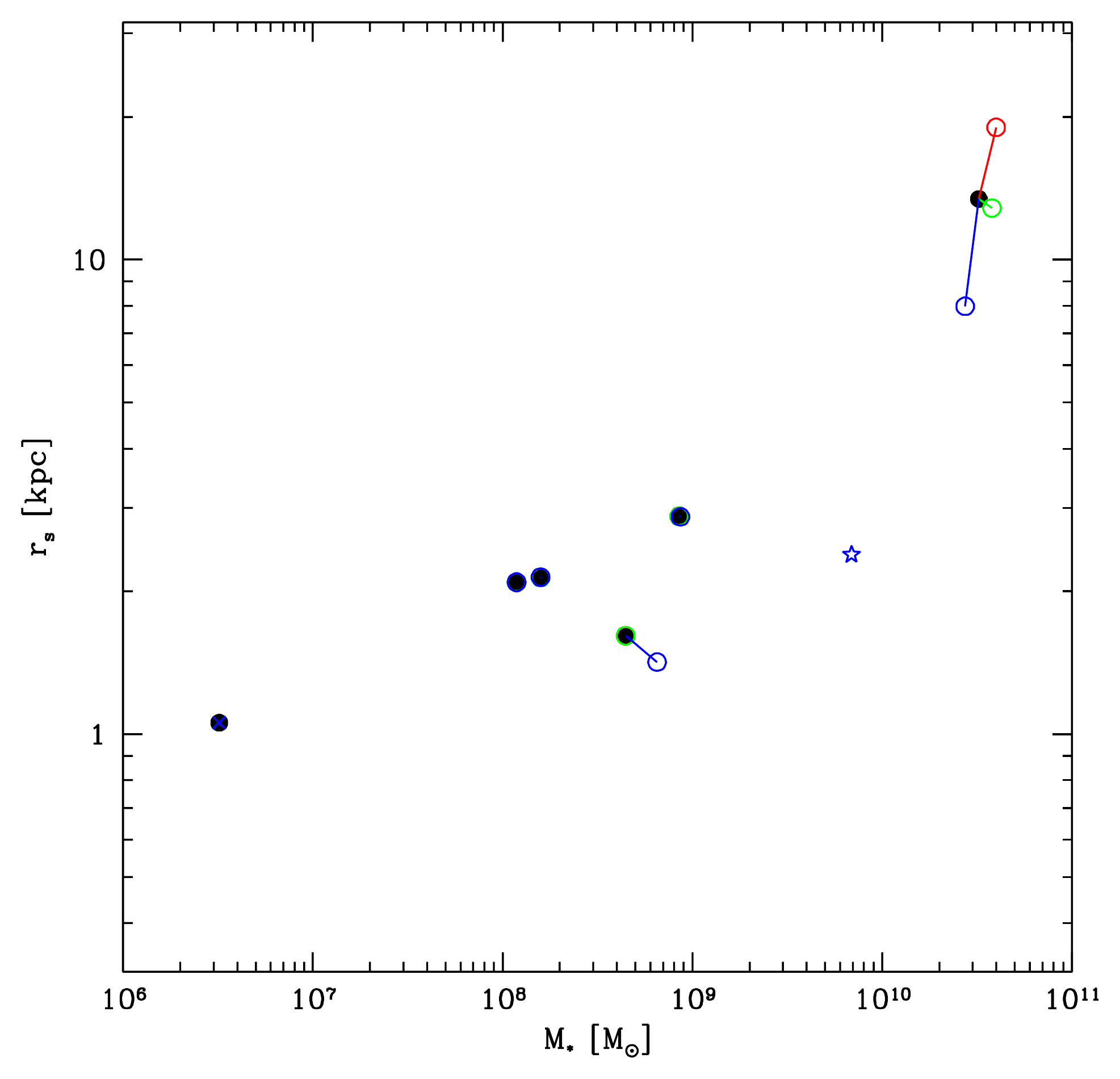}
 \end{center}
 \caption{Demonstration of convergence in galaxy stellar mass and stellar mass-weighted scale length. A single merger tree, with a final halo mass of $10^{12}M_\odot$ was run. Each point in the figure corresponds to a galaxy existing within that halo at $z=0$. Black points correspond to our standard set of numerical parameters as defined in the text. Red points show a run with increased tolerance in the ODE solver. Green and blue points show runs with increased tolerance in time stepping relative to a node's host and to itself respectively. Where the same galaxy exists in the original model, the galaxy is represented by an open circle connected to the original model galaxy by a line. If a galaxy no longer exists in a modified model it is shown by a cross, while if a galaxy exists in the modified model that was not present in the original model it is shown by a star.}
 \label{fig:Convergence}
\end{figure}

To test the numerical convergence in \glc\ we run a single $10^{12}M_\odot$ merger tree to $z=0$ as our base model. We repeat the calculation for the same merger tree with altered numerical parameters as will be described below. The results of this exercise are shown in Fig.~\ref{fig:Convergence}, in which we plot galaxies from this merger tree in the plane of total stellar mass and stellar mass-weighted scale length. The base model is shown by filled black circles. Galaxies in the comparison models which have a correspondent in the base model are indicated by open circles connected by a line to the base model galaxy. Where a base model galaxy is missing from the comparison model it is marked with a cross, while in cases where a comparison model galaxy is not present in the base model it is indicated by a star. The parameters adjusted in the three comparison models are:
\begin{description}
 \item [{\tt [odeToleranceRelative]}] \emph{Base model value:} 0.01; \emph{Comparison model value:} 0.001; \emph{Color:} red. This parameter specifies the accuracy requested from the ODE solver\footnote{We typically use a relatively large tolerance. Since the approximations made by semi-analytic models are not expected to be valid to high precision a higher tolerance is not generally warranted. Higher tolerances can be adopted if necessary of course.}---it is required to keep fractional errors in evolved quantities below this value during any given time step.

 \item [{\tt [timestepHostRelative]}] \emph{Base model value:} 0.1; \emph{Comparison model value:} 0.01; \emph{Color:} green. This parameter limits the time interval by which satellite nodes may be evolved beyond the time at which their host halo is currently located. Specifically, no satellite node is allowed to evolve beyond {\tt [timestepHostRelative]} times the cosmological expansion timescales, $H^{-1}(t)$, at the time of the host halo, before it must wait for the host node to catch up.

 \item [{\tt [timestepSimpleRelative]}] \emph{Base model value:} 0.1; \emph{Comparison model value:} 0.01; \emph{Color:} blue. This parameter, specified as a fraction of the cosmological expansion timescale, $H^{-1}(t)$, at the time of the node, limits the time step over which any node may be evolved before evolution is paused and any necessary post-processing (such as transferring gas driven out of a satellite galaxy) is performed.
\end{description}

The results of this convergence study, as shown in Fig.~\ref{fig:Convergence}, lead to several insights. First, the properties of half of the galaxies in this merger tree are almost perfectly stable to the specific changes in the numerical parameters that we have explored, indicating a high level of convergence. The other three galaxies show significant changes in their properties however. We begin by considering the most massive galaxy. When the ODE solver tolerance is improved (red symbols), this central galaxy gains slightly in mass and becomes significantly larger. By tracing the merging history of this galaxy in \glc, we can identify the cause of this as a merger with the lowest mass galaxy that was present in the base model. In the base model calculation, this lowest mass galaxy (located at a mass of around $3\times10^6M_\odot$ in Fig.~\ref{fig:Convergence}) did not merge with the central. However, the improved ODE tolerance resulted in a merging event being triggered which then significantly altered the further evolution of the central galaxy. Conversely, when the {\tt [timestepSimpleRelative]} parameter is reduced (blue symbols) we find a new galaxy appearing (indicated by the blue star). Here the opposite has happened---a galaxy which did merge in the base model failed to merge in the model with better timestep tolerance. This galaxy originally merged into the most massive galaxy in the tree. Consequently, that galaxy is now somewhat less massive and significantly smaller that in the base model. Finally, in this same model, the galaxy originally at $4.5\times10^8M_\odot$ is increased in mass and slightly smaller. Once again examining the history of this galaxy in \glc, we find that this change occurs because the galaxy in question has a disk which sits close to the stability threshold for bar formation (see \S\ref{sec:DiskStability}). Under our current implementation of this process, the timescale for mass transfer from disk to spheroid due to the bar instability is finite at the stability threshold, but infinite below that threshold. As such, properties of galaxies close to this threshold will be sensitive to small changes in numerical parameters in \glc. Fundamentally, this represents a limitation of the physical model adopted for the bar instability---clearly a more physically motivated model is desirable \citep{athanassoula_disc_2008}. While these effects make for substantial changes in the properties of the galaxy population of this particular merger tree, we find much less change when considering statistical samples of galaxies. Nevertheless, it is important to test for convergence in the quantities of interest for any particular study performed with \glc.

\begin{figure}
 \begin{center}
 \includegraphics[width=80mm]{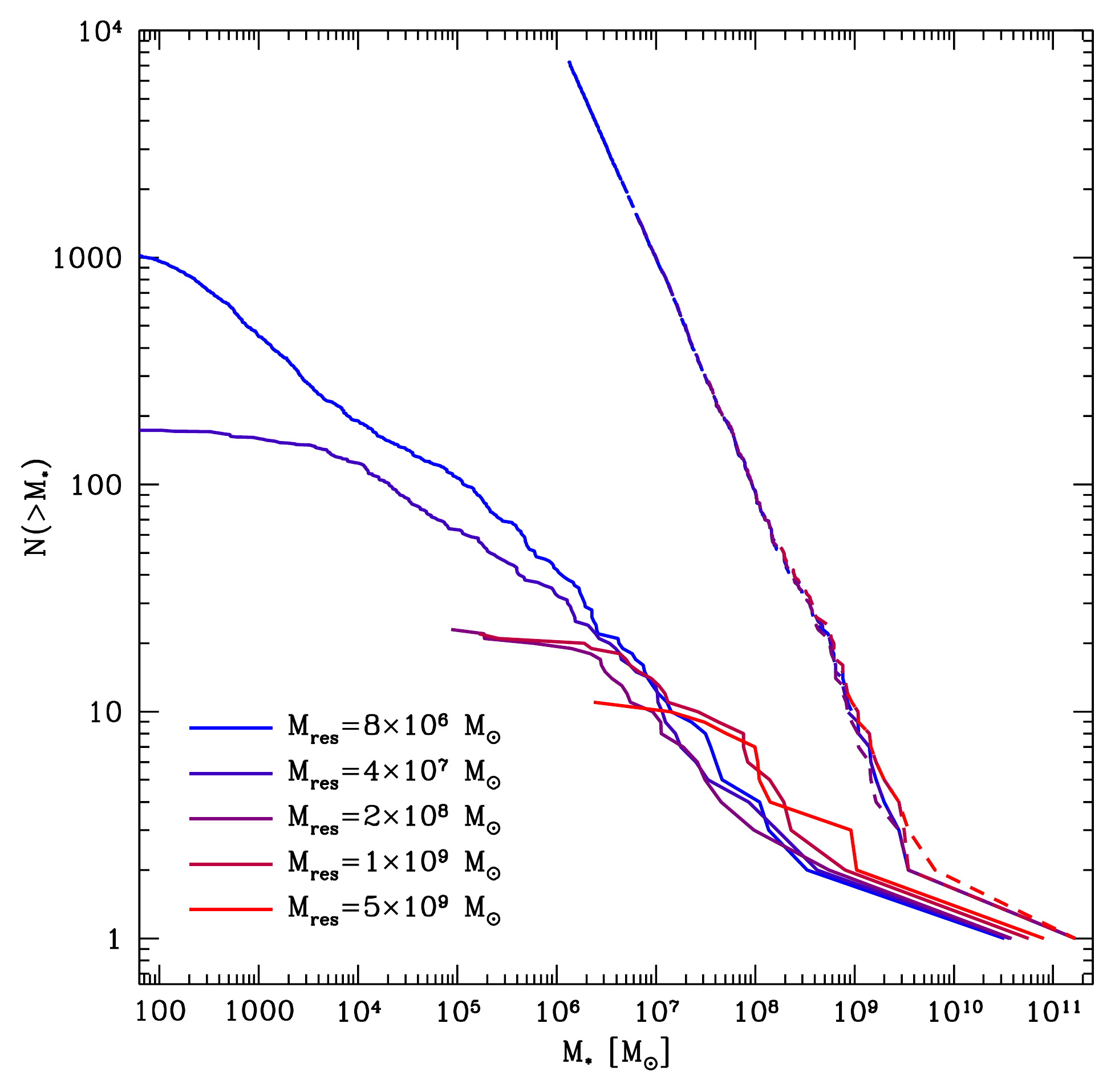}
 \end{center}
 \caption{Tests of convergence in the cumulative stellar mass function of galaxies formed in a merger tree with a final halo mass of $10^{12}M_\odot$ at $z=0$. Solid lines show the stellar mass function, while dashed lines indicate the (sub)halo mass function scaled by a factor of $\Omega_{\rm b}/\Omega_0$. The difference in slope between the two sets of mass functions is a consequence of the \protect\SNe-driven feedback included in this model. It can be seen that the (sub)halo mass function is well converged, while the galaxy stellar mass function is not. Note that black holes were not included in calculations used for this particular convergence study.}
 \label{fig:MassFunction}
\end{figure}

Figure~\ref{fig:MassFunction} shows another example of convergence. We take the same merger tree and re-simulate it while increasing the resolution of the tree branches\footnote{Since our trees are generated by a Monte Carlo method simply changing the resolution and rebuilding the tree would result in a very different tree (a single different branching would lead to an entirely different branching path). To make a fair comparison between the different resolutions, we therefore create a tree at the highest resolution considered and then prune away branches below the required resolution threshold.}. Each line in Fig.~\ref{fig:MassFunction} represents a factor 5 decrease in the minimum mass resolved in merger tree branches. Solid lines show the cumulative stellar mass function of galaxies, while dashed lines show the cumulative mass function of dark matter (sub)halos scaled by a factor $\Omega_{\rm b}/\Omega_0$. Clearly the (sub)halo mass function is well converged. The galaxy mass function, however, is not. For example, the stellar mass of the most massive galaxy changes by a factor of around 3 from the highest to lowest resolution runs. Of course, the results shown span a range of 625 in merger tree mass resolution. This resolution dependence is expected---as lower mass halos are resolved more gas gets locked up in low mass galaxies leaving less to form the massive galaxy. Convergence should only be reached once a physical suppression scale is reached. In the case of cold dark matter this will be a baryonic scale (since any cut-off in the \CDM\ power spectrum is expected to occur at very low masses) associated with a truncation in the cooling function or with the temperature of the \IGM\ and photoheating by an ionizing background. Even at the highest resolution shown, full convergence across the whole mass function is not reached. However, for resolutions of $2\times10^8M_\odot$ and better the mass of the most massive (central) galaxy is converged to better than 50\% which may be sufficient for many applications. Convergence in tree mass resolution will depend on the details of the implemented baryonic physics, and also on what galaxy samples/properties are being studied. This clearly demonstrates that a convergence test of the type carried out here should always be performed to ensure that results are not affected by the limited resolution of merger trees.

\subsubsection{Timing}

\begin{figure}
 \begin{center}
 \includegraphics[width=80mm]{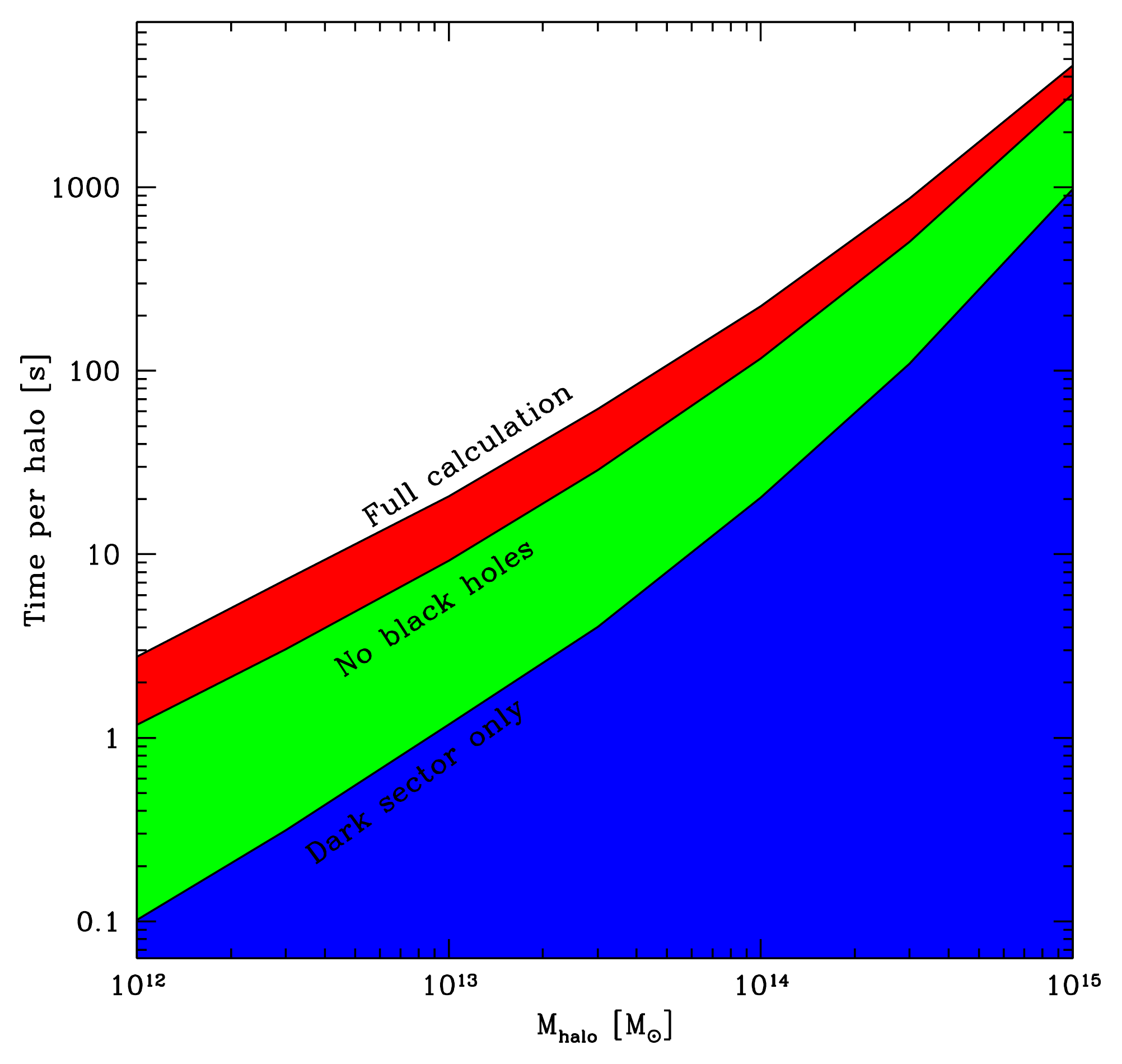}
 \end{center}
 \caption{Execution time per merger tree. (Run on a 2.7GHz AMD Opteron processor and compiled using standard optimizations.) The red region indicates the average time taken to evolve a tree in the example model to $z=0$. The green region indicates the time taken to run the same tree with black hole physics switched off. Finally, the blue region shows the time taken to evolve each tree when only dark sector physics are included (i.e. all baryonic physics is switched off).}
 \label{fig:Timing}
\end{figure}

\glc\ was designed with simplicity, modularity and flexibility as key design principles. Execution speed was not a high priority but, nevertheless, we have given significant attention to optimizing key areas of the code. To explore the time taken to run typical calculations we have evolved sets of merger trees of different masses in our example model and computed the mean time taken to evolve each tree. The results are shown in Fig.~\ref{fig:Timing}, where we show results for the full model, the same model but with black hole physics switched off and the same model but with all baryonic physics switched off.

Fig.~\ref{fig:Timing} shows that the time taken to evolve a tree scales close to, but somewhat faster than, linearly with halo mass. Since we use a fixed mass resolution in these merger trees, the number of nodes into which a tree is resolved will scale approximately with the mass of the halo, so this result is to be expected. While a Milky Way-mass halo can be evolved in around 2 seconds, a high mass cluster takes around 1.25 hours. This is not as fast as some other semi-analytic models, but, as we stated above, speed is not our primary concern, with accuracy, simplicity and modularity taken precedence. Excluding black holes from the calculation significantly reduces run time by a factor of 2 to 2.5. This is because black hole properties (particularly spin) can evolve on very short timescales making evolution of the ODE system computationally expensive. This is worsened by the fact that many black holes are close to the equilibrium spin at which spin-up by accretion and spin down from powering jets cancel, forcing small timesteps to maintain this balance. Finally, switching off all baryonic physics (which leaves tree building, and evolution of the dark matter halos including dynamical friction, and which can be useful for exploring the properties of dark matter models) reduces run time by a further factor of around 10.

\subsection{Galaxy Properties}

In this final subsection we compare several observational datasets with their equivalents as predicted by the example \glc\ model. We reiterate that this example model is not intended to represent the best fit to observational data attainable with \glc---a full search of the \glc\ parameter space is a significant undertaking which will be explored in a future paper. Instead, the results in this subsection are intended to illustrate that types of quantities which can be computed and to show that they are, at least, in broad agreement with current observational constraints. We give a brief description and discussion of each result. Much of the underlying physics behind the model results has been previously discussed by other authors, as mentioned below. In a few cases, we defer further study to future papers which will explore some of the key physical drivers in greater detail.

\begin{figure}
 \begin{center}
 \includegraphics[width=80mm]{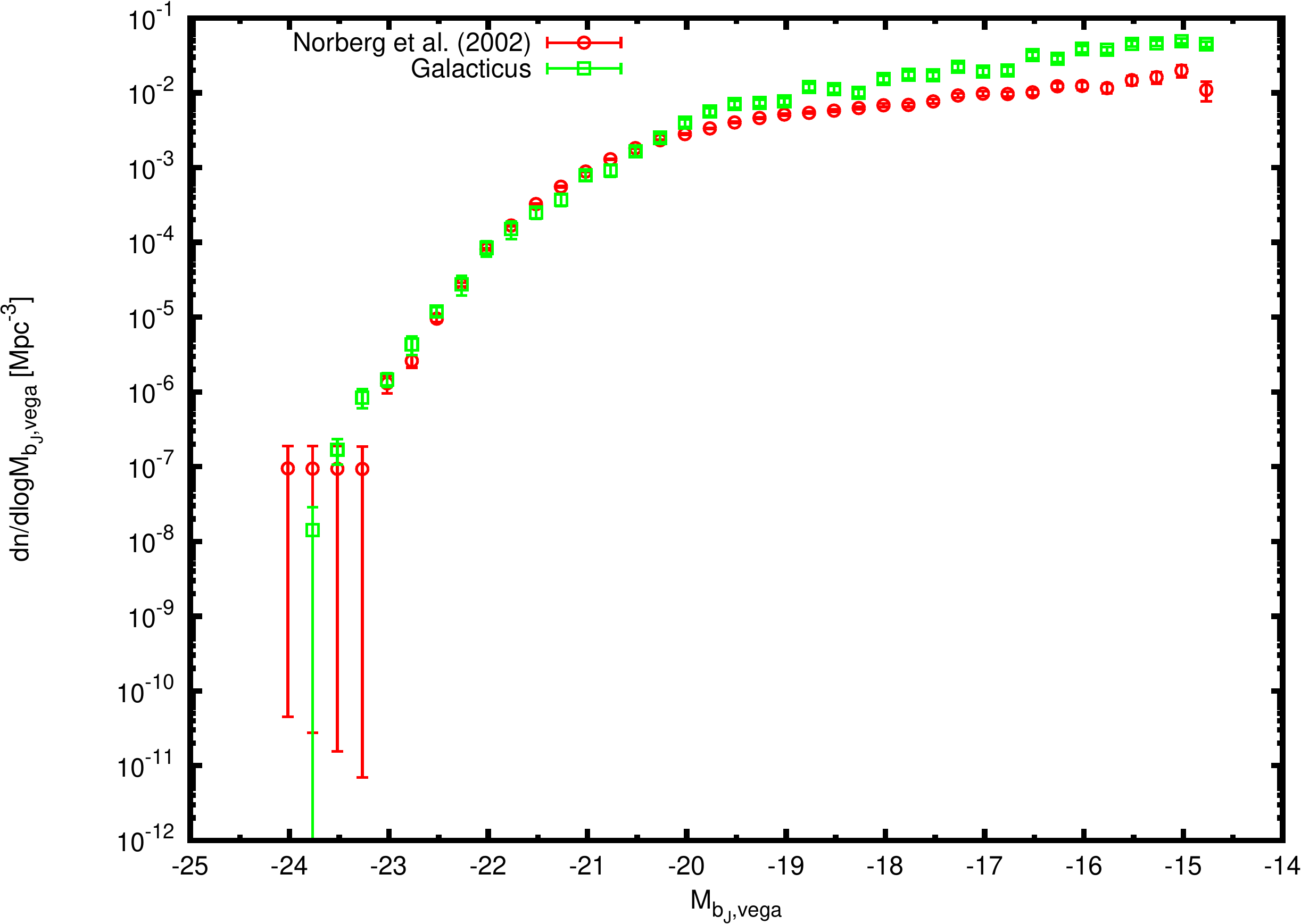}
 \end{center}
 \caption{The $z=0$ b$_{\rm J}$-band luminosity function. Red points indicate the observed luminosity function from the \protect\TdF\ \protect\citep{norberg_2df_2002}, while green points show results from our example \protect\glc\ model after including the effects of dust-extinction using the model of \protect\cite{ferrara_atlas_1999}. Error bars on the model points indicate the uncertainty due to the finite number of merger trees computed.}
 \label{fig:bJLF}
\end{figure}

\begin{figure}
 \begin{center}
 \includegraphics[width=80mm]{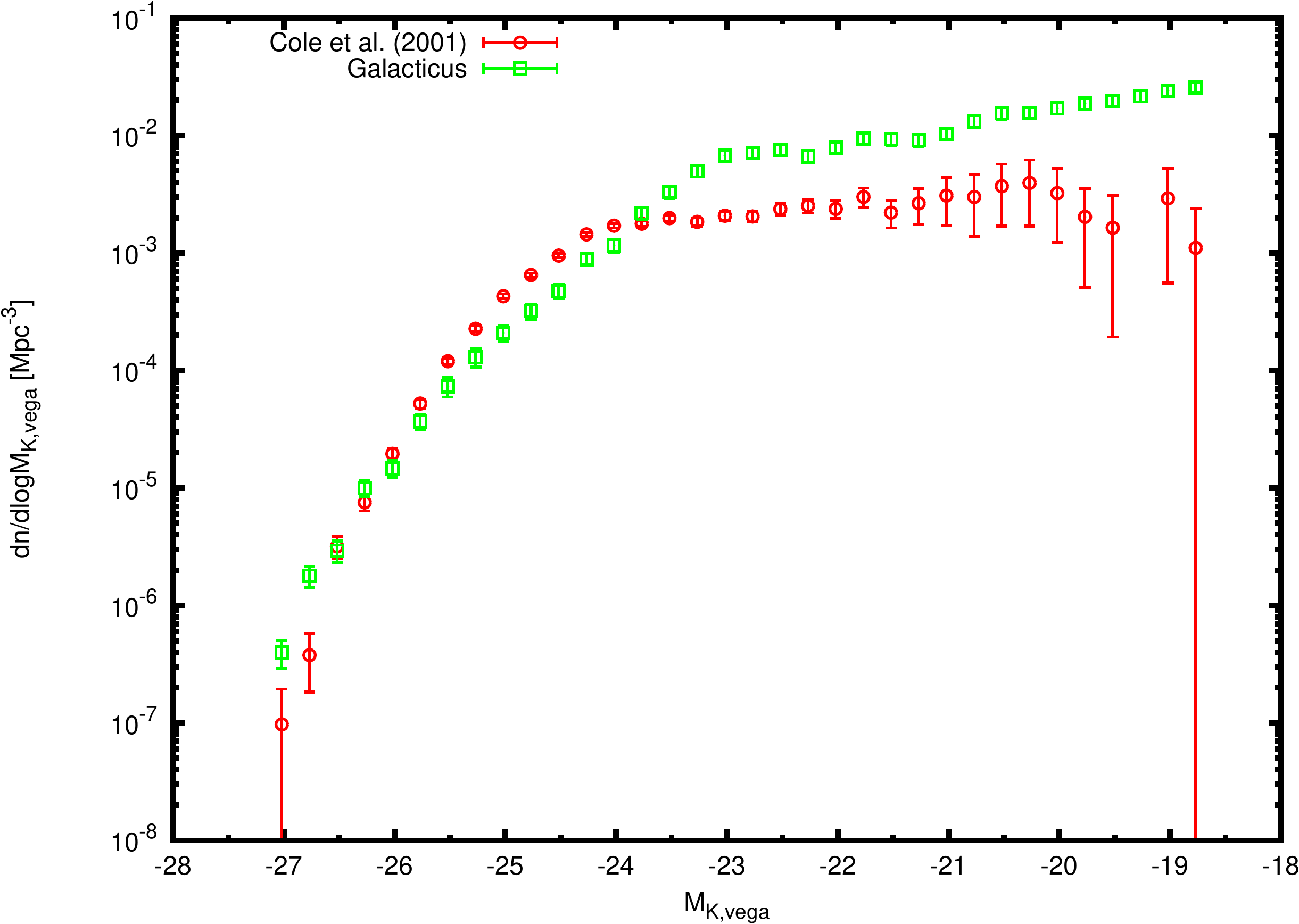}
 \end{center}
 \caption{The $z=0$ K-band luminosity function from the example model. Red points indicate data from the \protect\TdF +\protect\TMASS\ \protect\citep{cole_2df_2001}, while green points show results from our example \protect\glc\ model after including the effects of dust-extinction using the model of \protect\cite{ferrara_atlas_1999}. Error bars on the model points indicate the uncertainty due to the finite number of merger trees computed.}
 \label{fig:KLF}
\end{figure}

Figures~\ref{fig:bJLF} and \ref{fig:KLF} show b$_{\rm J}$ and K-band luminosity functions at $z=0$ derived from the \TdF\ and \TMASS\ galaxy surveys, with observational determinations shown by red points and results from the example \glc\ model show by green points. Dust extinction is included in the model galaxy magnitudes using the results of \cite{ferrara_atlas_1999}. The match to the bright end cut off in the luminosity functions is quite good, although the K-band prediction undershoots the knee of the observed luminosity function. However, the predicted faint end slope is significantly steeper than those observed (particularly in the K-band). This is a well known problem in semi-analytic models (e.g. \citealt{benson_what_2003}) and is closely linked to the inclusion of \SNe-feedback in such models. Stronger feedback would help to reconcile this discrepancy.

\begin{figure}
 \begin{center}
 \includegraphics[width=80mm]{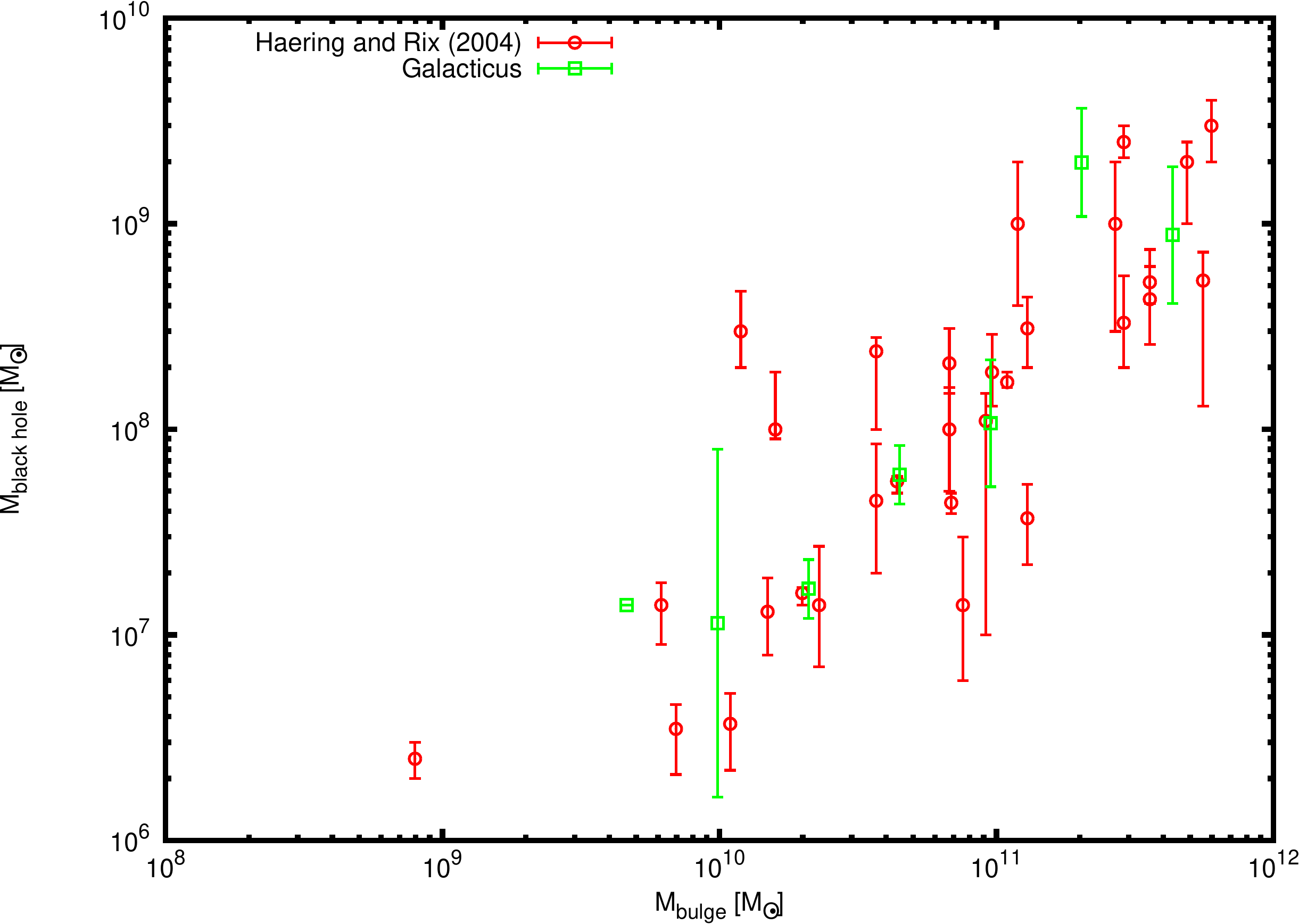}
 \end{center}
 \caption{The relation between supermassive black hole mass and galaxy bulge stellar mass. Data are taken from \protect\cite{haring_black_2004} and are shown by red points. Green points with error bars show the mean and 1-$\sigma$ dispersion from our example \protect\glc\ model.}
 \label{fig:BH}
\end{figure}

A key ingredient in producing the sharp cut off at the bright end of the galaxy luminosity function is the inclusion of \AGN\ feedback in the example \glc\ model \citep{croton_many_2006,bower_breakinghierarchy_2006}. As such, it is crucial that predicted black hole properties be reasonable. Figure~\ref{fig:BH} shows the relation between black hole mass and spheroid stellar mass from observations (red points) and from the example \glc\ model (green points with error bars; indicating the mean and 1-$\sigma$ dispersion at each spheroid mass). The normalization and slope of the model relation is in good agreement with that which is observed. The model displays little scatter in black hole mass at fixed stellar mass, except for at the highest masses (partially due to the limited number of such high mass spheroids in our model sample) and around $10^{10}M_\odot$ spheroid mass. In the model, this increase in scatter below $10^{10}M_\odot$ occurs because the regulating mechanism of \AGN\ feedback (which controls the relation between black hole and spheroid mass) breaks down.

\begin{figure}
 \begin{center}
 \includegraphics[width=80mm]{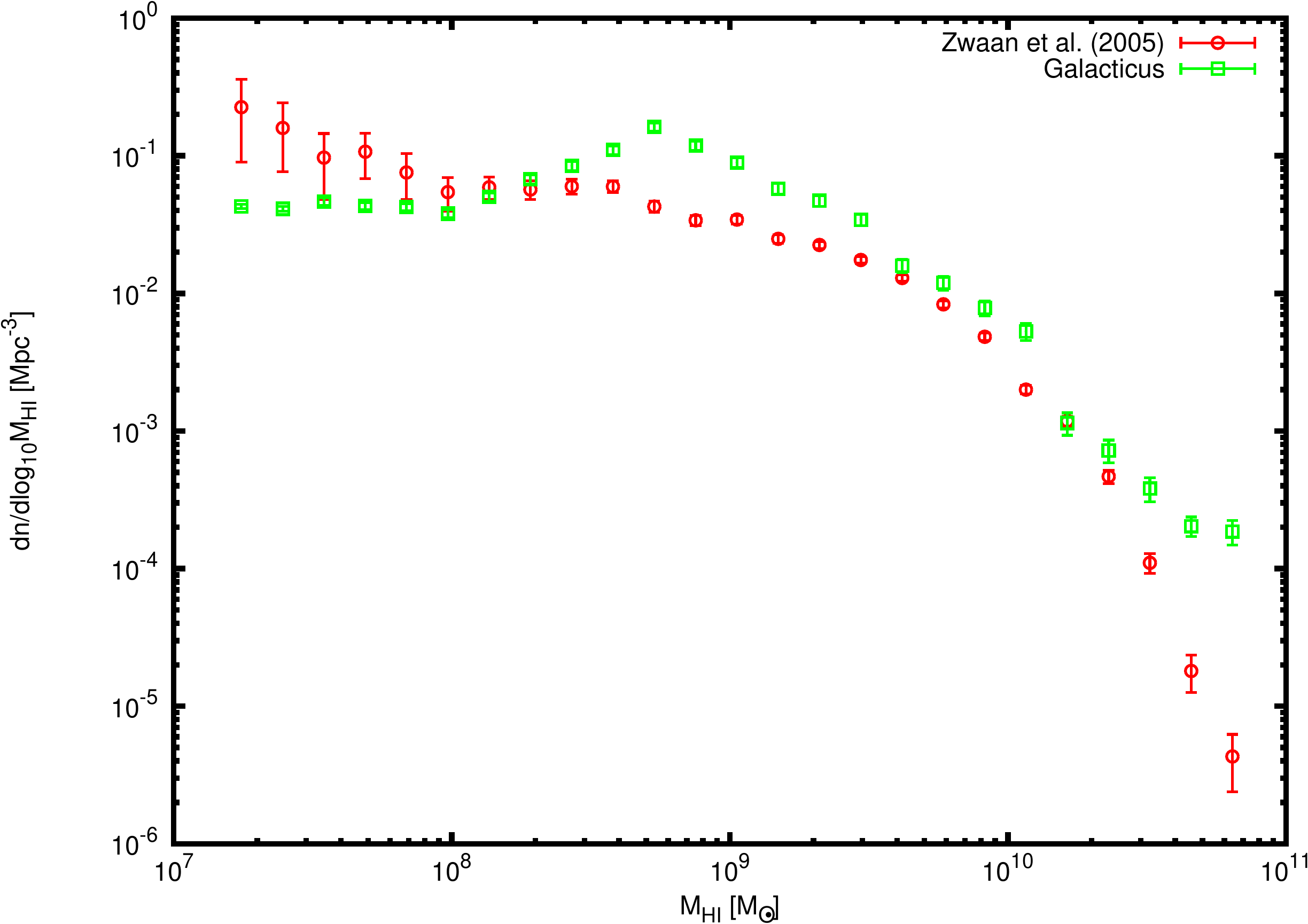}
 \end{center}
 \caption{The $z=0$ HI gas mass function. The green points show the mass function from the example \protect\glc\ model assuming a constant factor of $0.54$ to convert from total gas mass to HI mass \protect\citep{power_redshift_2010}. Red points indicate data from \protect\cite{zwaan_hipass_2005}.}
 \label{fig:HIMF}
\end{figure}

Figure~\ref{fig:HIMF} explores the gas content of galaxies in the example model, by comparing the model HI mass function with that observed by \cite{zwaan_hipass_2005}. While the overall normalization is approximately correct the shape of the predicted function is incorrect and, in particular, does not truncate sharply enough at high masses. Similar results were found and considered in greater detail by \cite{power_redshift_2010}.

\begin{figure}
 \begin{center}
 \includegraphics[width=80mm]{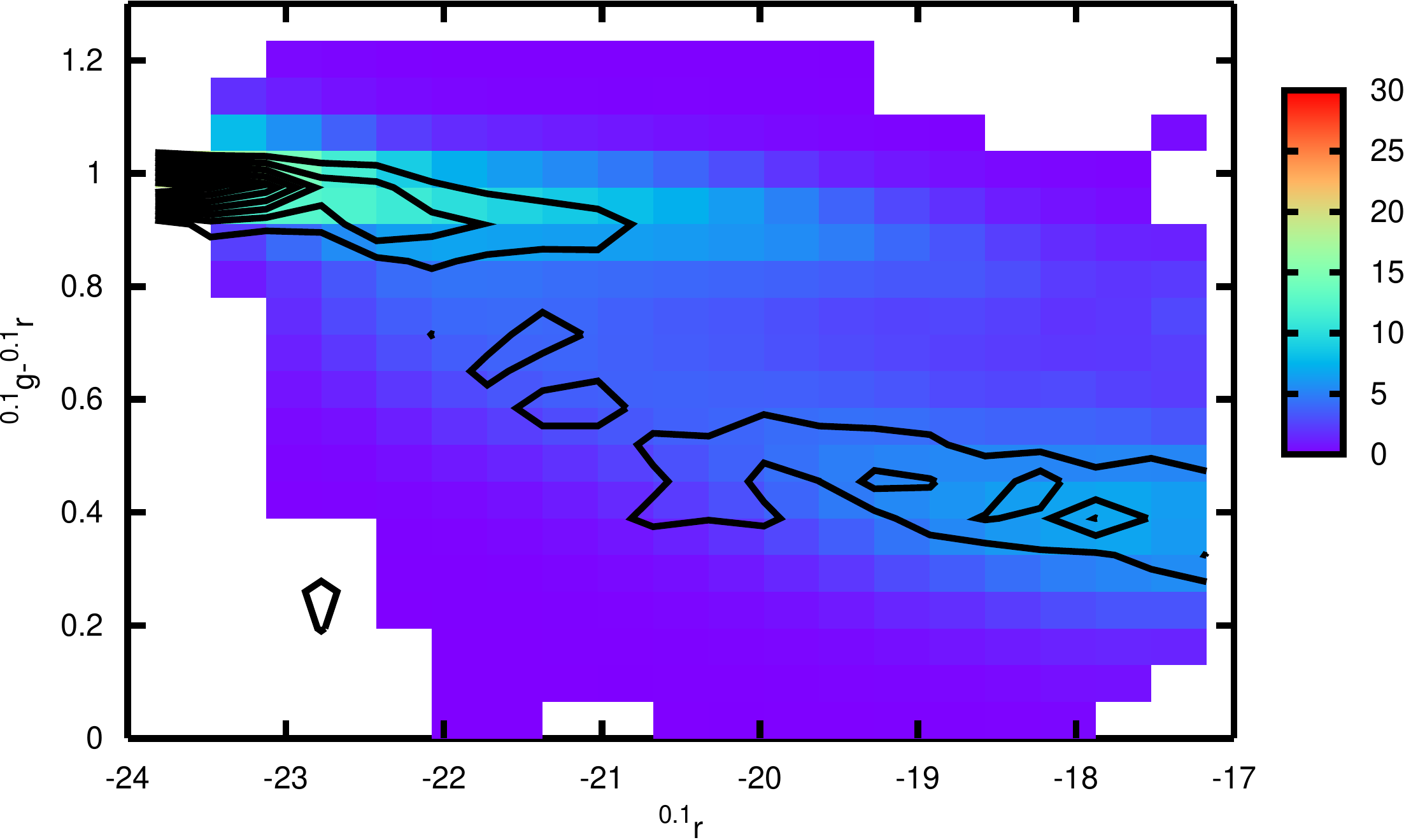}
 \end{center}
 \caption{$^{0.1}$g$-^{0.1}$r differential color distribution (normalized to unit area under the surface) for galaxies at $z=0.1$ as a function g-band absolute magnitude. The color scale shows data from the \protect\SDSS\ \protect\citep{weinmann_properties_2006}, while the black contours show the distribution from the example \protect\glc\ model. Ten contour levels are shown, linearly spaced between the minimum and maximum of the color scale.}
 \label{fig:Color}
\end{figure}

A key result from the \SDSS\ has been a clear demonstration of the dichotomy of the galaxy population, with most galaxies being part of a star forming ``blue cloud'' or a quiescent ``red sequence''. Figure~\ref{fig:Color} shows the \SDSS\ color-magnitude diagram (colored shading) derived by \cite{weinmann_properties_2006} which clearly shows these two populations. The black contours show the color-magnitude diagram from the example \glc\ model. This clearly shows the same two populations of galaxies, with approximately the correct median colors and transition luminosity. In the example model, \AGN\ feedback plays a key role in establishing this division, as has been found by other authors (\citealt{croton_many_2006,bower_breakinghierarchy_2006}; see also \citealt{font_colours_2008}).

\begin{figure}
 \begin{center}
 \includegraphics[width=80mm]{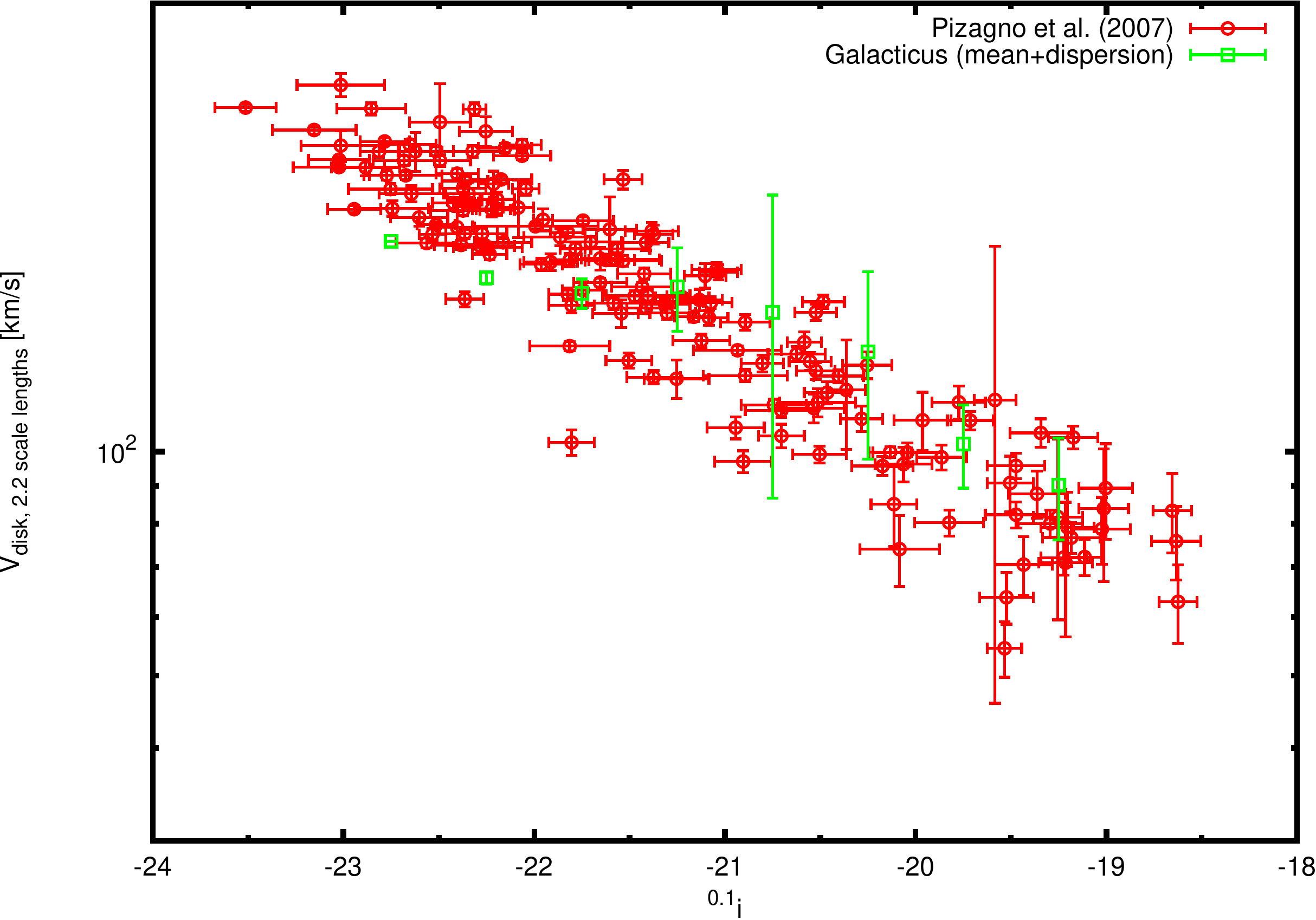}
 \end{center}
 \caption{The i-band Tully-Fisher relation from the \protect\SDSS\ \protect\citep{pizagno_tully-fisher_2007} is shown by red points, while results from the example \protect\glc\ model are indicated by green points. Model points show the mean disk circular velocity at each magnitude with error bars indicating the 1-$\sigma$ dispersion in model galaxies.}
 \label{fig:TF}
\end{figure}

The structural and dynamical state of model galaxies is explored in Figure~\ref{fig:TF}, in which we compare results from the example model to the i-band Tully-Fisher relation measured by \cite{pizagno_tully-fisher_2007}. The model is in moderately good agreement with the data---the slope at lower luminosities is approximately correct, but flattens at higher luminosities such that the model velocities are too low. Additionally, the dispersion in model galaxy velocities at intermediate luminosities is too large. The Tully-Fisher relation encodes significant constraints on the process of galaxy formation. As such, we intend to explore this aspect of the \glc\ model in greater detail in a future paper.

\begin{figure}
 \begin{center}
 \includegraphics[width=80mm]{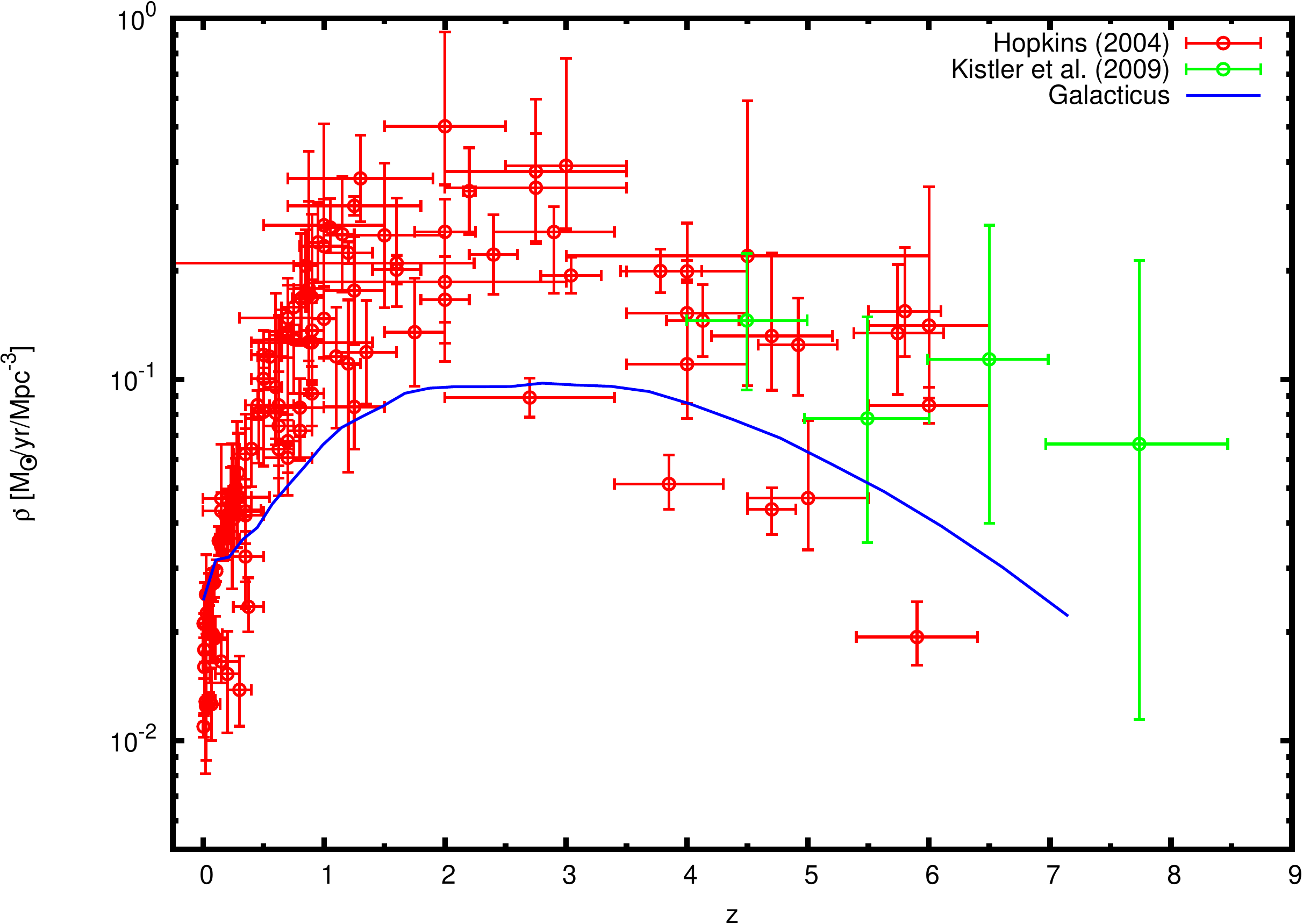}
 \end{center}
 \caption{The star formation rate per unit comoving volume in the Universe as a function of redshift. Red points show observational estimates from a variety of sources as compiled by \protect\cite{hopkins_evolution_2004} while green points show the star formation rate inferred from gamma ray bursts by \protect\cite{kistler_star_2009}. The blue line shows the result from the example \protect\glc\ model.}
 \label{fig:SFH}
\end{figure}

The volume averaged star formation rate density in the example model is shown in Fig.~\ref{fig:SFH} and is compared with a compilation of observational data. Clearly the model under-predicts the star formation rate at all redshifts above $z\approx 0.25$. This is a key failing of the example model, and therefore represents a key constraint for future parameter space studies with \glc. The origins of this underestimate of the star formation rate are not clear without further investigation, but could plausibly be due to over-zealous \AGN\ feedback at intermediate redshifts or simply a breakdown in the assumed scalings of star formation timescales (which are currently based on simple scaling relations rather than any physical model).

\begin{figure}
 \begin{center}
 \includegraphics[width=80mm]{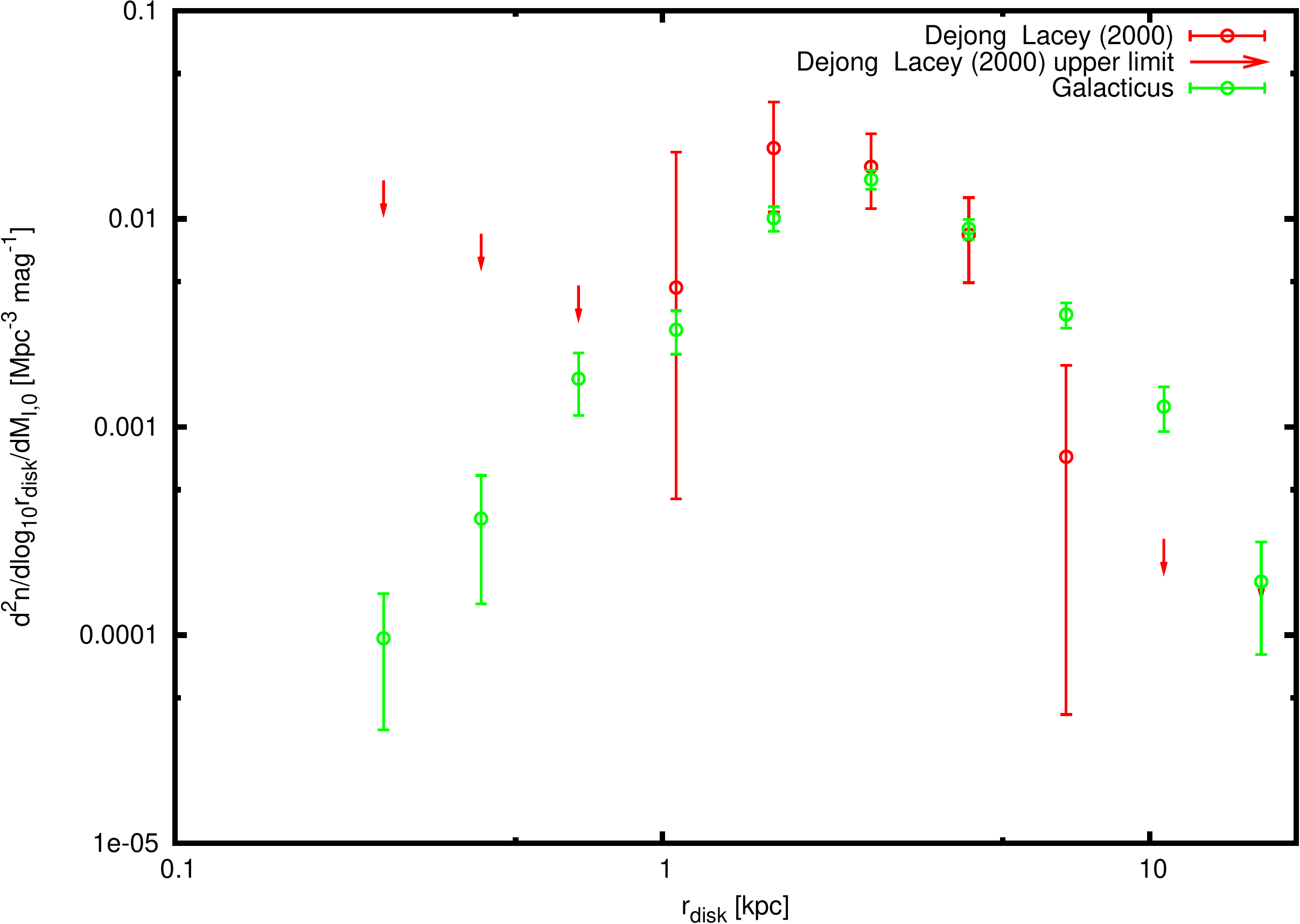}
 \end{center}
 \caption{Distribution of disk scale lengths for galaxies at $z=0$ selected by face-on I-band absolute magnitude, $-20 < M_{\rm I,0} < -19$. Red circles show data from \protect\cite{de_jong_local_2000} with upper limits indicated by red triangles. Green points show results from our example \protect\glc\ model.}
 \label{fig:Sizes}
\end{figure}

Finally, we show in Fig.~\ref{fig:Sizes} the sizes of galactic disks for galaxies in a narrow range of face-on I-band magnitude, $-20 < M_{\rm I,0} < -19$, and compared to the observational determination of \cite{de_jong_local_2000}. The peak of the distribution in the example model is in approximately the correct location, but the model predicts somewhat too many very large disks compared to the data (although, given the large errors on the data in this region, the difference between model and data is not very significant). Sizes of galaxies are a key property since they directly affect galactic dynamical timescales and, therefore, star formation rates.

\section{Summary}\label{sec:Summary}

We have described a new, free and open source semi-analytic model of galaxy formation, \glc, which can be downloaded from \href{http://sites.google.com/site/galacticusmodel/}{{\tt http://sites.google.com/site/galacticusmodel}}. A full manual documenting both the physics and technical implementation of \glc\ can also be found on the same website\footnote{This article specifically describes the current development version of \protect\glc, v0.9.0. The current stable version, v0.0.1, contains almost all of the same features.}. In particular in this article, we have focused on the technical and practical implementation of \glc\ and have detailed the numerous physical ingredients which make up the model. Additionally, we have explored basic properties of a specific example model, demonstrating numerical convergence and illustrating the type of quantities which \glc\ can compute and compare against observational data. \glc\ contains the majority of the physics incorporated into other major semi-analytic models (see \cite{benson_galaxy_2010} for a review). Missing physics, notably tidal and ram pressure mass loss from satellite galaxies \citep{benson_effects_2002,lanzoni_galics-_2005,font_colours_2008,henriques_tidal_2010} and the ability to use information on subhalos from N-body merger trees\footnote{\protect\glc\ can utilize simple merger trees from N-body simulations---software is provided, for example, to convert trees from the Millennium Simulation into \glc's format---but not those with subhalo information.}, is already being incorporated into the development version of \glc.

The \glc\ model can be employed for a wide variety of calculations related to galaxy formation studies, ranging from dark matter phenomenology, through observationally testable statistical properties of galaxy samples to the construction of mock galaxy samples. Its flexibility makes it easily adaptable to new problems and accepting of new physical prescriptions.

\glc\ requires only free and open source libraries and compilers to run and so is portable and practical to use. In designing \glc\ we placed an emphasis on simplicity, flexibility and extensibility to make it simple to use and maintain and to facilitate future development as required in the rapidly changing field of galaxy formation. We therefore hope that \glc\ will prove to be a valuable tool for a significant time, and we are already endeavoring to expand and improve upon this initial version.

\section*{Acknowledgments}

The author acknowledges the support of the Gordon \& Betty Moore Foundation.

\bibliographystyle{elsarticle-harv}
\bibliography{galacticusI}

\begin{thebibliography}{83}
\expandafter\ifx\csname natexlab\endcsname\relax\def\natexlab#1{#1}\fi
\expandafter\ifx\csname url\endcsname\relax
  \def\url#1{\texttt{#1}}\fi
\expandafter\ifx\csname urlprefix\endcsname\relax\def\urlprefix{URL }\fi

\bibitem[{Athanassoula(2008)}]{athanassoula_disc_2008}
Athanassoula, E., Oct. 2008. Disc instabilities and semi-analytic modelling of
  galaxy formation. Monthly Notices of the Royal Astronomical Society 390,
  L69--L72.
\newline\urlprefix\url{http://adsabs.harvard.edu/abs/2008MNRAS.390L..69A}

\bibitem[{Bardeen(1970)}]{bardeen_kerr_1970}
Bardeen, J.~M., Apr. 1970. Kerr metric black holes. Nature 226, 64.
\newline\urlprefix\url{http://adsabs.harvard.edu/abs/1970Natur.226...64B}

\bibitem[{Bardeen et~al.(1986)Bardeen, Bond, Kaiser, and
  Szalay}]{bardeen_statistics_1986}
Bardeen, J.~M., Bond, J.~R., Kaiser, N., Szalay, A.~S., May 1986. The
  statistics of peaks of gaussian random fields. Astrophysical Journal 304,
  15--61.
\newline\urlprefix\url{http://adsabs.harvard.edu/abs/1986ApJ...304...15B}

\bibitem[{Benson(2005)}]{benson_orbital_2005}
Benson, A.~J., Apr. 2005. Orbital parameters of infalling dark matter
  substructures. {MNRAS} 358, 551--562.
\newline\urlprefix\url{http://adsabs.harvard.edu/abs/2005MNRAS.358..551B}

\bibitem[{Benson(2010)}]{benson_galaxy_2010}
Benson, A.~J., Jun. 2010. Galaxy formation theory.
  {http://adsabs.harvard.edu/abs/2010arXiv1006.5394B}.
\newline\urlprefix\url{http://adsabs.harvard.edu/abs/2010arXiv1006.5394B}

\bibitem[{Benson and Babul(2009)}]{benson_maximum_2009}
Benson, A.~J., Babul, A., Aug. 2009. Maximum spin of black holes driving jets.
  Monthly Notices of the Royal Astronomical Society 397, 1302--1313.
\newline\urlprefix\url{http://adsabs.harvard.edu/abs/2009MNRAS.397.1302B}

\bibitem[{Benson and Bower(2010{\natexlab{a}})}]{benson_cold_2010}
Benson, A.~J., Bower, R., Apr. 2010{\natexlab{a}}. Cold mode accretion in
  galaxy formation. {http://adsabs.harvard.edu/abs/2010arXiv1004.1162B}.
\newline\urlprefix\url{http://adsabs.harvard.edu/abs/2010arXiv1004.1162B}

\bibitem[{Benson and Bower(2010{\natexlab{b}})}]{benson_galaxy_2010-1}
Benson, A.~J., Bower, R.~G., Feb. 2010{\natexlab{b}}. Galaxy formation spanning
  cosmic history. {http://adsabs.harvard.edu/abs/2010arXiv1003.0011B}.
\newline\urlprefix\url{http://adsabs.harvard.edu/abs/2010arXiv1003.0011B}

\bibitem[{Benson et~al.(2003)Benson, Bower, Frenk, Lacey, Baugh, and
  Cole}]{benson_what_2003}
Benson, A.~J., Bower, R.~G., Frenk, C.~S., Lacey, C.~G., Baugh, C.~M., Cole,
  S., Dec. 2003. What shapes the luminosity function of galaxies? Astrophysical
  Journal 599, 38--49.
\newline\urlprefix\url{http://adsabs.harvard.edu/abs/2003ApJ...599...38B}

\bibitem[{Benson et~al.(2002)Benson, Lacey, Baugh, Cole, and
  Frenk}]{benson_effects_2002}
Benson, A.~J., Lacey, C.~G., Baugh, C.~M., Cole, S., Frenk, C.~S., Jun. 2002.
  The effects of photoionization on galaxy formation - i. model and results at
  z=0. {MNRAS} 333, 156--176.
\newline\urlprefix\url{http://adsabs.harvard.edu/abs/2002MNRAS.333..156B}

\bibitem[{Benson et~al.(2001)Benson, Pearce, Frenk, Baugh, and
  Jenkins}]{benson_comparison_2001}
Benson, A.~J., Pearce, F.~R., Frenk, C.~S., Baugh, C.~M., Jenkins, A., 2001. A
  comparison of semi-analytic and smoothed particle hydrodynamics galaxy
  formation. Monthly Notices of the Royal Astronomical Society 320, 261--280.
\newline\urlprefix\url{http://adsabs.harvard.edu/abs/2001MNRAS.320..261B}

\bibitem[{Bertelli et~al.(2008)Bertelli, Girardi, Marigo, and
  Nasi}]{bertelli_scaled_2008}
Bertelli, G., Girardi, L., Marigo, P., Nasi, E., Jun. 2008. Scaled solar tracks
  and isochrones in a large region of the {Z-Y} plane. i. from the {ZAMS} to
  the {TP-AGB} end for 0.15-2.5 {{M}ȯ} stars. Astronomy and Astrophysics 484,
  815--830.
\newline\urlprefix\url{http://adsabs.harvard.edu/abs/2008A\%26A...484..815B}

\bibitem[{Bertelli et~al.(2009)Bertelli, Nasi, Girardi, and
  Marigo}]{bertelli_scaled_2009}
Bertelli, G., Nasi, E., Girardi, L., Marigo, P., Dec. 2009. Scaled solar tracks
  and isochrones in a large region of the {Z-Y} plane. {II.} from 2.5 to 20 mȯ
  stars. Astronomy and Astrophysics 508, 355--369.
\newline\urlprefix\url{http://adsabs.harvard.edu/abs/2009A\%26A...508..355B}

\bibitem[{Bett et~al.(2007)Bett, Eke, Frenk, Jenkins, Helly, and
  Navarro}]{bett_spin_2007}
Bett, P., Eke, V., Frenk, C.~S., Jenkins, A., Helly, J., Navarro, J., Mar.
  2007. The spin and shape of dark matter haloes in the millennium simulation
  of a lambda cold dark matter universe. {MNRAS} 376, 215--232.
\newline\urlprefix\url{http://adsabs.harvard.edu/abs/2007MNRAS.376..215B}

\bibitem[{Bower et~al.(2006)Bower, Benson, Malbon, Helly, Frenk, Baugh, Cole,
  and Lacey}]{bower_breakinghierarchy_2006}
Bower, R.~G., Benson, A.~J., Malbon, R., Helly, J.~C., Frenk, C.~S., Baugh,
  C.~M., Cole, S., Lacey, C.~G., Aug. 2006. Breaking the hierarchy of galaxy
  formation. {MNRAS} 370, 645--655.
\newline\urlprefix\url{http://adsabs.harvard.edu/abs/2006MNRAS.370..645B}

\bibitem[{Bower et~al.(2010)Bower, Vernon, Goldstein, Benson, Lacey, Baugh,
  Cole, and Frenk}]{bower_parameter_2010}
Bower, R.~G., Vernon, I., Goldstein, M., Benson, A.~J., Lacey, C.~G., Baugh,
  C.~M., Cole, S., Frenk, C.~S., 2010. The parameter space of galaxy formation.
  Monthly Notices of the Royal Astronomical Society, no--no.
\newline\urlprefix\url{http://adsabs.harvard.edu/doi/10.1111/j.1365-2966.2010.%
16991.x}

\bibitem[{{Boylan-Kolchin} et~al.(2008){Boylan-Kolchin}, Ma, and
  Quataert}]{boylan-kolchin_dynamical_2008}
{Boylan-Kolchin}, M., Ma, C., Quataert, E., 2008. Dynamical friction and galaxy
  merging time-scales. {MNRAS} 383, 93--101.
\newline\urlprefix\url{http://adsabs.harvard.edu/abs/2008MNRAS.383...93B}

\bibitem[{Bryan and Norman(1998)}]{bryan_statistical_1998}
Bryan, G.~L., Norman, M.~L., Mar. 1998. Statistical properties of {X-Ray}
  clusters: Analytic and numerical comparisons. The Astrophysical Journal 495,
  80.
\newline\urlprefix\url{http://adsabs.harvard.edu/abs/1998ApJ...495...80B}

\bibitem[{Bullock et~al.(2001)Bullock, Kolatt, Sigad, Somerville, Kravtsov,
  Klypin, Primack, and Dekel}]{bullock_profiles_2001}
Bullock, J.~S., Kolatt, T.~S., Sigad, Y., Somerville, R.~S., Kravtsov, A.~V.,
  Klypin, A.~A., Primack, J.~R., Dekel, A., Mar. 2001. Profiles of dark haloes:
  evolution, scatter and environment. Monthly Notices of the Royal Astronomical
  Society 321, 559--575.
\newline\urlprefix\url{http://adsabs.harvard.edu/abs/2001MNRAS.321..559B}

\bibitem[{Chabrier(2001)}]{chabrier_galactic_2001}
Chabrier, G., Jun. 2001. The galactic disk mass budget. i. stellar mass
  function and density. The Astrophysical Journal 554, 1274--1281.
\newline\urlprefix\url{http://adsabs.harvard.edu/abs/2001ApJ...554.1274C}

\bibitem[{Ciotti et~al.(2009)Ciotti, Ostriker, and
  Proga}]{ciotti_feedbackcentral_2009}
Ciotti, L., Ostriker, J.~P., Proga, D., Jul. 2009. Feedback from central black
  holes in elliptical galaxies. i. models with either radiative or mechanical
  feedback but not both. The Astrophysical Journal 699, 89--104.
\newline\urlprefix\url{http://adsabs.harvard.edu/abs/2009ApJ...699...89C}

\bibitem[{Cole et~al.(2000)Cole, Lacey, Baugh, and
  Frenk}]{cole_hierarchical_2000}
Cole, S., Lacey, C.~G., Baugh, C.~M., Frenk, C.~S., Nov. 2000. Hierarchical
  galaxy formation. {MNRAS} 319, 168--204.
\newline\urlprefix\url{http://adsabs.harvard.edu/abs/2000MNRAS.319..168C}

\bibitem[{Cole et~al.(2001)Cole, Norberg, Baugh, Frenk, {Bland-Hawthorn},
  Bridges, Cannon, Colless, Collins, Couch, Cross, Dalton, Propris, and
  Driver}]{cole_2df_2001}
Cole, S., Norberg, P., Baugh, C.~M., Frenk, C.~S., {Bland-Hawthorn}, J.,
  Bridges, T., Cannon, R., Colless, M., Collins, C., Couch, W., Cross, N.,
  Dalton, G., Propris, R.~D., Driver, S.~P., Sep. 2001. The {2dF} galaxy
  redshift survey: near-infrared galaxy luminosity functions. Monthly Notices
  of the Royal Astronomical Society 326, 255--273.
\newline\urlprefix\url{http://adsabs.harvard.edu/abs/2001MNRAS.326..255C}

\bibitem[{Conroy et~al.(2009)Conroy, Gunn, and White}]{conroy_propagation_2009}
Conroy, C., Gunn, J.~E., White, M., Jul. 2009. The propagation of uncertainties
  in stellar population synthesis modeling. i. the relevance of uncertain
  aspects of stellar evolution and the initial mass function to the derived
  physical properties of galaxies. Astrophysical Journal 699, 486--506.
\newline\urlprefix\url{http://adsabs.harvard.edu/abs/2009ApJ...699..486C}

\bibitem[{Croton et~al.(2006)Croton, Springel, White, Lucia, Frenk, Gao,
  Jenkins, Kauffmann, Navarro, and Yoshida}]{croton_many_2006}
Croton, D.~J., Springel, V., White, S. D.~M., Lucia, G.~D., Frenk, C.~S., Gao,
  L., Jenkins, A., Kauffmann, G., Navarro, J.~F., Yoshida, N., 2006. The many
  lives of active galactic nuclei: cooling flows, black holes and the
  luminosities and colours of galaxies. {MNRAS} 365, 11--28.
\newline\urlprefix\url{http://adsabs.harvard.edu/abs/2006MNRAS.365...11C}

\bibitem[{de~Jong and Lacey(2000)}]{de_jong_local_2000}
de~Jong, R.~S., Lacey, C., Dec. 2000. The local space density of {SB-SDM}
  galaxies as function of their scale size, surface brightness, and luminosity.
  The Astrophysical Journal 545, 781--797.
\newline\urlprefix\url{http://adsabs.harvard.edu/abs/2000ApJ...545..781D}

\bibitem[{Edgar(2004)}]{edgar_review_2004}
Edgar, R., Sep. 2004. A review of {Bondi-Hoyle-Lyttleton} accretion. New
  Astronomy Reviews 48~(10), 843--859.

\bibitem[{Efstathiou et~al.(1982)Efstathiou, Lake, and
  Negroponte}]{efstathiou_stability_1982}
Efstathiou, G., Lake, G., Negroponte, J., Jun. 1982. The stability and masses
  of disc galaxies. Monthly Notices of the Royal Astronomical Society 199,
  1069--1088.
\newline\urlprefix\url{http://adsabs.harvard.edu/abs/1982MNRAS.199.1069E}

\bibitem[{Eisenstein and Hu(1999)}]{eisenstein_power_1999}
Eisenstein, D.~J., Hu, W., Jan. 1999. Power spectra for cold dark matter and
  its variants. The Astrophysical Journal 511, 5--15.
\newline\urlprefix\url{http://adsabs.harvard.edu/abs/1999ApJ...511....5E}

\bibitem[{Ferrara et~al.(1999)Ferrara, Bianchi, Cimatti, and
  Giovanardi}]{ferrara_atlas_1999}
Ferrara, A., Bianchi, S., Cimatti, A., Giovanardi, C., Aug. 1999. An atlas of
  monte carlo models of dust extinction in galaxies for cosmological
  applications. Astrophysical Journal Supplement Series 123, 437--445.
\newline\urlprefix\url{http://adsabs.harvard.edu/abs/1999ApJS..123..437F}

\bibitem[{Font et~al.(2010)Font, Benson, Bower, and
  Frenk}]{font_modelingmilky_2010}
Font, A., Benson, A., Bower, R., Frenk, C., 2010. Modeling the milky way
  satellites with {Self-Consistent} reionization. in prep.

\bibitem[{Font et~al.(2008)Font, Bower, {McCarthy}, Benson, Frenk, Helly,
  Lacey, Baugh, and Cole}]{font_colours_2008}
Font, A.~S., Bower, R.~G., {McCarthy}, I.~G., Benson, A.~J., Frenk, C.~S.,
  Helly, J.~C., Lacey, C.~G., Baugh, C.~M., Cole, S., Oct. 2008. The colours of
  satellite galaxies in groups and clusters. {MNRAS} 389, 1619--1629.
\newline\urlprefix\url{http://adsabs.harvard.edu/abs/2008MNRAS.389.1619F}

\bibitem[{Gao et~al.(2008)Gao, Navarro, Cole, Frenk, White, Springel, Jenkins,
  and Neto}]{gao_redshift_2008}
Gao, L., Navarro, J.~F., Cole, S., Frenk, C.~S., White, S. D.~M., Springel, V.,
  Jenkins, A., Neto, A.~F., Jun. 2008. The redshift dependence of the structure
  of massive lambda cold dark matter haloes. {MNRAS} 387, 536--544.
\newline\urlprefix\url{http://adsabs.harvard.edu/abs/2008MNRAS.387..536G}

\bibitem[{Gnedin et~al.(2004)Gnedin, Kravtsov, Klypin, and
  Nagai}]{gnedin_response_2004}
Gnedin, O.~Y., Kravtsov, A.~V., Klypin, A.~A., Nagai, D., Nov. 2004. Response
  of dark matter halos to condensation of baryons: Cosmological simulations and
  improved adiabatic contraction model. Astrophysical Journal 616, 16--26.
\newline\urlprefix\url{http://adsabs.harvard.edu/abs/2004ApJ...616...16G}

\bibitem[{H\"aring and Rix(2004)}]{haring_black_2004}
H\"aring, N., Rix, H., Apr. 2004. On the black hole {Mass-Bulge} mass relation.
  Astrophysical Journal 604, L89--92.
\newline\urlprefix\url{http://adsabs.harvard.edu/abs/2004ApJ...604L..89H}

\bibitem[{Hatton et~al.(2003)Hatton, Devriendt, Ninin, Bouchet, Guiderdoni, and
  Vibert}]{hatton_galics-_2003}
Hatton, S., Devriendt, J. E.~G., Ninin, S., Bouchet, F.~R., Guiderdoni, B.,
  Vibert, D., Jul. 2003. {GALICS-} i. a hybrid n-body/semi-analytic model of
  hierarchical galaxy formation. {MNRAS} 343, 75--106.
\newline\urlprefix\url{http://adsabs.harvard.edu/abs/2003MNRAS.343...75H}

\bibitem[{Heger and Woosley(2002)}]{heger_nucleosynthetic_2002}
Heger, A., Woosley, S.~E., Mar. 2002. The nucleosynthetic signature of
  population {III}. Astrophysical Journal 567, 532--543.
\newline\urlprefix\url{http://adsabs.harvard.edu/abs/2002ApJ...567..532H}

\bibitem[{Helly et~al.(2003)Helly, Cole, Frenk, Baugh, Benson, Lacey, and
  Pearce}]{helly_comparison_2003}
Helly, J.~C., Cole, S., Frenk, C.~S., Baugh, C.~M., Benson, A., Lacey, C.,
  Pearce, F.~R., Feb. 2003. A comparison of gas dynamics in smooth particle
  hydrodynamics and semi-analytic models of galaxy formation. {MNRAS} 338,
  913--925.
\newline\urlprefix\url{http://adsabs.harvard.edu/abs/2003MNRAS.338..913H}

\bibitem[{Henriques and Thomas(2010)}]{henriques_tidal_2010}
Henriques, B. M.~B., Thomas, P.~A., Apr. 2010. Tidal disruption of satellite
  galaxies in a semi-analytic model of galaxy formation. Monthly Notices of the
  Royal Astronomical Society 403, 768--779.
\newline\urlprefix\url{http://adsabs.harvard.edu/abs/2010MNRAS.403..768H}

\bibitem[{Henriques et~al.(2009)Henriques, Thomas, Oliver, and
  Roseboom}]{henriques_monte_2009}
Henriques, B. M.~B., Thomas, P.~A., Oliver, S., Roseboom, I., Jun. 2009. Monte
  carlo markov chain parameter estimation in semi-analytic models of galaxy
  formation. Monthly Notices of the Royal Astronomical Society 396, 535--547.
\newline\urlprefix\url{http://adsabs.harvard.edu/abs/2009MNRAS.396..535H}

\bibitem[{Hernquist(1990)}]{hernquist_analytical_1990}
Hernquist, L., Jun. 1990. An analytical model for spherical galaxies and
  bulges. Astrophysical Journal 356, 359--364.
\newline\urlprefix\url{http://adsabs.harvard.edu/abs/1990ApJ...356..359H}

\bibitem[{Hopkins(2004)}]{hopkins_evolution_2004}
Hopkins, A.~M., Nov. 2004. On the evolution of star-forming galaxies.
  Astrophysical Journal 615, 209--221.
\newline\urlprefix\url{http://adsabs.harvard.edu/abs/2004ApJ...615..209H}

\bibitem[{Jiang et~al.(2008)Jiang, Jing, Faltenbacher, Lin, and
  Li}]{jiang_fitting_2008}
Jiang, C.~Y., Jing, Y.~P., Faltenbacher, A., Lin, W.~P., Li, C., Mar. 2008. A
  fitting formula for the merger timescale of galaxies in hierarchical
  clustering. {ApJ} 675, 1095--1105.
\newline\urlprefix\url{http://adsabs.harvard.edu/abs/2008ApJ...675.1095J}

\bibitem[{Kauffmann et~al.(1993)Kauffmann, White, and
  Guiderdoni}]{kauffmann_formation_1993}
Kauffmann, G., White, S. D.~M., Guiderdoni, B., Sep. 1993. The formation and
  evolution of galaxies within merging dark matter haloes. {MNRAS} 264, 201.
\newline\urlprefix\url{http://adsabs.harvard.edu/abs/1993MNRAS.264..201K}

\bibitem[{Kennicutt(1983)}]{kennicutt_rate_1983}
Kennicutt, R.~C., Sep. 1983. The rate of star formation in normal disk
  galaxies. The Astrophysical Journal 272, 54--67.
\newline\urlprefix\url{http://adsabs.harvard.edu/abs/1983ApJ...272...54K}

\bibitem[{Kistler et~al.(2009)Kistler, Yuksel, Beacom, Hopkins, and
  Wyithe}]{kistler_star_2009}
Kistler, M.~D., Yuksel, H., Beacom, J.~F., Hopkins, A.~M., Wyithe, J. S.~B.,
  Jun. 2009. The star formation rate in the reionization era as indicated by
  gamma-ray bursts. 0906.0590.
\newline\urlprefix\url{http://arxiv.org/abs/0906.0590}

\bibitem[{Komatsu et~al.(2010)Komatsu, Smith, Dunkley, Bennett, Gold, Hinshaw,
  Jarosik, Larson, Nolta, Page, Spergel, Halpern, Hill, Kogut, Limon, Meyer,
  Odegard, Tucker, Weiland, Wollack, and Wright}]{komatsu_seven-year_2010}
Komatsu, E., Smith, K.~M., Dunkley, J., Bennett, C.~L., Gold, B., Hinshaw, G.,
  Jarosik, N., Larson, D., Nolta, M.~R., Page, L., Spergel, D.~N., Halpern, M.,
  Hill, R.~S., Kogut, A., Limon, M., Meyer, S.~S., Odegard, N., Tucker, G.~S.,
  Weiland, J.~L., Wollack, E., Wright, E.~L., Jan. 2010. {Seven-Year} wilkinson
  microwave anisotropy probe {(WMAP)} observations: Cosmological
  interpretation. {http://adsabs.harvard.edu/abs/2010arXiv1001.4538K}.
\newline\urlprefix\url{http://adsabs.harvard.edu/abs/2010arXiv1001.4538K}

\bibitem[{Kroupa(2001)}]{kroupa_variation_2001}
Kroupa, P., Apr. 2001. On the variation of the initial mass function. Monthly
  Notices of the Royal Astronomical Society 322, 231--246.
\newline\urlprefix\url{http://adsabs.harvard.edu/abs/2001MNRAS.322..231K}

\bibitem[{Lacey and Cole(1993)}]{lacey_merger_1993}
Lacey, C., Cole, S., Jun. 1993. Merger rates in hierarchical models of galaxy
  formation. Monthly Notices of the Royal Astronomical Society 262, 627--649.
\newline\urlprefix\url{http://adsabs.harvard.edu/abs/1993MNRAS.262..627L}

\bibitem[{Lanzoni et~al.(2005)Lanzoni, Guiderdoni, Mamon, Devriendt, and
  Hatton}]{lanzoni_galics-_2005}
Lanzoni, B., Guiderdoni, B., Mamon, G.~A., Devriendt, J., Hatton, S., Aug.
  2005. {GALICS-} {VI.} modelling hierarchical galaxy formation in clusters.
  {MNRAS} 361, 369--384.
\newline\urlprefix\url{http://adsabs.harvard.edu/abs/2005MNRAS.361..369L}

\bibitem[{Leitherer et~al.(1992)Leitherer, Robert, and
  Drissen}]{leitherer_deposition_1992}
Leitherer, C., Robert, C., Drissen, L., Dec. 1992. Deposition of mass,
  momentum, and energy by massive stars into the interstellar medium. {ApJ}
  401, 596--617.
\newline\urlprefix\url{http://adsabs.harvard.edu/abs/1992ApJ...401..596L}

\bibitem[{Lu et~al.(2010)Lu, Kere\v{s}, Katz, Mo, Fardal, and
  Weinberg}]{lu_algorithms_2010}
Lu, Y., Kere\v{s}, D., Katz, N., Mo, H.~J., Fardal, M., Weinberg, M.~D., Aug.
  2010. On the algorithms of radiative cooling in semi-analytic models.
  {http://adsabs.harvard.edu/abs/2010arXiv1008.1075L}.
\newline\urlprefix\url{http://adsabs.harvard.edu/abs/2010arXiv1008.1075L}

\bibitem[{Madau(1995)}]{madau_radiative_1995}
Madau, P., Mar. 1995. Radiative transfer in a clumpy universe: The colors of
  high-redshift galaxies. The Astrophysical Journal 441, 18--27.
\newline\urlprefix\url{http://adsabs.harvard.edu/abs/1995ApJ...441...18M}

\bibitem[{Meier(2001)}]{meier_association_2001}
Meier, D.~L., Feb. 2001. The association of jet production with geometrically
  thick accretion flows and black hole rotation. The Astrophysical Journal 548,
  L9--L12.
\newline\urlprefix\url{http://adsabs.harvard.edu/abs/2001ApJ...548L...9M}

\bibitem[{Meiksin(2006)}]{meiksin_colour_2006}
Meiksin, A., Jan. 2006. Colour corrections for high-redshift objects due to
  intergalactic attenuation. Monthly Notices of the Royal Astronomical Society
  365, 807--812.
\newline\urlprefix\url{http://adsabs.harvard.edu/abs/2006MNRAS.365..807M}

\bibitem[{Miller and Scalo(1979)}]{miller_initial_1979}
Miller, G.~E., Scalo, J.~M., Nov. 1979. The initial mass function and stellar
  birthrate in the solar neighborhood. The Astrophysical Journal Supplement
  Series 41, 513--547.
\newline\urlprefix\url{http://adsabs.harvard.edu/abs/1979ApJS...41..513M}

\bibitem[{Monaco et~al.(2007)Monaco, Fontanot, and
  Taffoni}]{monaco_morgana_2007}
Monaco, P., Fontanot, F., Taffoni, G., Mar. 2007. The {MORGANA} model for the
  rise of galaxies and active nuclei. {MNRAS} 375, 1189--1219.
\newline\urlprefix\url{http://adsabs.harvard.edu/abs/2007MNRAS.375.1189M}

\bibitem[{Nagashima et~al.(2005)Nagashima, Lacey, Baugh, Frenk, and
  Cole}]{nagashima_metal_2005}
Nagashima, M., Lacey, C.~G., Baugh, C.~M., Frenk, C.~S., Cole, S., Apr. 2005.
  The metal enrichment of the intracluster medium in hierarchical galaxy
  formation models. Monthly Notices of the Royal Astronomical Society 358,
  1247--1266.
\newline\urlprefix\url{http://adsabs.harvard.edu/abs/2005MNRAS.358.1247N}

\bibitem[{Narayan and Yi(1994)}]{narayan_advection-dominated_1994}
Narayan, R., Yi, I., Jun. 1994. Advection-dominated accretion: A self-similar
  solution. Astrophysical Journal 428, L13--L16.
\newline\urlprefix\url{http://adsabs.harvard.edu/abs/1994ApJ...428L..13N}

\bibitem[{Navarro et~al.(1997)Navarro, Frenk, and
  White}]{navarro_universal_1997}
Navarro, J.~F., Frenk, C.~S., White, S. D.~M., Dec. 1997. A universal density
  profile from hierarchical clustering. {ApJ} 490, 493.
\newline\urlprefix\url{http://adsabs.harvard.edu/abs/1997ApJ...490..493N}

\bibitem[{Norberg et~al.(2002)Norberg, Cole, Baugh, Frenk, Baldry,
  {Bland-Hawthorn}, Bridges, Cannon, Colless, Collins, Couch, Cross, Dalton,
  Propris, Driver, Efstathiou, Ellis, Glazebrook, Jackson, Lahav, Lewis,
  Lumsden, Maddox, Madgwick, Peacock, Peterson, Sutherland, and
  Taylor}]{norberg_2df_2002}
Norberg, P., Cole, S., Baugh, C.~M., Frenk, C.~S., Baldry, I.,
  {Bland-Hawthorn}, J., Bridges, T., Cannon, R., Colless, M., Collins, C.,
  Couch, W., Cross, N. J.~G., Dalton, G., Propris, R.~D., Driver, S.~P.,
  Efstathiou, G., Ellis, R.~S., Glazebrook, K., Jackson, C., Lahav, O., Lewis,
  I., Lumsden, S., Maddox, S., Madgwick, D., Peacock, J.~A., Peterson, B.~A.,
  Sutherland, W., Taylor, K., Nov. 2002. The {2dF} galaxy redshift survey: the
  {bJ-band} galaxy luminosity function and survey selection function. Monthly
  Notices of the Royal Astronomical Society 336, 907--931.
\newline\urlprefix\url{http://adsabs.harvard.edu/abs/2002MNRAS.336..907N}

\bibitem[{Ostriker et~al.(2010)Ostriker, Choi, Ciotti, Novak, and
  Proga}]{ostriker_momentum_2010}
Ostriker, J.~P., Choi, E., Ciotti, L., Novak, G.~S., Proga, D., Apr. 2010.
  Momentum driving: which physical processes dominate {AGN} feedback?
  {http://adsabs.harvard.edu/abs/2010arXiv1004.2923O}.
\newline\urlprefix\url{http://adsabs.harvard.edu/abs/2010arXiv1004.2923O}

\bibitem[{Parkinson et~al.(2008)Parkinson, Cole, and
  Helly}]{parkinson_generating_2008}
Parkinson, H., Cole, S., Helly, J., Jan. 2008. Generating dark matter halo
  merger trees. Monthly Notices of the Royal Astronomical Society 383,
  557--564.
\newline\urlprefix\url{http://adsabs.harvard.edu/abs/2008MNRAS.383..557P}

\bibitem[{Percival(2005)}]{percival_cosmological_2005}
Percival, W.~J., Dec. 2005. Cosmological structure formation in a homogeneous
  dark energy background. Astronomy and Astrophysics 443, 819--830.
\newline\urlprefix\url{http://adsabs.harvard.edu/abs/2005A\%26A...443..819P}

\bibitem[{Pizagno et~al.(2007)Pizagno, Prada, Weinberg, Rix, Pogge, Grebel,
  Harbeck, Blanton, Brinkmann, and Gunn}]{pizagno_tully-fisher_2007}
Pizagno, J., Prada, F., Weinberg, D.~H., Rix, H., Pogge, R.~W., Grebel, E.~K.,
  Harbeck, D., Blanton, M., Brinkmann, J., Gunn, J.~E., Sep. 2007. The
  {Tully-Fisher} relation and its residuals for a broadly selected sample of
  galaxies. Astronomical Journal 134, 945--972.
\newline\urlprefix\url{http://adsabs.harvard.edu/abs/2007AJ....134..945P}

\bibitem[{Power et~al.(2010)Power, Baugh, and Lacey}]{power_redshift_2010}
Power, C., Baugh, C.~M., Lacey, C.~G., Jul. 2010. The redshift evolution of the
  mass function of cold gas in hierarchical galaxy formation models. Monthly
  Notices of the Royal Astronomical Society 406, 43--59.
\newline\urlprefix\url{http://adsabs.harvard.edu/abs/2010MNRAS.406...43P}

\bibitem[{Press and Schechter(1974)}]{press_formation_1974}
Press, W.~H., Schechter, P., Feb. 1974. Formation of galaxies and clusters of
  galaxies by {Self-Similar} gravitational condensation. Astrophysical Journal
  187, 425--438.
\newline\urlprefix\url{http://adsabs.harvard.edu/abs/1974ApJ...187..425P}

\bibitem[{Rezzolla et~al.(2008)Rezzolla, Barausse, Dorband, Pollney, Reisswig,
  Seiler, and Husa}]{rezzolla_final_2008}
Rezzolla, L., Barausse, E., Dorband, E.~N., Pollney, D., Reisswig, C., Seiler,
  J., Husa, S., Aug. 2008. Final spin from the coalescence of two black holes.
  Physical Review D 78, 44002.
\newline\urlprefix\url{http://adsabs.harvard.edu/abs/2008PhRvD..78d4002R}

\bibitem[{Salpeter(1955)}]{salpeter_luminosity_1955}
Salpeter, E.~E., Jan. 1955. The luminosity function and stellar evolution. The
  Astrophysical Journal 121, 161.
\newline\urlprefix\url{http://adsabs.harvard.edu/abs/1955ApJ...121..161S}

\bibitem[{Scalo(1986)}]{scalo_stellar_1986}
Scalo, J.~M., May 1986. The stellar initial mass function. Fundamentals of
  Cosmic Physics 11, 1--278.
\newline\urlprefix\url{http://adsabs.harvard.edu/abs/1986FCPh...11....1S}

\bibitem[{Shakura and Sunyaev(1973)}]{shakura_black_1973}
Shakura, N.~I., Sunyaev, R.~A., 1973. Black holes in binary systems.
  observational appearance. Astronomy and Astrophysics 24, 337--355.
\newline\urlprefix\url{http://ukads.nottingham.ac.uk/abs/1973A\%26A....24..337%
S}

\bibitem[{Shapley and Curtis(1921)}]{shapley__1921}
Shapley, H., Curtis, H.~D., 1921. Tech. Rep.~2.

\bibitem[{Sheth et~al.(2001)Sheth, Mo, and Tormen}]{sheth_ellipsoidal_2001}
Sheth, R.~K., Mo, H.~J., Tormen, G., May 2001. Ellipsoidal collapse and an
  improved model for the number and spatial distribution of dark matter haloes.
  Monthly Notices of the Royal Astronomical Society 323, 1--12.
\newline\urlprefix\url{http://adsabs.harvard.edu/abs/2001MNRAS.323....1S}

\bibitem[{Somerville et~al.(2008)Somerville, Hopkins, Cox, Robertson, and
  Hernquist}]{somerville_semi-analytic_2008}
Somerville, R.~S., Hopkins, P.~F., Cox, T.~J., Robertson, B.~E., Hernquist, L.,
  Dec. 2008. A semi-analytic model for the co-evolution of galaxies, black
  holes and active galactic nuclei. Monthly Notices of the Royal Astronomical
  Society 391, 481--506.
\newline\urlprefix\url{http://adsabs.harvard.edu/abs/2008MNRAS.391..481S}

\bibitem[{Springel et~al.(2005)Springel, White, Jenkins, Frenk, Yoshida, Gao,
  Navarro, Thacker, Croton, Helly, Peacock, Cole, Thomas, Couchman, Evrard,
  Colberg, and Pearce}]{springel_simulations_2005}
Springel, V., White, S. D.~M., Jenkins, A., Frenk, C.~S., Yoshida, N., Gao, L.,
  Navarro, J., Thacker, R., Croton, D., Helly, J., Peacock, J.~A., Cole, S.,
  Thomas, P., Couchman, H., Evrard, A., Colberg, J., Pearce, F., Jun. 2005.
  Simulations of the formation, evolution and clustering of galaxies and
  quasars. Nature 435~(7042), 629--636.
\newline\urlprefix\url{http://dx.doi.org/10.1038/nature03597}

\bibitem[{Stringer et~al.(2010)Stringer, Brooks, Benson, and
  Governato}]{stringer_analytic_2010}
Stringer, M.~J., Brooks, A.~M., Benson, A.~J., Governato, F., Jan. 2010.
  Analytic and numerical realisations of a disk galaxy.
  {http://adsabs.harvard.edu/abs/2010arXiv1001.0594S}.
\newline\urlprefix\url{http://adsabs.harvard.edu/abs/2010arXiv1001.0594S}

\bibitem[{Tinker et~al.(2010)Tinker, Robertson, Kravtsov, Klypin, Warren,
  Yepes, and Gottlober}]{tinker_large_2010}
Tinker, J.~L., Robertson, B.~E., Kravtsov, A.~V., Klypin, A., Warren, M.~S.,
  Yepes, G., Gottlober, S., Jan. 2010. The large scale bias of dark matter
  halos: Numerical calibration and model tests.
  {http://adsabs.harvard.edu/abs/2010arXiv1001.3162T}.
\newline\urlprefix\url{http://adsabs.harvard.edu/abs/2010arXiv1001.3162T}

\bibitem[{Wechsler et~al.(2002)Wechsler, Bullock, Primack, Kravtsov, and
  Dekel}]{wechsler_concentrations_2002}
Wechsler, R.~H., Bullock, J.~S., Primack, J.~R., Kravtsov, A.~V., Dekel, A.,
  Mar. 2002. Concentrations of dark halos from their assembly histories. The
  Astrophysical Journal 568, 52--70.
\newline\urlprefix\url{http://adsabs.harvard.edu/abs/2002ApJ...568...52W}

\bibitem[{Weinmann et~al.(2006)Weinmann, van~den Bosch, Yang, and
  Mo}]{weinmann_properties_2006}
Weinmann, S.~M., van~den Bosch, F.~C., Yang, X., Mo, H.~J., Feb. 2006.
  Properties of galaxy groups in the sloan digital sky survey - i. the
  dependence of colour, star formation and morphology on halo mass. Monthly
  Notices of the Royal Astronomical Society 366, 2--28.
\newline\urlprefix\url{http://adsabs.harvard.edu/abs/2006MNRAS.366....2W}

\bibitem[{White and Frenk(1991)}]{white_galaxy_1991}
White, S. D.~M., Frenk, C.~S., Sep. 1991. Galaxy formation through hierarchical
  clustering. Astrophysical Journal 379, 52--79.
\newline\urlprefix\url{http://adsabs.harvard.edu/abs/1991ApJ...379...52W}

\bibitem[{White and Rees(1978)}]{white_core_1978}
White, S. D.~M., Rees, M.~J., May 1978. Core condensation in heavy halos - a
  two-stage theory for galaxy formation and clustering. {MNRAS} 183, 341--358.
\newline\urlprefix\url{http://adsabs.harvard.edu/abs/1978MNRAS.183..341W}

\bibitem[{Yoshida et~al.(2002)Yoshida, Stoehr, Springel, and
  White}]{yoshida_gas_2002}
Yoshida, N., Stoehr, F., Springel, V., White, S. D.~M., Sep. 2002. Gas cooling
  in simulations of the formation of the galaxy population. Monthly Notices of
  the Royal Astronomical Society 335, 762--772.
\newline\urlprefix\url{http://adsabs.harvard.edu/abs/2002MNRAS.335..762Y}

\bibitem[{Zwaan et~al.(2005)Zwaan, Meyer, {Staveley-Smith}, and
  Webster}]{zwaan_hipass_2005}
Zwaan, M.~A., Meyer, M.~J., {Staveley-Smith}, L., Webster, R.~L., May 2005. The
  {HIPASS} catalogue: {$\Omega$\_{\rm} {HI}} and environmental effects on the
  {HI} mass function of galaxies. Monthly Notices of the Royal Astronomical
  Society 359, L30--L34.
\newline\urlprefix\url{http://adsabs.harvard.edu/abs/2005MNRAS.359L..30Z}

\end{thebibliography}

\end{document}